\documentclass[preprint,12pt]{elsarticle}

\usepackage{natbib}

\usepackage{mydef}
\usepackage{amssymb}
\usepackage{amsmath}
\usepackage{graphicx}
\usepackage{multirow}
\usepackage{mdframed}
\usepackage{enumitem}
\usepackage{amsthm}
\usepackage{afterpage}
\usepackage{setspace}

\usepackage{mdframed}
\usepackage{booktabs}
\usepackage{listings}
\usepackage{caption}
\usepackage[hidelinks]{hyperref}
\definecolor{tdc_color}{RGB}{10,128,122}
\definecolor{dc_color}{RGB}{230, 245, 244}
\definecolor{ds_color}{RGB}{195, 230, 227}
\definecolor{ms_color}{RGB}{150, 214, 209}

\theoremstyle{definition}
\newtheorem{definition}{Definition}

\hypersetup{
    colorlinks=true,
    linkcolor=tdc_color,
    filecolor=tdc_color,      
    urlcolor=tdc_color,
    citecolor=tdc_color
}

\journal{Information Fusion}

\begin{document}

\newlist{priorlist_gene}{enumerate}{1}
\setlist[priorlist_gene,1]{label=Prior G\arabic*, ref=Prior G\arabic*}  %
\newlist{priorlist_protein}{enumerate}{1}
\setlist[priorlist_protein,1]{label=Prior P\arabic*, ref=Prior P\arabic*}  %
\newlist{priorlist_singlecell}{enumerate}{1}
\setlist[priorlist_singlecell,1]{label=Prior S\arabic*, ref=Prior S\arabic*}  %

\begin{frontmatter}

\title{
  A Review of BioTree Construction in the Context of Information Fusion: Priors, Methods, Applications and Trends
  }

\author[1,2]{Zelin Zang}
\author[2]{Yongjie Xu} %
\author[2]{Chenrui Duan} %
\author[2]{Yue Yuan} %
\author[1,3]{Jinlin Wu} %
\author[1,3,4]{Zhen Lei$^\dagger$}
\author[2]{Stan Z. Li$^\dagger$} %

\affiliation[1]{Centre for Artificial Intelligence and Robotics (CAIR); HKISI-CAS}
\affiliation[2]{AI Division; School of Engineering; Westlake University; Hangzhou; 310030; China}
\affiliation[3]{State Key Laboratory of Multimodal Artificial Intelligence Systems (MAIS); Institute of Automation; Chinese Academy of Sciences (CASIA)}
\affiliation[4]{School of Artificial Intelligence; University of Chinese Academy of Sciences (UCAS)}

\begin{abstract}

\normalsize
\begin{spacing}{0.9}
{
    \footnotesize
    \vspace{-0.2cm}
    Biological tree (BioTree) analysis is a foundational tool in biology, enabling the exploration of evolutionary and differentiation relationships among organisms, genes, and cells. Traditional tree construction methods, while instrumental in early research, face significant challenges in handling the growing complexity and scale of modern biological data, particularly in integrating multimodal datasets. Advances in deep learning (DL) offer transformative opportunities by enabling the fusion of biological prior knowledge with data-driven models. These approaches address key limitations of traditional methods, facilitating the construction of more accurate and interpretable BioTrees.
    This review highlights critical biological priors essential for phylogenetic and differentiation tree analyses and explores strategies for integrating these priors into DL models to enhance accuracy and interpretability. Additionally, the review systematically examines commonly used data modalities and databases, offering a valuable resource for developing and evaluating multimodal fusion models. Traditional tree construction methods are critically assessed, focusing on their biological assumptions, technical limitations, and scalability issues. Recent advancements in DL-based tree generation methods are reviewed, emphasizing their innovative approaches to multimodal integration and prior knowledge incorporation. 
    Finally, the review discusses diverse applications of BioTrees in various biological disciplines, from phylogenetics to developmental biology, and outlines future trends in leveraging DL to advance BioTree research. By addressing the challenges of data complexity and prior knowledge integration, this review aims to inspire interdisciplinary innovation at the intersection of biology and DL. 
}
\end{spacing}

\end{abstract}

\begin{graphicalabstract}
  \begin{figure*}[h]
    \begin{center}
      \includegraphics[width=0.81\linewidth]{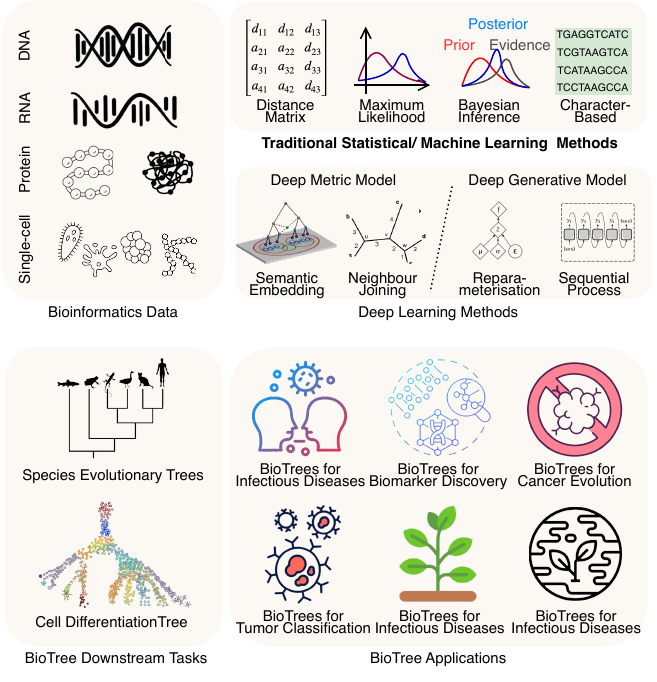}
    \end{center}
    \caption{\textbf{Graphical Abstract. Overview of methodologies, data types, and applications for biological tree-based research.} Summary of bioinformatics data types, methodological advancements, and applications reviewed in this study. Covers traditional and deep learning methods for biological tree construction and their roles in species evolution, cell differentiation, and disease research.}
    \label{fig_intro}
  \end{figure*}
\end{graphicalabstract}

\begin{highlights}
  \item Biological tree analysis reveals relationships among organisms, genes, and cells.
  \item Traditional methods struggle with large-scale multimodal data.
  \item DL integrates biological priors and multimodal data, enhancing accuracy.
  \item Explores advancements, applications, and future trends in BioTree research.
\end{highlights}

\end{frontmatter}

\clearpage
\tableofcontents 
\clearpage

\section{Backgrounds} \label{sec_background}
Biological tree (BioTree) analysis methods are fundamental tools in biological research, playing a crucial role in revealing evolutionary and differentiation relationships among organisms \citep{hedges2002origin, schiotz2024serial}, genes \citep{yamamoto2023autophagy, martinez2020compendium}, and cells \citep{leone2020metabolism, armingol2024diversification}. These methods are widely used in phylogenetics, developmental biology \citep{basra2024cotton}, and ecology \citep{catford2022addressing}, helping scientists gain a deeper understanding of the origins and maintenance mechanisms of biodiversity~(as shwon in Figure.~\ref{fig_intro}). In phylogenetics, BioTree analysis involves constructing phylogenetic trees to uncover evolutionary relationships between organisms, providing a basis for taxonomists to classify and name species \citep{Cavalli1967, Nei1987, dylus2024inference, grigoriadis2024conipher}. In \textit{developmental biology and stem cell research}, differentiation tree analysis helps researchers trace cell differentiation processes, elucidating how stem cells generate various specialized cell types \citep{trapnell2014dynamics, domcke2023reference}.
{\color{black}Moreover, BioTree analysis is not only central to species classification but also pivotal in advancing modern biomedical research, such as in deciphering disease mechanisms, facilitating cell regeneration, and tailoring personalized medicine strategies. In an era marked by an unprecedented surge in biological data complexity and volume, the limitations of traditional methods become increasingly evident, necessitating the development of more efficient and scalable BioTree construction techniques.}

\begin{figure*}[h]
\begin{center}
    \includegraphics[width=0.9\linewidth]{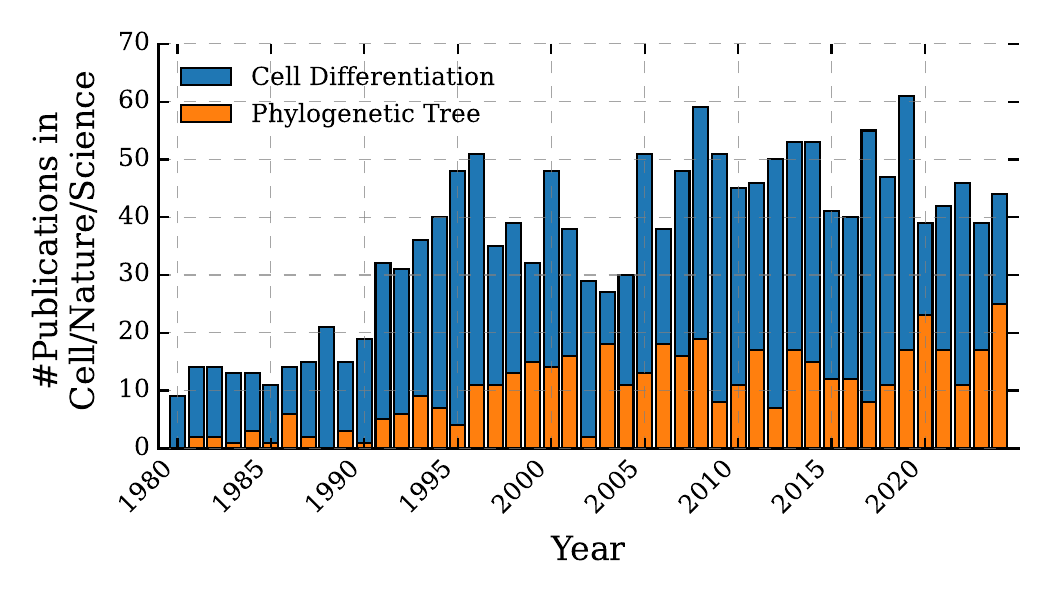}
\end{center}
\vspace{-1cm}
\caption{\textbf{\color{black}Number of publications in Cell, Nature, and Science related to cell differentiation and phylogenetic tree from 1980 to 2025.} \color{black}Publications were retrieved from the Web of Science Core Collection using the following search queries:  
(a) \textbf{Phylogenetic Tree} (469 publications): TS=(``phylogenetic tree" OR ``evolutionary tree" OR ``tree of life" OR ``phylogenetic analysis" OR ``tree-based" OR ``phylogenetic reconstruction" OR ``phylogenetic relationship" OR ``evolutionary relationships").
(b) \textbf{Cell Differentiation} (1689 publications): TS=(``cell differentiation" OR ``cellular differentiation" OR ``differentiation of cells" OR ``trajectory inference" OR ``lineage inference" OR ``pseudotime inference" OR ``cell lineage" OR ``cell fate").
The blue bars represent publications related to cell differentiation, while the orange bars represent those related to phylogenetic tree. The data shows an increasing trend in both fields, with cell differentiation seeing a more pronounced growth.}
\label{fig_intro_fig}
\end{figure*}

\begin{figure*}[h]
    \begin{center}
        \includegraphics[width=0.9\linewidth]{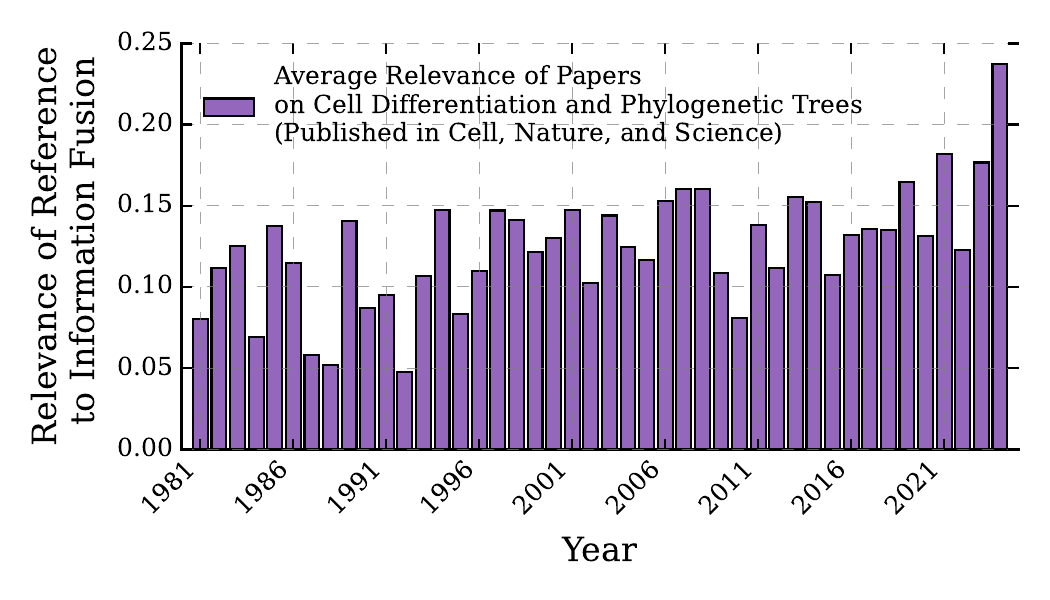}
    \end{center}
    \vspace{-1cm}
    \caption{\textbf{\color{black}Relevance Analysis of Papers on Cell Differentiation and Phylogenetic Trees to ``Information Fusion" Topic~(Published in Cell, Nature, and Science).} \color{black} This analysis evaluates the relevance of papers to the topic of ``Information Fusion" using the DeepSeek-70B large language model. Each bar represents the average relevance score for papers published in a given year, showcasing trends in how research aligns with the ``Information Fusion" theme over time. Relevance score `0' means the paper is not relevant to the topic, while `1' indicates high relevance. The code for DeepSeek-70B analysis is available at \url{https://github.com/zangzelin/code_info_fusion_biotree}.}
    \label{fig_intro_fig_re}
    \end{figure*}

{\color{black} To further substantiate that BioTree methods are gaining increasing attention in the mainstream scientific community, we conducted a bibliometric analysis of publications in leading scientific journals. By systematically searching through these journals, we identified over 2,000 research articles directly related to BioTree methodologies~(in the supplementary meterial). The annual distribution of these publications is shown in Figure.\ref{fig_intro_fig} , demonstrating a steady growth in interest over the past decade. Moreover, the relationship between BioTree methods and the field of information fusion has been strengthening in recent years. By leveraging the large language model~(DeepSeekR1 70B~\cite{guo2025deepseek}), we analyzed the relevance of these articles to information fusion. The results, summarized in Figure.~\ref{fig_intro_fig_re} , reveal a significant proportion of studies integrating BioTree approaches with advanced data fusion techniques.  This trend underscores the pivotal role of BioTree methods in synthesizing and interpreting complex biological datasets~\cite{nagalakshmi2008transcriptional}, further solidifying their importance in modern scientific research~\cite{svensson2018single,xia2019spatial, subramanian2020multiomics}. }

However, traditional tree construction methods, while instrumental in early research, have limitations that are increasingly apparent \citep{Delsuc2019, chen2023tbtools}. In \textit{phylogenetic analyses}, traditional methods perform well on small-scale datasets \citep{huelsenbeck1996combining, wiens2006missing}. However, as modern biological data grow in size and complexity, these methods struggle with accuracy and efficiency due to reliance on heuristic algorithms and predefined modeling assumptions \citep{Yang2012, Szollosi2020}. For \textit{differentiation analyses} in cell differentiation processes, current methods primarily rely on data representation and employ dimensionality reduction and visualization methods for lineage inference \citep{Stuart2019, wang2023theoretical}. While these visualization-based methods provide rough estimates of developmental lineages, they are inadequate for generating accurate tree structures and performing downstream tasks such as target discovery, especially when dealing with multimodal and temporal data \citep{Macaulay2017, zheng2024multi, Wagner2020}.

Two critical challenges for the further development of BioTree analysis are as follows,
\textbf{(a) How to fuse biological prior knowledge with data-driven learning approaches.} The construction of BioTrees heavily depends on biological prior knowledge, such as evolutionary laws and genomic functional modules. This prior knowledge provides biologically meaningful constraints for models. {\color{black}One major challenge lies in effectively integrating these rich biological priors into deep learning models, thereby enhancing both the interpretability and accuracy of the resulting BioTrees while maintaining model flexibility.}
\textbf{(b) How to effectively integrate information from multiple data modalities.} Modern high-throughput technologies produce multimodal data with rich complementarities and complex correlations. These data modalities often exhibit inconsistent dimensions, varying noise levels, and semantic heterogeneity. {\color{black}Addressing this challenge requires the development of unified frameworks capable of reconciling the diverse characteristics of multimodal data—a crucial step for advancing research in genomics, transcriptomics, and cell differentiation pathways, and for overcoming existing research bottlenecks.}

The rapid advancement of {\color{black} DL in recent years \citep{angermueller2016deep, ma2018using, stahl2023protein, hong2023cross, li2024casformer} offers new opportunities to address these challenges. DL models have the potential to incorporate biological prior knowledge into data-driven methods through well-designed loss functions \citep{xu2023dm, kathail2024current} and techniques like knowledge embedding \citep{elhamod2023discovering, chen2024simba},} graph neural networks \citep{zhou2023phylogfn, duanphylogen, peng2023stgnnks}, and attention mechanisms \citep{szalata2024transformers, ly2024treeformer}. These approaches enhance interpretability and accuracy by embedding complex biological priors \citep{han2018openke}. Additionally, DL excels at handling multimodal information fusion \citep{ma2023multimodal}, offering sophisticated methods to integrate diverse data modalities despite differences in dimensionality, noise levels, and semantics. Models like multimodal autoencoders and transformers facilitate unified representations of heterogeneous data, enabling comprehensive analysis in BioTree construction. Notably, DL enables phylogenetic tree and differentiation tree problems to be abstracted into a unified scientific framework. Despite focusing on different scales—phylogenetic trees on macro-level evolutionary relationships and differentiation trees on micro-level cell pathways—they can be addressed using similar models and methodologies. This unification provides a solid foundation for integrating multimodal data and biological prior knowledge, offering new perspectives for BioTree analysis.
To better understand and analyze this emerging trend, we present this review, which comprehensively explores the intersection of DL and BioTree analysis, focusing on the integration of biological prior knowledge and multimodal data fusion~(as shwon in Figure.~\ref{fig_intro}). The main contributions of this review are as follows.
\begin{enumerate}
    \vspace{-0.3cm}
    \item \textbf{Systematically review commonly used data modalities and databases.} We systematically review the data formats and databases commonly used in BioTree analysis, providing comprehensive data resources for testing and developing new information fusion models (Section~\ref{sec_notions} \& Section~\ref{sec_dataset}).
    \vspace{-0.3cm}
    \item \textbf{Summarize the key biological prior knowledge in BioTree analysis} To foster interdisciplinary understanding between DL researchers and biologists, we first summarize the commonly used biological prior knowledge in phylogenetic and differentiation tree analyses, helping to establish a deeper interdisciplinary foundation (Section~\ref{sec_prior}).
    \vspace{-0.3cm}
    \item \textbf{Critically analyze traditional BioTree construction methods.} We conduct a comprehensive review of traditional tree generation methods, analyzing their underlying biological priors, technical solutions, and characteristics, and summarizing their limitations in practical applications (Section~\ref{sec_classical}).
    \vspace{-0.3cm}
    \item \textbf{Review DL-based BioTree construction methods.} We review current DL-based tree generation methods, summarizing recent advancements and existing challenges, providing a holistic perspective on current research directions (Section~\ref{sec_deep}).
    \vspace{-0.3cm}
    \item \textbf{Summarize the extensive applications of BioTrees.} We summarize the broad applications of BioTrees, highlighting their importance in phylogenetics, developmental biology, medicine, and ecology (Section~\ref{sec_applications}).
    \vspace{-0.3cm}
    \item \textbf{Discuss future research directions.} Finally, we discuss potential future directions for using DL in BioTree research, proposing possible research methods and trends to guide further exploration in this field (Section~\ref{sec_limitations}).
\end{enumerate}

\section{Fundamental Concepts of BioTree Construction} \label{sec_notions}
In order to provide a solid foundation for the subsequent in-depth discussion on the fusion of biological prior knowledge with multimodal data in BioTree construction, we begin with an overview of the key notations and basic concepts used in BioTree analysis. These basics are essential for understanding the intricacies of the subsequent chapters. 

\subsection{Fundamental Data Types in BioTree Construction}

Multimodal biological data play a crucial role in constructing and analyzing BioTrees and provide the raw materials necessary for effective information fusion. In this subsection, we introduce the essential data types commonly used in BioTree analysis, including gene sequences, protein sequences, RNA sequences, morphological characteristics, and single-cell data.

\begin{mdframed}[hidealllines=true,backgroundcolor=tdc_color!5]
    \begin{itemize}
        \label{term_Data}
        \setlength\itemsep{0em}
        \renewcommand{\thempfootnote}{$\star$}
        \item \textit{Gene Sequences:} Gene sequences are the order of nucleotides in DNA or RNA that encode genetic information. They are one of the most commonly used data types in phylogenetic analysis \citep{Sanger1977,Li2017}.
        \item \textit{Protein Sequences:} Protein sequences are chains of amino acids that build and regulate physiological processes in organisms. They are critical for studying the evolution of protein functions \citep{Doolittle1981,Alberts2002}.
        \item \textit{RNA Sequences:} RNA sequences are the nucleotide sequences in RNA molecules that convey and regulate genetic information, particularly significant in studying gene expression regulation and non-coding RNA \citep{Sharp1985,Cech2014}.
        \item \textit{Morphological Characteristics:} Morphological characteristics refer to the physical or structural traits of organisms, often used in phenotypic studies and classification within phylogenetic analysis \citep{Hennig1966,Rieppel1988}.
        \item \textit{Single-Cell Data:} Single-cell data are sequencing or analytical data obtained from individual cells, typically used to study cell differentiation, development processes, and the cellular basis of diseases \citep{stuart2019comprehensive}.
    \end{itemize}
\end{mdframed}

\subsection{Fundamental Algorithms and Models in BioTree Construction}

The construction and analysis of BioTrees require various algorithms and models that contribute to the accuracy and efficiency of tree construction. In this subsection, we discuss key algorithms and models used in phylogenetic studies, such as heuristic algorithms, maximum likelihood methods, Bayesian inference, deep learning models, and clustering algorithms.

\begin{mdframed}[hidealllines=true,backgroundcolor=tdc_color!5]
    \begin{itemize}
        \label{term_Algorithms}
        \setlength\itemsep{0em}
        \renewcommand{\thempfootnote}{$\star$}
        \item \textit{Heuristic Algorithms:} Heuristic algorithms are optimization methods based on empirical rules, often used to quickly generate approximate solutions but may be limited when applied to large-scale datasets \citep{zang_hybrid_2019}.
        \item \textit{Maximum Likelihood:} Maximum likelihood is a statistical method that estimates model parameters by maximizing the likelihood function given observed data, commonly used in constructing phylogenetic trees \citep{shen2020investigation}.
        \item \textit{Bayesian Inference:} Bayesian inference is a statistical method that updates the posterior distribution of parameters based on prior distribution and observed data, used for parameter estimation and model selection \citep{Huelsenbeck2001}.
        \item \textit{Deep Learning Models:} Deep learning models are machine learning models composed of multiple layers of neural networks, excelling at handling complex pattern recognition tasks and widely applied in BioTree construction \citep{jumper2021highly}.
        \item \textit{Clustering Algorithms:} Clustering algorithms partition a dataset into multiple groups or clusters, making data points within the same cluster more similar. They have important applications in biological data classification and phylogenetic tree construction \citep{ikotun2023k}.
    \end{itemize} 
\end{mdframed}

\subsection{Fundamental Tree Concepts in BioTree Construction}

Understanding key concepts related to tree structures is fundamental for interpreting the evolutionary relationships represented in BioTrees. This subsection introduces essential tree concepts such as common ancestors, nodes, branches, resolution, lineages, and tree balance.

\begin{mdframed}[hidealllines=true,backgroundcolor=tdc_color!5]
    \begin{itemize}
        \label{term_Concepts} 
        \setlength\itemsep{0em}
        \renewcommand{\thempfootnote}{$\star$}
        \item \textit{Common Ancestor:} A common ancestor is the earliest shared ancestor of multiple descendant species in an evolutionary tree, representing a key node in phylogenetic analysis \citep{Maddison2018}.
        \item \textit{Node:} A node is a point in a phylogenetic tree representing a species or evolutionary event, often used to denote the starting or ending point of divergence or evolutionary pathways \citep{Felsenstein2004}.
        \item \textit{Branch:} A branch is a line in a phylogenetic tree that represents the relationship between an ancestor and its descendants in the evolutionary process \citep{Felsenstein2004}.
        \item \textit{Resolution:} Resolution is the ability to distinguish between different organisms in a phylogenetic tree. High resolution means a finer distinction of evolutionary relationships \citep{Hillis2019}.
        \item \textit{Lineage:} A lineage is a continuous pathway of evolutionary events from an ancestor to its descendants, commonly used to study the evolutionary history of species or cells \citep{Maddison2018}.
        \item \textit{Tree Balance:} Tree balance describes the symmetry of branch lengths or structures in a phylogenetic tree, where a balanced tree often indicates a more uniform evolutionary process \citep{Blum2006}.
    \end{itemize}
\end{mdframed}

\subsection{Fundamental Mathematical and Statistical Concepts in BioTree Construction}

Mathematical and statistical methods form the backbone of BioTree construction and analysis. This subsection highlights important concepts such as evolutionary distance, support values, topology, evidence lower bound (ELBO), and Kullback-Leibler (KL) divergence, which are critical for interpreting results accurately.

\begin{mdframed}[hidealllines=true,backgroundcolor=tdc_color!5]
    \begin{itemize}
        \label{term_Mathematical}
        \setlength\itemsep{0em}
        \renewcommand{\thempfootnote}{$\star$}
        \item \textit{Evolutionary Distance:} Evolutionary distance is a measure of the difference between two species or genes on an evolutionary tree, typically calculated based on gene sequence differences \citep{Nei1987}.
        \item \textit{Support Values:} Support values are a measure of the reliability of branches in a phylogenetic tree, often obtained through bootstrap resampling \citep{Felsenstein1985}.
        \item \textit{Topology:} Topology is the arrangement of branches and nodes in a phylogenetic tree, determining how evolutionary relationships are presented \citep{Semple2003}.
        \item \textit{Evidence Lower Bound (ELBO):} ELBO is a key metric in variational Bayesian inference, used to approximate the lower bound of the model's log-likelihood \citep{Blei2017}.
        \item \textit{Kullback-Leibler (KL) Divergence:} KL divergence is an asymmetric measure of the difference between two probability distributions, often used in the design of loss functions in deep learning models \citep{Kullback1951}.
    \end{itemize}
\end{mdframed}

\section{Datasets of BioTree Construction} \label{sec_dataset}
\subsection{Datasets Used in BioTree Construction} \label{sec_notions_data}

To advance BioTree research and enable effective information fusion, it is essential to understand the various biological data modalities and datasets commonly used in the field\cite{albahri2023systematic}. In this section, we provide an overview of gene-related, protein-related, single-cell, and image-based datasets. Each category offers unique insights into genetic variation, protein structure and function, cellular heterogeneity, and biodiversity. Each category offers unique insights—genetic variation, protein structures and functions, cellular heterogeneity, and morphological characteristics—that are complementary. Integrating these diverse datasets is crucial for constructing comprehensive biological trees and achieving effective information fusion in BioTree research\cite{chen2023information,qian2023survey}.

\subsubsection{Gene Datasets}

Gene datasets, comprising DNA and RNA sequences, are fundamental for understanding the genetic basis of life and the evolutionary relationships among organisms \citep{Consortium2001HumanGenome}. These datasets are obtained through sequencing technologies and play a pivotal role in constructing phylogenetic trees and analyzing genetic diversity.

\begin{mdframed}[hidealllines=true,backgroundcolor=tdc_color!5]
    \begin{itemize}
        \item \textit{Data Collection and Technologies:} The collection of gene data begins with the extraction of DNA or RNA from biological samples such as tissues, blood, or cell cultures \citep{waits2005noninvasive}. For DNA sequencing, the extracted DNA is fragmented and adapters are ligated for amplification and sequencing \citep{ansorge2009next}. RNA sequencing involves isolating mRNA and reverse-transcribing it into complementary DNA (cDNA) \citep{ozsolak2011rna}. Common sequencing technologies include Sanger sequencing \citep{Sanger1977}, Next-Generation Sequencing (NGS) \citep{Mardis2008}, and Third-Generation Sequencing (TGS) technologies like Oxford Nanopore and PacBio \citep{Eid2009, Jain2016}.
        \item \textit{Data Format:} The final output is typically raw sequence data. A DNA sequence is represented as a string \(x^\text{g}\) over the alphabet \(\Sigma = \{A, C, G, T\}\), corresponding to the four nucleotides. An example of a DNA sequence is:
              \begin{equation}
                  x^\text{g} = \texttt{ATCGGCTAAGT...}
              \end{equation}
              where each letter represents one of the four nucleotides.
        \item \textit{Relevance to BioTree Construction:} Gene sequences are essential for constructing phylogenetic trees as they provide the genetic information needed to assess evolutionary relationships and genetic divergence among species.
    \end{itemize}
\end{mdframed}

\subsubsection{Protein Datasets}

Protein datasets, including amino acid sequences and three-dimensional structures, are critical for understanding protein function and evolution, which are important aspects of BioTree analysis \citep{Alberts2002}.

\begin{mdframed}[hidealllines=true,backgroundcolor=tdc_color!5]
    \begin{itemize}
        \item \textit{Data Collection and Technologies:} Protein data are obtained through techniques like mass spectrometry for sequencing and X-ray crystallography or cryo-electron microscopy for structural analysis \citep{Aebersold2003MassSpectrometry, Rhodes2006Crystallography}.
        \item \textit{Data Format:} Protein sequences are represented as strings \(x^\text{p} = \{s_1, s_2, \dots, s_n\}\), where each \(s_i\) is an amino acid from the set of 20 standard amino acids. Structural data are stored in formats like PDB, containing atomic coordinates. A protein sequence example:
              \begin{equation}
                  x^\text{p} = \texttt{MTEYKLVVVGAGGVGKSALTIQL...}
              \end{equation}
              with each character denoting an amino acid using the standard single-letter code.
        \item \textit{Relevance to BioTree Construction:} Protein data enable the study of evolutionary relationships at the protein level, offering insights into functional divergence and adaptation.
    \end{itemize}
\end{mdframed}

\subsubsection{Single-Cell Datasets}

Single-cell datasets allow researchers to explore cellular heterogeneity and are essential for constructing cell differentiation trees in BioTree analysis \citep{Zheng2017}.

\begin{mdframed}[hidealllines=true,backgroundcolor=tdc_color!5]
    \begin{itemize}
        \item \textit{Data Collection and Technologies:} Single-cell data are obtained using technologies like single-cell RNA sequencing (scRNA-seq), which profiles gene expression at the individual cell level \citep{Kolodziejczyk2015}. Advanced techniques like CITE-seq and ASAP-seq integrate multiple omics layers, providing a more comprehensive view of cellular states \citep{Stoeckius2017, Mimitou2021}.
        \item \textit{Data Format:} Data are typically stored in formats that capture the high dimensionality of single-cell measurements, such as expression matrices where rows represent genes and columns represent individual cells.An expression matrix example:
              \begin{equation}
                  \begin{array}{l|cccc}
                                    & \text{Cell}_1 & \text{Cell}_2 & \text{Cell}_3 & \dots  \\
                      \hline
                      \text{Gene}_1 & 5             & 0             & 3             & \dots  \\
                      \text{Gene}_2 & 2             & 6             & 0             & \dots  \\
                      \text{Gene}_3 & 0             & 1             & 4             & \dots  \\
                      \vdots        & \vdots        & \vdots        & \vdots        & \ddots \\
                  \end{array}
              \end{equation}
              where rows represent genes, columns represent individual cells, and the values indicate expression levels.
        \item \textit{Relevance to BioTree Construction:} Single-cell data are crucial for constructing differentiation trees, as they provide detailed information on cell states and transitions during development or disease progression.
    \end{itemize}
\end{mdframed}

\subsection{Commonly Used Dataset for BioTree Construction} \label{sec_notions_DataSet}
BioTree construction is fundamental in deciphering the evolutionary relationships and functional dynamics among biological entities. Different datasets contribute uniquely: gene datasets provide genetic blueprints, protein datasets reveal functional mechanisms, single-cell data uncover cellular diversity, and image-based datasets offer morphological insights. 

\subsubsection{Gene Datasets: Foundations for Exploring Genetic Variation}
Gene-related datasets are foundational for exploring genetic variation, gene expression, and genomic annotations. The \textit{dbSNP} database \citep{dbSNP} provides an extensive collection of over 150 million single nucleotide polymorphisms (SNPs) and is integral to studies of genetic variation and genome-wide association studies. Similarly, the \textit{Gene Expression Omnibus (GEO)} \citep{GEO} offers a vast repository of gene expression datasets, allowing researchers to explore gene regulation and expression patterns across different species and conditions.

The \textit{Human Microbiome Project (HMP)} \citep{HMPConsortium} is another crucial resource, advancing our understanding of the microbial communities associated with human health and disease. Meanwhile, the \textit{Genotype-Tissue Expression (GTEx) Project} \citep{GTExProject} provides gene expression data across various human tissues, helping to uncover the relationship between genetic variation and gene expression. Furthermore, large-scale efforts like \textit{The Cancer Genome Atlas (TCGA)} \citep{TCGA} have significantly contributed to cancer research by offering comprehensive genomic profiles of multiple cancer types, aiding in the identification of molecular alterations. In population genetics, the \textit{1000 Genomes Project} \citep{1000GenomesProject} has been instrumental in providing whole-genome sequencing data from diverse populations, essential for understanding global genetic diversity.
Other key datasets include \textit{Ensembl Genomes} \citep{EnsemblGenomes}, which offers genome annotations across multiple species, and the \textit{Genome Aggregation Database (gnomAD)} \citep{gnomAD}, which aggregates exome and genome data, providing crucial allele frequency information for variant interpretation in both research and clinical contexts.

\subsubsection{Protein Datasets: Insights into Structure and Function}
Understanding protein structure, function, and interactions is central to many biological processes, and protein-related datasets are critical in this context. The \textit{Protein Data Bank (PDB)} \citep{PDB} is a fundamental resource containing a vast collection of 3D structures of proteins and nucleic acids, making it indispensable for structural biology and drug discovery efforts. Additionally, \textit{PeptideAtlas} \citep{PeptideAtlas} curates peptides identified through mass spectrometry, supporting large-scale proteomics research and protein expression studies.

For the study of protein-protein interactions, the \textit{STRING} database \citep{STRING} provides essential data on known and predicted interactions, facilitating the construction of protein interaction networks. \textit{UniProt} \citep{UniProtConsortium}, the most comprehensive protein sequence and functional information repository, is critical for protein annotation and functional studies, offering insights into the biological roles of proteins across species.

\subsubsection{Single-Cell Datasets: Unveiling Cellular Heterogeneity and Dynamics}
The emergence of single-cell datasets has revolutionized the understanding of cellular heterogeneity and dynamic processes at the single-cell level. Single-cell transcriptomics, particularly from \textit{10x Genomics} \citep{TenXGenomics}, provides high-resolution gene expression data, enabling in-depth analyses of individual cell populations and their roles in tissue development and disease. The \textit{Human Cell Atlas (HCA)} \citep{HCAConsortium}, aiming to create comprehensive reference maps of all human cells, serves as a vital resource for exploring cellular states and types, contributing to our understanding of human biology at an unprecedented scale.

\subsubsection{Image-Based Datasets: Integrating Morphological Insights into BioTree Construction}
Image-based datasets are pivotal for integrating computational methods \cite{zang_boosting_2023} with biological research, particularly in biodiversity and taxonomy studies. For example, the \textit{iNaturalist 2021 Dataset (iNat21)} \citep{iNat21} leverages citizen science by compiling millions of organism images, making it an invaluable tool for biodiversity monitoring and species identification. DNA barcoding entries from \textit{BIOSCAN-1M} \citep{BIOSCAN1M} further enhance biodiversity research by enabling the mapping of global species diversity, supporting ecological studies and species discovery.
The \textit{Encyclopedia of Life (EOL)} \citep{EOL} aggregates taxonomic data, including images, to aid in biodiversity conservation efforts, while the \textit{TREEOFLIFE-10M} dataset \citep{Stevens2024BioCLIP} integrates image data with phylogenetic information, fostering advancements in computational biology and evolutionary studies.

The collection and integration of these diverse datasets have dramatically accelerated advancements in biological research. Gene-related datasets have facilitated the exploration of genetic variation and gene expression, while protein-related datasets provide critical insights into protein function and structure. Single-cell datasets have uncovered the complexity of cellular heterogeneity, and image-based datasets are instrumental in biodiversity monitoring and species identification. Together, these resources continue to drive discoveries in genomics, proteomics, and evolutionary biology, offering unprecedented opportunities for future research across multiple disciplines.

\begin{table*}[t]
    \centering
    \footnotesize
    \caption{\textbf{Overview of Key Datasets for Biological Research.} REF means Reference.}
    \begin{tabular}{p{1.0cm}|p{2.5cm}|p{1.5cm}|p{1.0cm}|p{5.5cm}}
        \toprule
                 & \textbf{Dataset Name} & \textbf{\#Entries} & \textbf{REF}        & \textbf{URL}                               \\
        \midrule
        \multirow{8}{*}{Gene}
                 & dbSNP                 & 150M               & \cite{dbSNP}              & \url{https://www.ncbi.nlm.nih.gov/snp/}    \\  
                 & GEO                   & 100k               & \cite{GEO}                & \url{https://www.ncbi.nlm.nih.gov/geo/}    \\
                 & HMP                   & 2.2k               & \cite{HMPConsortium}      & \url{https://hmpdacc.org/}                 \\
                 & GTEx Project          & 17k                & \cite{GTExProject}        & \url{https://gtexportal.org/}              \\
                 & TCGA                  & 20k                & \cite{TCGA}               & \url{https://www.cancer.gov/tcga}          \\
                 & Genomes Project       & 2,504              & \cite{1000GenomesProject} & \url{https://www.internationalgenome.org/} \\
                 & Ensembl Genomes       & 200k               & \cite{EnsemblGenomes}     & \url{https://ensemblgenomes.org/}          \\
                 & gnomAD                & 125k               & \cite{gnomAD}             & \url{https://gnomad.broadinstitute.org/}   \\
        \midrule
        \multirow{4}{*}{Protein}
                 & Protein Data Bank     & 180k               & \cite{PDB}                & \url{https://www.rcsb.org/}                \\
                 & PeptideAtlas          & 2M                 & \cite{PeptideAtlas}       & \url{http://www.peptideatlas.org/}         \\
                 & STRING                & 9.6M               & \cite{STRING}             & \url{https://string-db.org/}               \\
                 & UniProt               & 564M               & \cite{UniProtConsortium}  & \url{https://www.uniprot.org/}             \\
        \midrule
        {Single} & 10x Genomics          & 1.3M               & \cite{TenXGenomics}       & \url{https://www.10xgenomics.com/}         \\
        Cell     & Human Cell Atlas      & 2B                 & \cite{HCAConsortium}      & \url{https://www.humancellatlas.org/}      \\
        \midrule
        \multirow{4}{*}{Image}
                 & iNat21                & 2.7M               & \cite{iNat21}             & \url{https://www.inaturalist.org/}         \\
                 & BIOSCAN-1M            & 1M                 & \cite{BIOSCAN1M}          & \url{https://www.bioscan.org/}             \\
                 & EOL                   & 6.6M               & \cite{EOL}                & \url{https://eol.org/}                     \\
                 & TREEOFLIFE-10M        & 10.4M              & \cite{Stevens2024BioCLIP} & \url{https://imageomics.github.io/bioclip} \\
        \bottomrule
    \end{tabular}
    \label{tab_datasets}
\end{table*}

\section{Problem Definition of BioTree Construction} \label{sec_def}

\begin{definition}[Tree Construction Problem]
    Given a set of biological entities \( S = \{s_1, s_2, \dots, s_n\} \) (e.g., species, genes, or cells) and their corresponding attribute data \( A = \{a_1, a_2, \dots, a_n\} \), the goal is to construct a tree \( T = (V, E, L) \) that satisfies:
    \begin{itemize}
        \item \( V = \{v_1, v_2, \dots, v_m\} \): Nodes include biological entities \( S \) and inferred states (e.g., ancestors), with \( S \subseteq V \).
        \item \( E = \{e_1, e_2, \dots, e_{m-1}\} \): Edges represent relationships between nodes, forming a connected, acyclic graph.
        \item \( L: E \rightarrow \mathbb{R}^+ \): Assigns positive weights to edges, indicating evolutionary distance, time, or differentiation progression.
    \end{itemize}
    The tree must have a unique root node \( v_{\text{root}} \), representing the initial state (e.g., common ancestor). The objective function \( F(T) \) optimizes criteria like maximum likelihood, parsimony, or minimal total branch length, guided by prior knowledge.
\end{definition}

The tree \( T \) must be a connected acyclic graph (i.e., a tree), and it typically includes a unique root node \( v_{\text{root}} \) representing the common ancestor or initial state. The goal of constructing the tree is to optimize an objective function \( F(T) \), which may involve maximizing likelihood under a specific model, minimizing parsimony (the total number of evolutionary changes), or minimizing the total branch length, depending on the specific application.

\begin{definition}[Phylogenetic Tree Construction]
    When \( S \) represents species, genes, or proteins, and \( L(e_k) \) represents evolutionary distance or divergence time, the tree \( T \) is called a phylogenetic tree. The objective function \( F(T) \) may maximize likelihood under evolutionary models or minimize parsimony or total branch length.
\end{definition}

\begin{definition}[Differentiation Tree Construction]
    When \( S \) represents cells or developmental states, and \( L(e_k) \) represents differentiation progression, the tree \( T \) describes differentiation pathways. The objective \( F(T) \) aims to capture parsimonious or biologically consistent cell state transitions.
\end{definition}

Prior knowledge, such as evolutionary models for phylogenetic trees or developmental biology for differentiation trees, guides the construction process, ensuring \( T \) reflects underlying biological processes accurately.

\section{Information Fusion Prior Knowledge For BioTree Construction} \label{sec_prior}
Incorporating biological prior knowledge into models is essential for enhancing the accuracy, interpretability, and biological relevance of BioTree analyses. Biological systems are inherently complex, and purely data-driven learning approaches often struggle to capture the intricate mechanisms and patterns underlying these systems. By integrating prior knowledge—such as evolutionary relationships, functional genomic modules, and protein structure information—into data-driven frameworks, models can achieve a more robust representation of biological realities, reducing uncertainty and bias during the inference process.

The fusion of prior knowledge with data-driven methods not only strengthens model resilience against high-dimensional and multimodal data challenges but also significantly enhances the interpretability and usability of the results. To provide a comprehensive understanding, this section organizes and categorizes prior knowledge critical to BioTree construction.

\begin{table*}[t]
    \centering
    \scriptsize
    \caption{Summary of Prior Knowledge for Phylogenetic Tree Construction: Gene Data}
    \begin{tabular}{c|p{2.5cm} p{2cm} p{4cm} p{2cm}}
        \toprule
        \textbf{Prior} & \textbf{Descriptions}                & \textbf{Prior Form}                                          & \textbf{Knowledge Involved}                                                                 & \textbf{References}                                              \\
        \midrule
        G1             & Conserved Genomic Regions            & Indicator function \( I(x_i^g, x_j^g) \)                     & Regions that are relatively unchanged across species, indicating evolutionary relationships & \cite{Mount2004}, \cite{Notredame2007}, \cite{Altschul1990}      \\ \midrule
        G2             & Evolutionary Substitution            & Transition probability matrix \( P(t) \)                     & Describes probabilistic changes in nucleotide sequences over time                           & \cite{Felsenstein1981}, \cite{Kimura1980}, \cite{tavare1986some} \\ \midrule
        G3             & Genomic Linear Order of Genes        & Permutation vector \( \pi \)                                 & Specific order of genes along chromosomes, providing clues about evolutionary relationships & \cite{Saitou1987}, \cite{Fitch1971}                              \\ \midrule
        G4             & Ancestral Relationship Information   & Ancestral matrix \( A \)                                     & Known or inferred relationships between species based on shared ancestors                   & \cite{Maddison2007}, \cite{Ronquist2003}                         \\ \midrule
        G5             & Sequence Homology Information        & Similarity matrix \( H \)                                    & Shared ancestry between pairs of genes or sequences, critical for accurate inference        & \cite{Thompson1994}, \cite{Smith1981}                            \\ \midrule
        G6             & Gene Duplication and Loss Events     & Probabilistic model \( P(T \mid \text{duplication, loss}) \) & Models gene duplication and loss events, impacting tree topology                            & \cite{Hahn2009}, \cite{Gu2005}                                   \\ \midrule
        G7             & Taxonomic Classification Constraints & Taxonomy tree \( \mathcal{T} \)                              & Known hierarchical relationships among species, ensuring consistency with classification    & \cite{Faith1992}, \cite{Hennig1965}, \cite{Swofford1996}         \\
        \bottomrule
    \end{tabular}
    \label{tab_gene_data_prior_knowledge}
\end{table*}

\begin{table*}[t]
    \centering
    \scriptsize
    \caption{Summary of Prior Knowledge for Phylogenetic Tree Construction: Protein Structure and Sequence Data}
    \begin{tabular}{c|p{2.5cm} p{2cm} p{4cm} p{2cm}}
        \toprule
        \textbf{Prior} & \textbf{Descriptions}                           & \textbf{Prior Form}                      & \textbf{Knowledge Involved}                                                                  & \textbf{References}                      \\
        \midrule
        P1             & Conserved Protein Domains                       & Indicator function \( I(d_i^p, d_j^p) \) & Conserved regions within protein sequences, indicating functional importance                 & \cite{Murzin1995}, \cite{Marchler2011}   \\ \midrule
        P2             & Evolutionary Models for Amino Acid Substitution & Substitution matrix \( Q \)              & Describes the rate of amino acid substitutions over evolutionary time                        & \cite{Jones1992}, \cite{Dayhoff1978}     \\ \midrule
        P3             & Protein Secondary Structure Information         & Similarity matrix \( S \)                & Conserved secondary structures like alpha-helices and beta-sheets                            & \cite{Kabsch1983}, \cite{Chothia1984}    \\ \midrule
        P4             & Tertiary Structure Conservation                 & RMSD (Root-Mean-Square Deviation)        & 3D structure, which is often more conserved than the primary sequence                        & \cite{Sali1994}                          \\ \midrule
        P5             & Functional Site Conservation                    & Function \( F(x_i^p, x_j^p) \)           & Conservation of critical functional sites in proteins                                        & \cite{Bartlett2002}, \cite{Thornton2000} \\ \midrule
        P6             & Protein Family Classification                   & Classification \( \mathcal{C} \)         & Groups proteins based on sequence and structural similarity, reflecting evolutionary origins & \cite{Bateman2002}, \cite{Finn2016}      \\ \midrule
        P7             & Co-Evolutionary Relationships                   & Co-evolution matrix \( C \)              & Captures the functional interdependencies of proteins through co-evolution                   & \cite{Gobel1994}, \cite{Marks2011}       \\
        \bottomrule
    \end{tabular}
    \label{tab_protein_structure_prior_knowledge}
\end{table*}

\begin{table*}[t]
    \centering
    \scriptsize
    \caption{Summary of Prior Knowledge for Phylogenetic Tree Construction: Single-Cell Multimodal Data}
    \begin{tabular}{c|p{2.5cm} p{2cm} p{4cm} p{2cm}}
        \toprule
        \textbf{Prior} & \textbf{Descriptions}           & \textbf{Prior Form}               & \textbf{Knowledge Involved}                                                           & \textbf{References}                          \\
        \midrule
        S1             & Gene Expression Profiles        & Expression matrix \( E \)         & Abundance of mRNA transcripts in single cells, indicating functional state            & \cite{Trapnell2014}, \cite{Qiu2017}          \\ \midrule
        S2             & RNA Velocity                    & Velocity vector \( v_i^c \)       & Estimates the future state of individual cells based on RNA transcriptional changes   & \cite{LaManno2018}, \cite{Bergen2020}        \\ \midrule
        S3             & Cell Type-Specific Marker Genes & Binary matrix \( B \)             & Genes uniquely expressed in specific cell types, used to identify cell identity       & \cite{Tirosh2016}, \cite{Plass2018}          \\ \midrule
        S4             & Pseudotime Ordering             & Pseudotime scalar \( \tau_i^c \)  & Orders cells along a continuous trajectory representing differentiation progress      & \cite{Trapnell2014}, \cite{Haghverdi2016}    \\
        \bottomrule
    \end{tabular}
    \label{tab_single_cell_prior_knowledge}
\end{table*}

\subsection{Prior Knowledge for Gene Phylogenetic Tree Construction}
When constructing phylogenetic trees using gene sequence data, leveraging prior knowledge is fundamental to enhancing the accuracy, reliability, and interpretability of inferred trees. This section organizes and describes seven key types of prior knowledge, emphasizing their complementary roles and providing formal mathematical representations with references.

\begin{priorlist_gene}
    \item \label{priorG_ConservedGenomicRegions} 
    \textbf{Conserved Genomic Regions} \\
    Conserved regions \citep{Mount2004, leypold2021evolutionary} are gene sequences that remain relatively unchanged across species due to strong selective pressure, indicating their critical role in evolutionary relationships\citep{Altschul1990, Notredame2007}. These regions can be represented using an indicator function \( I(x_i^g, x_j^g) \):
    \[
        I(x_i^g, x_j^g) = \begin{cases}
            1, & \text{if sequences } x_i^g \text{ and } x_j^g \text{ share conserved regions} \\
            0, & \text{otherwise}.
        \end{cases}
    \]
    The similarity between these regions is quantified as:
    \[
        d_{\text{conserved}}(x_i^g, x_j^g) = \sum_{k=1}^{L} I(x_{i,k}^g, x_{j,k}^g) \cdot d(x_{i,k}^g, x_{j,k}^g),
    \]
    where \( L \) is the sequence length, and \( d(x_{i,k}^g, x_{j,k}^g) \) is a distance metric like Hamming or Jukes-Cantor distance. This analysis focuses on regions critical to divergence, complementing broader evolutionary models.

    \item \textbf{Evolutionary Substitution Models} \label{priorG_Evolutionary_Substitution} \\
    Substitution models describe nucleotide changes over time, providing probabilistic frameworks for evolutionary inference \cite{Felsenstein1981, lucaci2023evolutionary}. For instance, the JC69 model assumes equal substitution probabilities and constant mutation rates, represented by the transition matrix \( P(t) \):
    \[
        P(t) = \frac{1}{4} + \frac{3}{4}e^{-\mu t} \cdot I,
    \]
    where \( \mu \) is the mutation rate, and \( I \) is the identity matrix. These models complement conserved regions by estimating distances where sequence variability is significant.

    \item \textbf{Genomic Linear Order of Genes} \label{priorG_LinearOrder} \\
    The order of genes on chromosomes provides context for phylogenetic relationships, particularly when conserved across species \cite{Saitou1987,Fitch1971,kouvelis2023evolution}. This can be modeled as a permutation vector \( \pi \), with similarity calculated as:
    \[
        d_{\text{linear}}(x_i^g, x_j^g) = \sum_{k=1}^{n} \delta(\pi_i(k), \pi_j(k)),
    \]
    where \( \delta \) is the Kronecker delta function, equal to 1 if gene order matches at position \( k \). This perspective complements substitution models by incorporating structural genome features.

    \item \textbf{Ancestral Relationship Information} \label{priorG_AncestralRelationship} \\
    Ancestral information, often derived from fossil records or historical data, informs phylogenetic trees by encoding known relationships\citep{Maddison2007, lewanski2024era}. This can be formalized using an ancestral matrix \( A \), where \( A_{ij} \) denotes the probability of a shared ancestor between species \( i \) and \( j \):
    \[
        P(\text{Tree} \mid A) = \prod_{i,j} P(\text{Tree} \mid A_{ij}) \cdot P(\text{Tree}),
    \]
    ensuring robustness when reconstructing tree topologies for well-documented clades.

    \item \textbf{Sequence Homology Information} \label{priorG_Homology} \\
    Homology reflects shared ancestry between genes or sequences, with orthologs arising from speciation and paralogs from duplication~\citep{Thompson1994, Smith1981}. Homology scores \( H_{ij} \) can be transformed into a distance metric:
    \[
        d_{\text{homology}}(x_i^g, x_j^g) = -\log(H_{ij}),
    \]
    enabling accurate evolutionary analysis, especially for complex gene families.

    \item \textbf{Gene Duplication and Loss Events} \label{priorG_GeneDuplicationLoss} \\
    Duplication and loss events shape gene family evolution and tree topology \citep{Hahn2009, Gu2005}. Probabilistic models capture these events:
    \[
        P(T \mid \text{duplication, loss}) = \prod_{d \in D} p_d \cdot \prod_{l \in L} p_l,
    \]
    where \( p_d \) and \( p_l \) are duplication and loss probabilities, respectively. This framework complements homology analysis in evolutionary studies.

    \item \textbf{Taxonomic Classification Constraints} \label{priorG_TaxonomicClassification} \\
    Taxonomic hierarchies provide a priori classifications to ensure phylogenetic consistency \citep{Hillis1992, Felsenstein2004}. Represented as a tree \( \mathcal{T} \), taxonomic constraints refine tree construction:
    \[
        P(\text{Tree} \mid \mathcal{T}) = P(\text{Tree} \mid \text{Taxonomic Constraints}) \cdot P(\text{Tree}),
    \]
    integrating established classifications while allowing inference in incomplete scenarios.
\end{priorlist_gene}

\subsection{Prior Knowledge for Protein Structure \& Sequence Phylogenetic Tree Construction}
When constructing phylogenetic trees using protein sequences and structures, leveraging prior knowledge at multiple levels—such as sequence conservation, secondary structure, and three-dimensional topology—significantly enhances the accuracy and biological relevance of the resulting trees. This section categorizes and formalizes these layers of prior knowledge, highlighting their complementary roles in phylogenetic inference.

\begin{priorlist_protein}
    \item \textbf{Conserved Protein Domains.} \label{priorP_ConservedProteinDomains}
    Conserved protein domains are specific regions within protein sequences that are preserved across different species due to their critical functional roles\citep{Marchler2011,wang2023conserved}. These domains are often associated with essential biological functions and exhibit lower variability over evolutionary time. The conservation of these domains can be represented using an indicator function \( I(d_i^p, d_j^p) \), where:
    \begin{equation}
        I(d_i^p, d_j^p) = \begin{cases}
            1, & \text{if domains } d_i^p \text{ and } d_j^p \text{ are conserved}, \\
            0, & \text{otherwise}.
        \end{cases}
    \end{equation}
    The similarity between conserved domains is then quantified as:
    \begin{equation}
        d_{\text{domain}}(x_i^p, x_j^p) = \sum_{k=1}^{M} I(d_{i,k}^p, d_{j,k}^p) \cdot d(d_{i,k}^p, d_{j,k}^p),
    \end{equation}
    where \( M \) is the number of domains and \( d(d_{i,k}^p, d_{j,k}^p) \) represents the distance metric between corresponding domains. These conserved regions provide a basis for understanding functional constraints and complement substitution models by focusing on stable features of evolutionary significance.

    \item \textbf{Evolutionary Models for Amino Acid Substitution.} \label{priorP_AminoAcid}
    Substitution models describe the changes in protein sequences over time, taking into account the biochemical properties of amino acids and the probabilities of specific substitutions \citep{Jones1992, alamdari2023protein}. For instance, the JTT model uses a substitution rate matrix \( Q \) to estimate the likelihood of one amino acid being replaced by another. The probability of substitution over time is given by:
    \begin{equation}
        P(t) = e^{Qt},
    \end{equation}
    where \( t \) represents evolutionary time. These models are particularly effective when combined with conserved domain information, as they estimate variability while accounting for underlying conservation patterns.

    \item \textbf{Protein Secondary Structure Information.} \label{priorP_SecondaryStructure}
    Secondary structures~\citep{Kabsch1983, jiang2023explainable}, such as alpha-helices and beta-sheets, are conserved when critical to protein function. These elements can be represented in a matrix \( S \), where \( S_{ij}^p \) quantifies the similarity between secondary structures of proteins \( x_i^p \) and \( x_j^p \). The structural similarity is calculated as:
    \begin{equation}
        d_{\text{secondary}}(x_i^p, x_j^p) = \sum_{k=1}^{L} S(x_{i,k}^p, x_{j,k}^p),
    \end{equation}
    where \( L \) is the length of the aligned sequences. Incorporating this structural layer ensures that functional constraints are reflected in the tree construction process.

    \item \textbf{Tertiary Structure Conservation.} \label{priorP_TertiaryStructure}
    Tertiary structures provide a higher-order perspective on evolutionary relationships \citep{Sali1994,wang2021structural}, as structural features tend to be conserved more than sequences. The similarity between 3D structures can be quantified using the root-mean-square deviation (RMSD):
    \begin{equation}
        d_{\text{tertiary}}(x_i^p, x_j^p) = \text{RMSD}(x_i^p, x_j^p),
    \end{equation}
    where a smaller RMSD indicates greater structural similarity. This metric is particularly useful when sequence similarity is low but structural preservation is evident.

    \item \textbf{Functional Site Conservation.} \label{priorP_FunctionalSites}
    Functional sites \citep{Bartlett2002, hoie2022predicting}, such as enzyme active sites or ligand-binding sites, are highly conserved due to their role in protein function. These sites can be compared across proteins using a similarity function \( F(x_i^p, x_j^p) \), which measures the correspondence between residues involved in the functional site:
    \begin{equation}
        d_{\text{functional}}(x_i^p, x_j^p) = \sum_{k=1}^{N} F(x_{i,k}^p, x_{j,k}^p),
    \end{equation}
    where \( N \) is the number of residues in the functional site. Including this information ensures that phylogenetic trees capture the functional constraints critical to evolutionary processes.

    \item \textbf{Protein Family Classification.} \label{priorP_ProteinFamily}
    Proteins are often grouped into families based on shared sequence and structural features \citep{abu2023innovative, Finn2016}. These classifications can constrain phylogenetic tree topologies to align with established family groupings. Given a classification \( \mathcal{C} \), tree construction can be influenced as:
    \begin{equation}
        P(\text{Tree} \mid \mathcal{C}) = \prod_{\text{family } i \in \mathcal{C}} P(\text{Tree} \mid i),
    \end{equation}
    ensuring consistency with known evolutionary relationships.

    \item \textbf{Co-Evolutionary Relationships.} \label{priorP_CoEvolution}
    Co-evolution between proteins or domains  \citep{ciampi2022co, Marks2011} can reveal functional interdependencies within biological pathways. Co-evolutionary signals are captured in a matrix \( C \), where \( C_{ij}^p \) reflects the strength of co-evolution between proteins \( x_i^p \) and \( x_j^p \). The similarity is represented as:
    \begin{equation}
        d_{\text{co-evolution}}(x_i^p, x_j^p) = -\log(C_{ij}^p),
    \end{equation}
    with stronger co-evolutionary signals corresponding to higher \( C_{ij}^p \). This perspective enhances the tree's ability to reflect functional and evolutionary interdependencies.
\end{priorlist_protein}

\subsection{Prior Knowledge for Single-Cell Differentiation Tree Construction}
When constructing cell differentiation trees using single-cell multimodal data, leveraging prior knowledge is crucial for accurately modeling the complex processes of cellular differentiation. These types of prior knowledge operate across multiple dimensions—static, dynamic, and temporal—and collectively enhance our ability to build robust differentiation trees. This section discusses key types of prior knowledge, providing biological context and formal mathematical descriptions, along with relevant references.

\begin{priorlist_singlecell}
    \item \textit{Gene Expression Profiles.} \label{priorS_GeneExpression}
    Gene expression profiles provide static insights into a cell's functional state by measuring mRNA transcript abundance \citep{poulin2020classification, Qiu2017}. These profiles are critical for identifying cellular identity and differentiation status. Represented as a matrix \( E \), where \( E_{ij}^c \) denotes the expression level of gene \( j \) in cell \( i \), the similarity between cells can be quantified by:
    \begin{equation}
        d_{\text{expression}}(c_i, c_j) = \sum_{k=1}^{G} \left( E_{ik}^c - E_{jk}^c \right)^2,
    \end{equation}
    where \( G \) is the total number of genes. This metric captures differences in gene expression patterns and establishes a foundation for further dynamic analysis using RNA velocity .

    \item \textit{RNA Velocity.} \label{priorS_RNAVelocity}
    RNA velocity \citep{bergen2021rna, Bergen2020} extends the static insights from gene expression profiles by introducing a dynamic layer, estimating the future transcriptional states of cells based on spliced and unspliced mRNA ratios. Represented as a vector \( v_i^c \) for each cell \( i \), RNA velocity quantifies differentiation directionality:
    \begin{equation}
        d_{\text{velocity}}(c_i, c_j) = \| v_i^c - v_j^c \|,
    \end{equation}
    where \( \| \cdot \| \) denotes the Euclidean norm. This dynamic information complements gene expression data by predicting future states, enhancing the resolution of differentiation trajectories.

    \item \textit{Cell Type-Specific Marker Genes.} \label{priorS_MarkerGenes}
    Marker genes \citep{perez2022single, Plass2018} are uniquely or highly expressed in specific cell types and are crucial for distinguishing cellular identities. Encoded as a binary matrix \( B \), where \( B_{ij}^c = 1 \) if marker gene \( j \) is expressed in cell \( i \), the similarity between cells can be computed as:
    \begin{equation}
        d_{\text{markers}}(c_i, c_j) = \sum_{k=1}^{M} \left| B_{ik}^c - B_{jk}^c \right|,
    \end{equation}
    where \( M \) is the total number of marker genes. Marker genes also serve as a bridge to temporal analyses like pseudotime ordering by anchoring cellular identities in differentiation pathways.

    \item \textit{Pseudotime Ordering.} \label{priorS_Pseudotime}
    Pseudotime ordering \citep{li2022trasig, mathur2024constrained} adds a temporal perspective by arranging cells along a continuous trajectory that represents differentiation progress. Represented as a scalar \( \tau_i^c \) for each cell \( i \), pseudotime facilitates the comparison of cells in their differentiation timeline:
    \begin{equation}
        d_{\text{pseudotime}}(c_i, c_j) = \left| \tau_i^c - \tau_j^c \right|.
    \end{equation}
    This temporal metric, informed by marker genes, captures the progression of differentiation and provides a comprehensive framework for visualizing cellular pathways .
\end{priorlist_singlecell}

\section{Classical BioTree Construction Methods} \label{sec_classical}

The construction of biological trees has been a fundamental approach in understanding evolutionary relationships, functional similarities, and lineage hierarchies among biological entities. Classical methods have laid the foundation for this field, offering a variety of algorithms tailored for different data types, including genomic sequences, protein structures, and single-cell data. In the following subsections, we systematically review these classical methods, highlighting their key principles, applications, and limitations, providing a comprehensive understanding of their historical context and impact on modern advancements.

\subsection{Classical General BioTree Construction Methods}
\begin{figure*}[t]
	\begin{center}
		\includegraphics[width=0.99\linewidth]{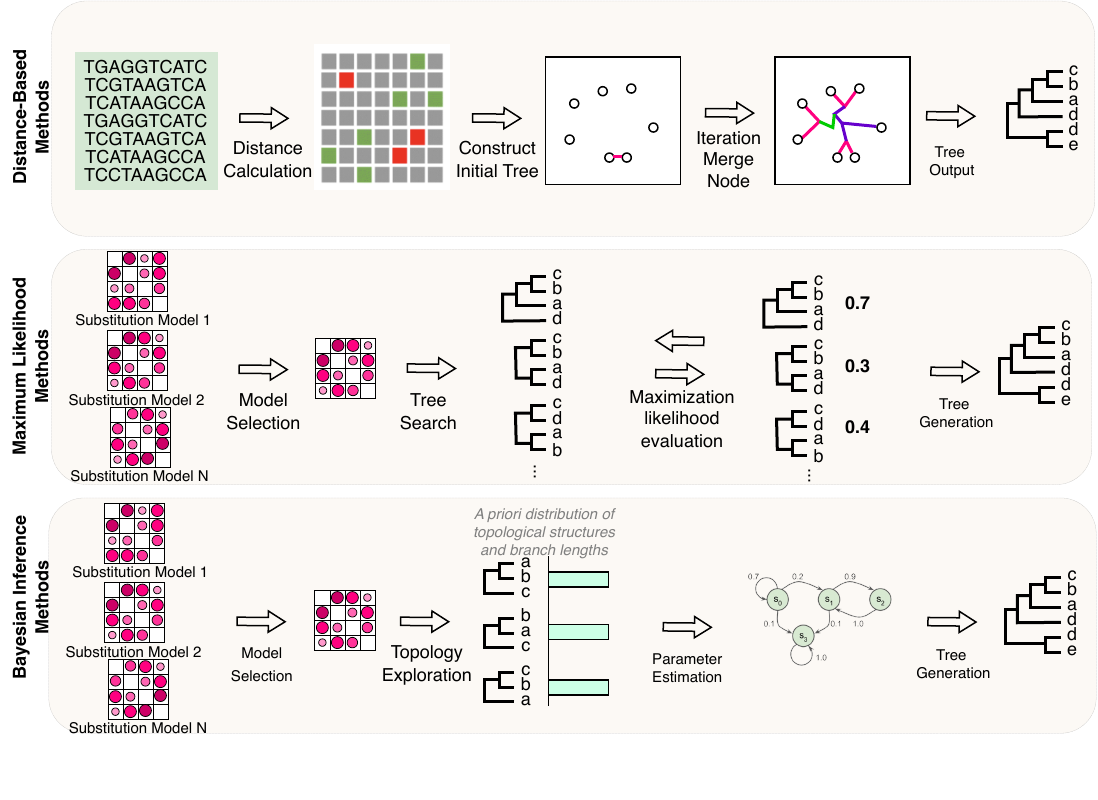}
	\end{center}
	\caption{\textbf{Overview of General BioTree Construction Methods.} Methods are categorized based on the type of input data, their capability to address conflicts between gene and species trees, and specific application contexts.}
	\label{fig_method}
\end{figure*}

\begin{figure*}[t]
	\begin{center}
		\includegraphics[width=0.99\linewidth]{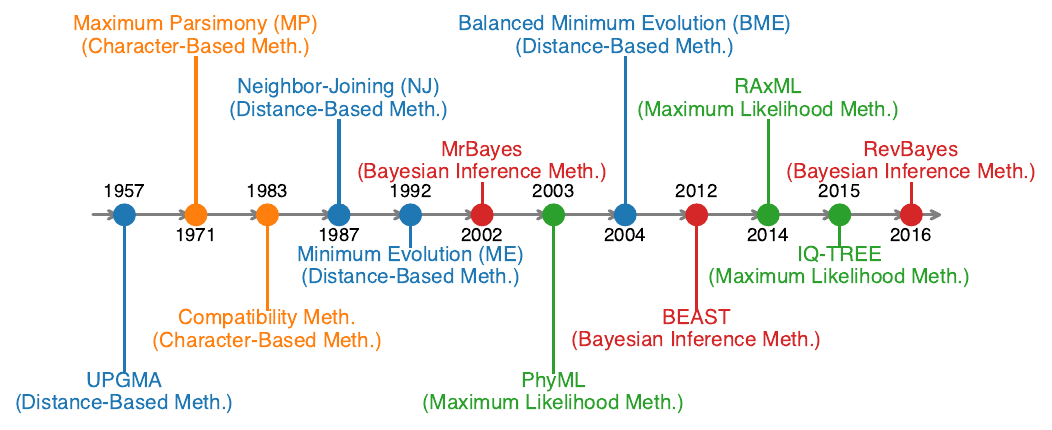}
	\end{center}
	\caption{\textbf{Timeline of General BioTree Construction Methods.} The timeline illustrates the progression of tree construction methods in phylogenetics from 1957 to 2016, grouped into feature-based, distance-based, Bayesian inference, and maximum likelihood methods. Different colors represent distinct categories.}
	\label{fig_method_classi_general}
\end{figure*}

General BioTree Construction Methods are broadly divided into three categories: \textit{feature-based methods}, \textit{distance-based methods}, and \textit{Bayesian inference methods}(Figure\ref{fig_method_classi_general}). These methods represent the foundational approaches to phylogenetic analysis, each addressing specific challenges such as computational efficiency, model flexibility, and the integration of prior knowledge. Among these, the concept of information fusion plays a pivotal role, as modern approaches increasingly emphasize the need to integrate diverse data sources—such as genetic sequences, evolutionary distances, and probabilistic models—to achieve a more holistic and accurate representation of phylogenetic relationships. Below, we detail these categories, their contributions, and how information fusion enhances their effectiveness in addressing complex biological questions.

\paragraph{Distance-Based Methods.} Among the earliest and computationally efficient techniques, distance-based methods rely on pairwise distance matrices derived from genetic or evolutionary sequences. Methods like \textit{UPGMA}\cite{stefan1996multiple} (Unweighted Pair Group Method with Arithmetic Mean) assume a constant rate of evolution (the molecular clock hypothesis), producing rooted trees \citep{Michener1957}. However, this assumption often does not align with biological reality, leading to potential biases. \textit{Neighbor-Joining} \cite{Saitou1987} (\textit{NJ}) eliminates the constant-rate assumption by constructing unrooted trees that minimize total branch lengths \citep{Saitou1987}. Further refinements, such as \textit{Minimum Evolution} (\textit{ME}) and \textit{Balanced Minimum Evolution} (\textit{BME})\cite{Desper2004}, optimize tree topology for both computational efficiency and accuracy \citep{Rzhetsky1992, Desper2004}. Despite their advantages, these methods reduce complex sequence data to pairwise distances, which may result in information loss. Therefore, distance-based methods are most effective for preliminary analyses or when computational resources are constrained.

\paragraph{Maximum Likelihood Methods.} \textit{Maximum Likelihood} (\textit{ML}) methods \cite{izquierdo-carrasco_algorithms_2011} provide a statistically rigorous framework for phylogenetic tree estimation. These methods optimize the likelihood of observing given sequence data under a specified evolutionary model. The process involves model selection, tree topology exploration, and branch length optimization. Tools like \textit{RAxML} \cite{Stamatakis2014} (Randomized Axelerated Maximum Likelihood) handle large datasets with high efficiency \citep{Stamatakis2014}, while \textit{PhyML} (Phylogenetic Maximum Likelihood) balances computational speed with accuracy \citep{Guindon2003,guindon2010new}. \textit{IQ-TREE} enhances usability by integrating automated model selection and ultrafast bootstrap methods \citep{Nguyen2015}. Although ML methods are robust and flexible, they are computationally intensive and require careful model selection to prevent bias. These methods are ideal for detailed phylogenetic studies when computational resources and domain expertise are available.

\paragraph{Bayesian Inference Methods.} \textit{Bayesian Inference} (\textit{BI}) methods integrate prior knowledge with observed data to estimate the posterior probability of phylogenetic trees. Key steps include model selection, posterior distribution sampling via Markov Chain Monte Carlo (\textit{MCMC}), and parameter estimation. Tools like \textit{MrBayes} offer broad model support \citep{Ronquist2012}, while \textit{BEAST} focuses on divergence time estimation \citep{Drummond2012}. \textit{RevBayes} provides flexibility for complex evolutionary process modeling \citep{Hohna2016}. The incorporation of prior distributions enables these methods to guide the tree estimation process effectively. However, their reliance on extensive \textit{MCMC} sampling makes them computationally demanding. BI methods are particularly valuable for comprehensive studies requiring rigorous probabilistic interpretation.

\subsection{Classical Gene-Based Phylogenetic BioTree Construction Methods}
\begin{figure*}[t]
	\begin{center}
		\includegraphics[width=0.95\linewidth]{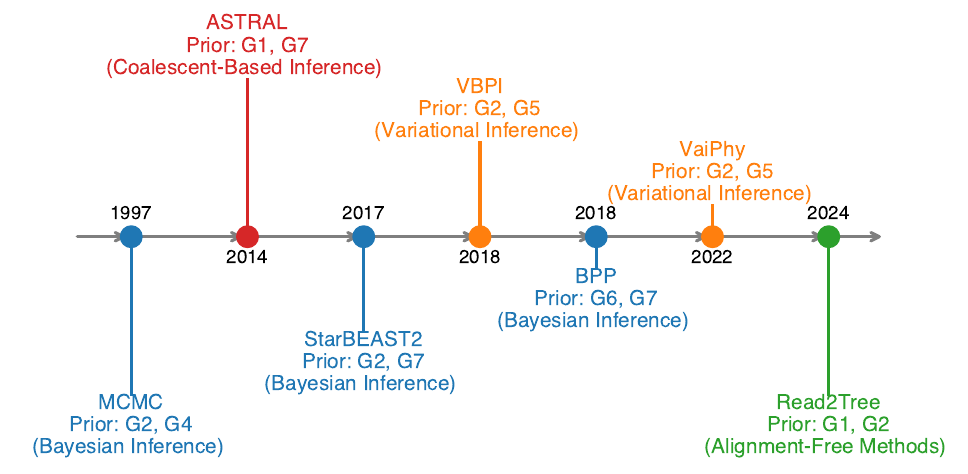}
	\end{center}
	\caption{\textbf{The timeline of Classical Gene-Based BioTree Construction Methods.} The figure shows the development of gene-based tree construction methods in phylogenetics from 1994 to 2022, categorized into Bayesian inference, coalescent-based methods, and alignment-free methods. Different colors indicate different categories.}
	\label{fig_method_gene}
\end{figure*}

\begin{table*}[t]
    \centering
    \scriptsize
    \caption{Overview of the Classical Gene-based Tree Construction Mehtods.}
    \begin{tabular}{p{1.8cm}|p{6cm} p{1.3cm} p{3cm}}
        \toprule
        \textbf{Method Name} & \textbf{Description}                                                                                                          & \textbf{Ref.}              & \textbf{URL}                                    \\
        \midrule
        ASTRAL               & A coalescent-based method for estimating species trees from multiple gene trees, known for its high accuracy                  & \citep{mirarab_astral_2014}     & \url{https://github.com/smirarab/ASTRAL/}       \\ \hline
        StarBEAST2           & A faster Bayesian method for species tree construction with accurate substitution rate estimates                                 & \citep{ogilvie_starbeast2_2017} & \url{https://github.com/genomescale/starbeast2} \\ \hline
        VBPI                 & A variational framework for Bayesian phylogenetic analysis, using stochastic gradient ascent for posterior estimation         & \citep{zhang_variational_2018}  & \url{https://github.com/tyuxie/VBPI-SIBranch}   \\ \hline
        BPP                  & A method using genomic sequences and multispecies coalescent for species tree estimation                                      & \citep{flouri_species_2018}     & \url{https://github.com/bpp/}                   \\ \hline
        VaiPhy               & A variational inference-based algorithm for approximate posterior inference in phylogeny                                      & \citep{koptagel_vaiphy_2022}    & \url{https://github.com/Lagergren-Lab/VaiPhy}   \\ \hline
        Read2Tree            & A method to infer phylogenetic trees directly from raw sequencing reads, bypassing traditional genome assembly and annotation & \citep{dylus_inference_2024}    & \url{https://github.com/DessimozLab/read2tree}  \\
        \bottomrule
    \end{tabular}
    \label{tab_classical_gene}
\end{table*}

Gene-Based BioTree construction methods have seen significant advancements in recent years, particularly in Bayesian inference and alignment-free approaches (Table~\ref{tab_classical_gene} and Figure~\ref{fig_method_gene}). These advancements address critical challenges such as computational efficiency, accuracy, and scalability.

Bayesian inference, originating from the \textit{Markov Chain Monte Carlo (MCMC)} framework, facilitates the estimation of posterior distributions for evolutionary trees. This method incorporates \textit{Evolutionary Substitution Models (G2)}, such as the Jukes-Cantor model, to capture sequence evolutionary relationships \citep{Felsenstein1981, Kimura1980}, and leverages \textit{Ancestral Relationship Information (G4)} for species tree estimation \citep{Maddison2007, Ronquist2003}. Despite its effectiveness in modeling complex evolutionary scenarios, MCMC's computational cost escalates significantly with larger datasets.

To overcome these limitations, coalescent-based methods like \textit{ASTRAL} were introduced, integrating multiple gene trees to infer species trees while addressing incomplete lineage sorting (ILS) \citep{mirarab_astral_2014}. By utilizing \textit{Conserved Genomic Regions (G1)} and \textit{Taxonomic Classification Constraints (G7)}, \textit{ASTRAL} enhances computational efficiency and maintains high accuracy \citep{Mount2004, Notredame2007}. These methods are instrumental in analyzing large-scale genomic data.

Variational inference (VI) methods provide further improvements in computational scalability. For instance, \textit{VBPI} optimizes phylogenetic inference using graphical models and stochastic gradient ascent, significantly reducing runtime compared to MCMC while preserving accuracy \citep{zhang_variational_2018}. This method uses the transition probability matrix \( P(t) \) within \textit{Evolutionary Substitution Models (G2)} to describe probabilistic changes in sequences over time.

Building on this, \textit{VaiPhy} refines VI approaches with efficient sampling strategies, such as SLANTIS proposal distributions, avoiding costly auto-differentiation operations \citep{koptagel_vaiphy_2022}. It effectively combines \textit{Evolutionary Substitution Models (G2)} and \textit{Sequence Homology Information (G5)} to achieve scalable and accurate inference for large datasets.

In parallel, coalescent-based Bayesian methods like \textit{BPP} have enhanced multilocus species tree estimation, integrating \textit{Gene Duplication and Loss Events (G6)} to address incomplete lineage sorting and gene flow \citep{flouri_species_2018}. Similarly, \textit{StarBEAST2} improves the integration of taxonomic constraints and substitution rate models, achieving higher accuracy and faster inference \citep{ogilvie_starbeast2_2017}.

Alignment-free methods, exemplified by \textit{Read2Tree}, bypass traditional alignment steps, directly inferring trees from raw sequencing data \citep{dylus_inference_2024}. Utilizing \textit{Genomic Linear Order of Genes (G3)} and \textit{Conserved Genomic Regions (G1)}, these methods minimize computational overhead while maintaining robust performance on diverse genomic datasets \citep{Saitou1987, Mount2004}.

The complementarity of Bayesian inference and alignment-free approaches highlights their potential for integration. Bayesian methods, with their robust probabilistic frameworks, address uncertainties in evolutionary modeling, while alignment-free techniques provide computationally efficient solutions for large-scale analyses. Future research should focus on hybrid methods that leverage multi-layered prior knowledge, aiming to enhance both accuracy and efficiency in phylogenetic inference.

\subsection{Classical Protein-Based Phylogenetic BioTree Construction Methods}

\begin{figure*}[t]
	\begin{center}
		\includegraphics[width=0.99\linewidth]{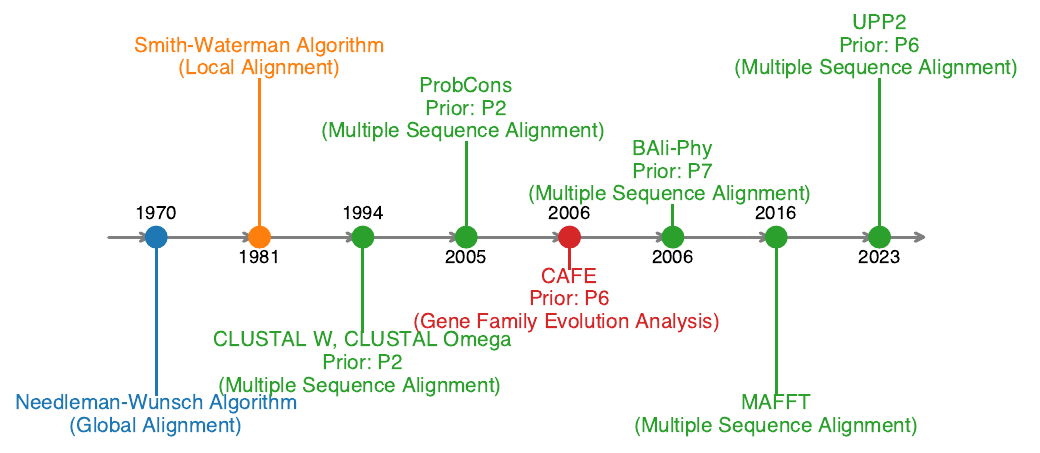}
	\end{center}
	\caption{\textit{The timeline of classical protein sequence-based phylogenetic tree construction methods.} The figure shows the development of protein sequence-based tree construction methods in phylogenetics from 1970 to 2023, categorized into sequence alignment, multiple sequence alignment, and gene family evolution methods. Different colors indicate different categories.}
	\label{fig_method_protein_seq}
\end{figure*}

\begin{figure*}[t]
	\begin{center}
		\includegraphics[width=0.99\linewidth]{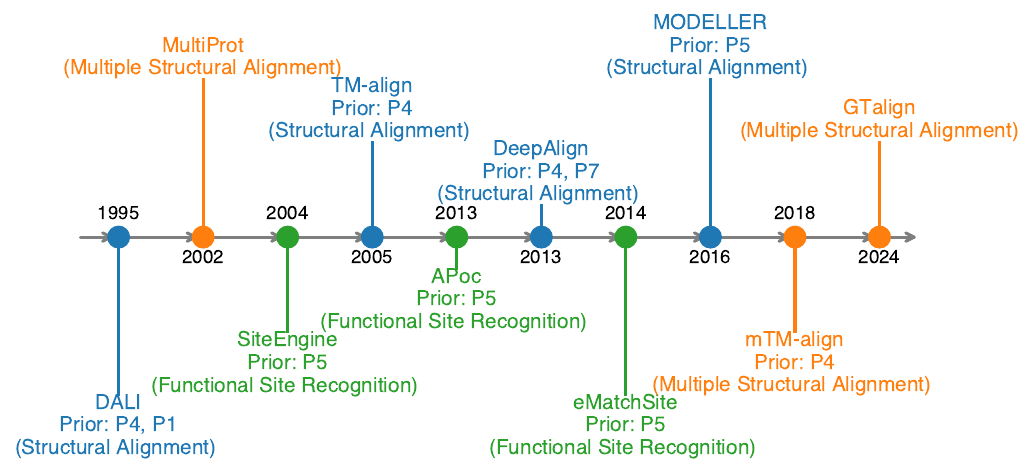}
	\end{center}
	\caption{\textit{The timeline of Classical Protein Structural Based Phylogenetic Tree Construction Methods.} The figure shows the development of protein structural alignment methods in phylogenetics, categorized into distance matrix-based alignment, multiple structure alignment, and functional site recognition methods. Different colors indicate different categories.}
	\label{fig_method_protein_str}
\end{figure*}

\begin{table*}[t]
    \centering
    \scriptsize
    \caption{ Overview of classical protein-based sequence alignment methods.}
    \begin{tabular}{p{1.8cm}|p{6cm} p{1.3cm} p{3cm}}
        \toprule
        \textbf{Method Name}   & \textbf{Description}                                                                             & \textbf{Ref.}                     & \textbf{URL}                                       \\
        \midrule
        \textbf{CLUSTAL Omega} & Fast multiple sequence alignment for large datasets using an advanced algorithm.                 & \citep{sievers_clustal_2014}           & N/A                                                \\\midrule
        \textbf{CLUSTAL W}     & Progressive multiple sequence alignment with position-specific gap penalties.                    & \citep{thompson_clustal_1994}          & N/A                                                \\\midrule
        \textbf{ProbCons}      & Probabilistic multiple sequence alignment using hidden Markov models for higher accuracy.        & \citep{do_probcons_2005}               & N/A                                                \\\midrule
        \textbf{CAFE}          & Gene family evolution modeling with random birth and death process.                              & \citep{de_bie_cafe_2006}               & N/A                                                \\\midrule
        \textbf{BAli-Phy}      & Bayesian sequence alignment and phylogenetic inference in one framework.                         & \citep{suchard_bali-phy_2006}          & \url{https://www.bali-phy.org/}                    \\\midrule
        \textbf{MAFFT}         & Fast multiple sequence alignment with sensitivity for remote homologs.                           & \citep{katoh_simple_2016}              & \url{http://www.blast2go.de}                       \\\midrule
        \textbf{UPP2}          & Ultra-large sequence alignment with phylogeny-aware profiles and HMMs for fragmentary sequences. & \citep{park_upp2_2023}                 & \url{https://github.com/gillichu/sepp}             \\ 
        \bottomrule
    \end{tabular}
    \label{tab_classical_protein_Seq_methods}
\end{table*}

\begin{table*}[t]
    \centering
    \scriptsize
    \caption{Overview of classical protein structural based tree construction methods.}
    \begin{tabular}{p{1.8cm}|p{6cm} p{1.3cm} p{3cm}}
        \toprule
        \textbf{Method Name}   & \textbf{Description}                                                                             & \textbf{Ref.}                     & \textbf{URL}                                       \\
        \midrule
        \textbf{DALI}          & Distance matrix-based structural alignment for detecting global and local similarities.          & \citep{holm_dali_1995}                 & N/A                                                \\ \midrule
        \textbf{MultiProt}     & Multiple structure alignment using geometric cores, suitable for partial alignments.             & \citep{shatsky_multiprot_2002}         & N/A                                                \\ \midrule
        \textbf{SiteEngine}    & Functional site recognition by comparing protein surface binding sites.                          & \citep{shulman-peleg_recognition_2004} & \url{https://bio.tools/siteengine}                 \\ \midrule
        \textbf{TM-align}      & TM-score-based pairwise structural alignment with high speed and accuracy.                       & \citep{zhang_tm-align_2005}            & \url{https://zhanggroup.org/TM-align/}             \\ \midrule
        \textbf{APoc}          & Large-scale structural comparison for identifying pockets on protein surfaces.                   & \citep{gao_apoc_2013}                  & \url{http://cssb.biology.gatech.edu/APoc}          \\ \midrule
        \textbf{DeepAlign}     & Protein structure alignment combining spatial proximity with evolutionary information.           & \citep{wang_protein_2013}              & \url{https://github.com/realbigws/DeepAlign}       \\ \midrule
        \textbf{eMatchSite}    & Binding site alignment tolerant to structural distortions in protein models.                     & \citep{brylinski_ematchsite_2014}      & \url{http://www.brylinski.org/ematchsite}          \\ \midrule
        \textbf{MODELLER}      & Comparative protein structure modeling based on sequence alignment with templates.               & \citep{webb_comparative_nodate_2016}   & \url{https://salilab.org/modeller/}                \\ \midrule
        \textbf{mTM-align}     & Extension of TM-align for multiple structure alignment with improved accuracy and speed.         & \citep{dong_mtm-align_2018}            & \url{https://github.com/CSB5/CaDRReS}              \\ \midrule
        \textbf{GTalign}       & Spatial index-driven multiple structure alignment with high efficiency for large datasets.       & \citep{margelevicius_gtalign_2024}     & \url{https://github.com/openCONTRABASS/CONTRABASS} \\
        \bottomrule
    \end{tabular}
    \label{tab_classical_protein_structural_methods}
\end{table*}

Protein-based phylogenetic tree construction methods are pivotal in understanding the evolutionary relationships and functional characteristics of proteins. These methods leverage the unique attributes of proteins, including their amino acid sequences and three-dimensional (3D) structures, to gain insights beyond what gene-based approaches can provide. By integrating sequence alignment techniques with structural analysis, protein-based approaches offer a complementary perspective that enhances our ability to uncover evolutionary patterns and functional insights. In the following sections, we provide a detailed exploration of classical methods, categorizing them into sequence-based and structure-based approaches. We examine their respective advantages, limitations, and the prior knowledge required for effective implementation, while also discussing the potential directions for future advancements in this field.

\subsubsection{Classical Protein Sequence Based Phylogenetic BioTree Construction Methods}

As shown in table \ref{tab_classical_protein_Seq_methods} and Figure.~\ref{fig_method_protein_seq}, sequence-based protein analysis methods are widely used for inferring evolutionary relationships and functional annotation. These methods utilize protein sequence information to reveal biological functions and evolutionary histories by comparing sequence similarities. \textit{Global and local alignments} are the most fundamental sequence alignment methods. The \textit{Needleman-Wunsch algorithm} \citep{needleman_general_1970} is a classical global alignment algorithm that uses dynamic programming to find the optimal global alignment path between two protein sequences, suitable for sequences of similar length and high similarity. However, its computational cost is high, making it less practical for large datasets. In contrast, the \textit{Smith-Waterman algorithm} \citep{smith_identification_1981} is designed for local alignment, capable of identifying the most similar local regions between sequences, making it suitable for sequences of different lengths or those that are only partially similar. Although it provides flexibility when dealing with highly divergent sequences, its computational overhead is similarly high.

Multiple sequence alignment methods are crucial for studying the similarity between multiple protein sequences. \textit{CLUSTAL W} \citep{thompson_clustal_1994} and \textit{CLUSTAL Omega} \citep{sievers_clustal_2014} are representative progressive multiple sequence alignment methods that optimize alignments using techniques such as progressive weighting and position-specific gap penalties, making them suitable for large-scale sequence datasets. These methods use prior knowledge of \textit{Evolutionary Models for Amino Acid Substitution (P2)}, such as substitution matrices (e.g., the JTT matrix or Dayhoff matrix) \citep{Jones1992, Dayhoff1978}, to model evolutionary relationships and guide the alignment process. However, they may lead to suboptimal alignments when dealing with sequences containing many insertions or deletions (indels). In contrast, \textit{ProbCons} \citep{do_probcons_2005} uses a {probabilistic consistency-based model} that also relies on \textit{evolutionary models (P2)}, but with a more sophisticated approach to account for sequence divergence, effectively capturing complex interactions between sequences during alignment and demonstrating higher accuracy. Nevertheless, the computational complexity of these statistical and probabilistic models remains a significant challenge when handling very large datasets.

The \textit{MAFFT} program \citep{katoh_simple_2016} introduces a new feature that addresses the issue of over-alignment, where unrelated segments are erroneously aligned. Traditional \textit{MAFFT} is known for its sensitivity in aligning conserved regions in remote homologs, but this sensitivity can lead to over-alignment, especially with low-quality or noisy sequences. The improved \textit{MAFFT} uses a variable scoring matrix for different pairs of sequences (or groups) within a single multiple sequence alignment, based on the global similarity of each pair. This approach reduces over-alignment and improves the overall reliability of the alignment, especially in databases increasingly populated by noisy sequences.

Similarly, \textit{UPP2} \citep{park_upp2_2023} is an advancement of the Ultra-large multiple sequence alignment method that deals with fragmentary sequences using an ensemble of Hidden Markov Models (eHMMs) to represent an estimated alignment on the full-length sequences in the input, and then adds the remaining sequences using selected HMMs from the ensemble. It significantly improves accuracy, especially in datasets with substantial sequence length heterogeneity. The use of \textit{Phylogeny-aware Profiles (P6)} as prior knowledge allows \textit{UPP2} to adaptively handle large datasets with varying sequence lengths, which makes it particularly effective in handling incomplete or highly divergent sequences, compared to other leading MSA methods.

Beyond sequence alignment, tools for gene family evolution and evolutionary analysis, such as \textit{CAFE} \citep{de_bie_cafe_2006} and \textit{BAli-Phy} \citep{suchard_bali-phy_2006}, play important roles in studying gene family expansion, gene loss, and protein functional evolution. \textit{CAFE} models gene family evolution by simulating a random birth and death process for gene family size, aiding in the study of gene family dynamics. This method incorporates \textit{Protein Family Classification (P6)} as prior knowledge to define gene family groups and model their evolutionary trajectories based on sequence and structural similarities \citep{Bateman2002, Finn2016}. However, its effectiveness heavily depends on the accuracy of the input phylogenetic tree. \textit{BAli-Phy}, on the other hand, integrates sequence alignment and phylogenetic inference within a \textit{Bayesian framework}, using priors like \textit{Co-Evolutionary Relationships (P7)} that capture the interdependencies between proteins through co-evolution \citep{Gobel1994, Marks2011}. This integration reduces the biases that may arise from separate analyses but has high computational complexity, limiting its application to large-scale datasets.

\subsubsection{Classical Protein Structure Based Phylogenetic BioTree Construction Methods}

As shown in table \ref{tab_classical_protein_structural_methods} and Figure.~\ref{fig_method_protein_str}, protein structure analysis is a critical component of bioinformatics, as it provides deeper insights into protein function, interactions, and evolutionary relationships that sequence-based methods alone cannot offer. Unlike sequence-based methods that rely solely on primary amino acid sequences, structure-based methods utilize three-dimensional (3D) structural information of proteins to capture more complex evolutionary and functional relationships. These methods typically require prior knowledge, such as conserved tertiary structures and functional site conservation. The following content discusses the development of structural alignment methods, functional site recognition techniques, and structural comparison algorithms in chronological and logical order, along with their applications.

\paragraph{Development of Protein Structural Alignment Methods.}
Early structural alignment methods, such as \textit{DALI} \citep{holm_dali_1995}, used distance matrix-based alignment to compare protein structures, aiming to detect both global and local structural similarities. {DALI} implemented a network-based tool for protein structure comparison, leveraging prior knowledge of \textit{Tertiary Structure Conservation (P4)} and \textit{Conserved Protein Domains (P1)} to effectively identify remote homologs and functionally similar proteins. {DALI} laid the foundation for the field of protein structural alignment, especially in uncovering distant evolutionary relationships that are not easily detectable by sequence analysis alone. However, its computational complexity limits its application to large-scale datasets.

As the demand for computational efficiency grew, \textit{TM-align} \citep{zhang_tm-align_2005} was introduced. {TM-align} uses the TM-score rotation matrix combined with dynamic programming to achieve optimal pairwise structural alignment, offering higher speed and better alignment accuracy than {DALI} and CE methods. {TM-align} focuses on \textit{Tertiary Structure Conservation (P4)} (e.g., RMSD) to ensure that alignments reflect conserved 3D structures. Its significant computational efficiency and accuracy have led to its widespread use in practical applications, particularly for rapid and precise comparison of large protein structure databases.

With the need for multiple protein structure alignments, the \textit{MultiProt} algorithm \citep{shatsky_multiprot_2002} provided a solution for multiple structural alignments. Unlike the previous methods, {MultiProt} identifies common geometric cores among proteins without requiring all molecules to participate in the alignment. Its advantage lies in handling highly variable datasets, especially in scenarios involving diverse structures and partial alignments. However, its computational cost increases significantly with larger data size and complexity.

In the 2010s, to address the growing number of protein structures and improve the accuracy of multiple alignments, \textit{mTM-align} \citep{dong_mtm-align_2018} was developed. {mTM-align} is an extension of the {TM-align} method, designed to tackle the challenge of aligning more than two protein structures simultaneously. This method retains the advantages of \textit{Tertiary Structure Conservation (P4)} and has been benchmarked on widely used datasets, demonstrating consistent superiority in alignment accuracy and computational efficiency. It is particularly useful for large-scale proteomic datasets where accurate and rapid multiple structural alignments are critical.

The most recent multiple structure alignment method, \textit{GTalign} \citep{margelevicius_gtalign_2024}, employs a spatial index-driven strategy to achieve optimal superposition at high speeds. {GTalign} focuses on providing rapid and accurate structural comparisons using its spatial indexing approach. Its high efficiency in parallel processing and rapid computation makes it highly applicable in modern biological research, especially when dealing with large-scale datasets. However, the requirement for pre-indexing structures can pose a challenge when new data is frequently added to the analysis pipeline.

\paragraph{Development of Functional Site Recognition Techniques.}
Functional site recognition is another critical aspect of structure-based protein analysis. The early method, \textit{SiteEngine} \citep{shulman-peleg_recognition_2004}, identifies regions on one protein surface that are similar to a binding site on another protein. {SiteEngine} does not require sequence or fold similarities; instead, it uses prior knowledge in the form of \textit{Functional Site Conservation (P5)} to recognize similar binding sites. This method is particularly advantageous for predicting molecular interactions and aiding in drug discovery. However, its dependency on high-quality protein structures can limit its application in cases where experimental data is sparse or noisy.

The \textit{APoc} method \citep{gao_apoc_2013} is another tool designed for large-scale structural comparison, particularly for identifying pockets on protein surfaces. {APoc} uses a scoring function called the Pocket Similarity Score (PS-score) to measure the similarity between different protein pockets and employs statistical models to assess the significance of these similarities. It leverages \textit{Functional Site Conservation (P5)} to enhance its predictive power in classifying ligand-binding sites and predicting protein molecular function. While robust, its performance is influenced by the quality of input data, especially when the structures are predicted models rather than experimentally determined ones.

\textit{eMatchSite} \citep{brylinski_ematchsite_2014} introduced a new sequence order-independent method for binding site alignment in protein models, capable of constructing accurate local alignments. {eMatchSite} shows high tolerance to structural distortions in weakly homologous protein models and uses \textit{Functional Site Conservation (P5)} as prior knowledge, providing new perspectives for studying drug-protein interaction networks, especially in system-level applications such as polypharmacology and rational drug repositioning.

\paragraph{Comparative Modeling and Other Methods.}
\textit{MODELLER} \citep{webb_comparative_nodate_2016} is a traditional tool for comparative protein structure modeling. It predicts 3D structures based on sequence alignment with known templates and uses \textit{Tertiary Structure Conservation (P4)} as key prior knowledge. While effective for modeling proteins with known homologs, {MODELLER}'s performance diminishes for novel proteins without suitable templates.

The \textit{DeepAlign} method \citep{wang_protein_2013} takes a different approach by combining spatial proximity with evolutionary information and hydrogen-bonding similarity, providing a more comprehensive alignment perspective that accounts for both geometric and evolutionary constraints.

\subsection{Classical Single-Cell-Based Lineage BioTree Construction Methods}
\begin{table*}[t]
    \centering
    \tiny
    \caption{Overview of Dimensionality Reduction, Probabilistic, and RNA Velocity-based Methods for Trajectory and Pseudotime Inference.}
    \begin{tabular}{p{1.8cm}|p{5.0cm} p{1.0cm} p{4.5cm}}
        \toprule
        \textbf{Method Name}    & \textbf{Description}                                                                         & \textbf{Ref.}                & \textbf{URL}                                               \\
        \midrule
        \textbf{TSCAN}          & Clusters cells based on gene expression and constructs an MST for trajectory identification. & \citep{ji_tscan_2016}             & \url{https://github.com/zji90/TSCAN}                       \\
        \textbf{Monocle 2}      & Enhances Monocle with a reversed graph embedding for linear and trajectories.                & \citep{qiu_reversed_2017}         & \url{https://cole-trapnell-lab.github.io/monocle-release/} \\
        \textbf{FORKS}          & Infers bifurcating and linear trajectories using Steiner trees.                              & \citep{sharma_forks_2017}         & \url{https://github.com/macsharma/FORKS}                   \\
        \textbf{Scanpy}         & Offers a framework for single-cell analysis, including trajectory methods.                   & \citep{wolf2018scanpy}            & \url{https://scanpy.readthedocs.io/}                       \\
        \textbf{Seurat}         & Comprehensive tool for single-cell RNA-seq trajectory inference.                             & \citep{stuart2019integrative}     & \url{https://satijalab.org/seurat/}                        \\
        \textbf{PAGA}           & Creates an abstracted graph of cellular relationships to refine trajectories.                & \citep{wolf_paga_2019}            & \url{https://github.com/theislab/paga}                     \\
        \textbf{Monocle 3}      & Combines Monocle 2, UMAP, and PAGA for managing complex branching trajectories.              & \citep{cao_single-cell_2019}      & \url{https://cole-trapnell-lab.github.io/monocle3}         \\
        \textbf{SoptSC}         & Constructs a cell similarity graph for pseudotemporal ordering.                              & \citep{wang_cell_2019}            & \url{https://github.com/WangShuxiong/SoptSC}               \\
        \textbf{Waddington-OT}  & Applies optimal transport to infer trajectories from scRNA-seq data.                         & \citep{schiebinger2019optimal}    & \url{https://github.com/zsteve/gWOT}                       \\
        \textbf{PoincaréMaps}   & Estimates pseudotime using hyperbolic distances in hyperbolic space.                         & \citep{klimovskaia2020poincare}   & \url{https://github.com/facebookresearch/PoincareMaps}     \\
        \textbf{VIA}            & Employs random walks and MCMC simulations for trajectory reconstruction.                     & \citep{stassen_generalized_2021}  & \url{https://github.com/ShobiStassen/VIA}                  \\
        \textbf{LineageOT}      & Models lineage progression using optimal transport theory.                                   & \citep{forrow2021lineageot}       & \url{https://github.com/aforr/LineageOT}                   \\
        \textbf{GeneTrajectory} & Uses optimal transport metrics to infer gene trajectories.                                   & \citep{qu_gene_2024}              & \url{https://github.com/KlugerLab/GeneTrajectory}          \\
        \midrule
        \textbf{SCUBA}          & Bifurcation analysis for trajectory inference in gene space.                                 & \citep{marco_bifurcation_2014}    & \url{https://github.com/gcyuan/SCUBA}                      \\
        \textbf{BGP}            & Estimates branching times for individual genes.                                              & \citep{boukouvalas_bgp_2018}      & \url{https://github.com/ManchesterBioinference/BranchedGP} \\
        \textbf{CSHMMs}         & Extends probabilistic methods to continuous trajectories.                                    & \citep{lin_continuous-state_2019} & \url{http://www.andrew.cmu.edu/user/chiehl1/CSHMM/}        \\
        \textbf{Ouija}          & Models gene expression along pseudotemporal trajectories.                                    & \citep{campbell_descriptive_2019} & \url{https://github.com/kieranrcampbell/ouija}             \\
        \midrule
        \textbf{RNA velocity}   & Analyzes spliced and unspliced transcripts to capture transcriptional dynamics.              & \citep{la2018rna}                 & \url{http://velocyto.org/}                                 \\
        \textbf{scVelo}         & Generalizes RNA velocity analysis to diverse kinetics.                                       & \citep{bergen_generalizing_2020}  & \url{https://scvelo.readthedocs.io/}                       \\
        \textbf{CellRank}       & Integrates RNA velocity with pseudotime inference to identify lineage drivers.               & \citep{lange_cellrank_2022}       & \url{https://cellrank.readthedocs.io/}                     \\
        \textbf{TFvelo}         & Integrates gene regulatory data to extend RNA velocity analysis.                             & \citep{li_tfvelo_2024}            & \url{https://github.com/xiaoyeye/TFvelo}                   \\
        \bottomrule
    \end{tabular}
    \label{tab_classical_sc_lineage}
\end{table*}

\begin{table*}[h]
    \centering
    \scriptsize
    \caption{Overview of Classical Single-cell Lineage Inference \& Tree Construction Methods.}
    \begin{tabular}{p{1.5cm}|p{5cm} p{1cm} p{4.5cm}}
        \toprule
        \textbf{Method Name}     & \textbf{Description}                                                                                                                                     & \textbf{Ref}               & \textbf{URL}                                                        \\ \midrule
        \textbf{cellTree}        & Uses a probabilistic framework to model gene expression data and construct a tree-like structure outlining hierarchical differentiation.                 & \cite{duverle_celltree_2016}     & \url{https://github.com/tidwall/celltree}                           \\ \midrule
        \textbf{Slingshot}       & Constructs lineage trees by embedding cells into a reduced dimensional space and connecting clusters through minimum spanning trees.                     & \cite{street_slingshot_2018}     & \url{https://github.com/kstreet13/slingshot}                        \\ \midrule
        \textbf{Monocle DDRTree} & Builds a tree structure representing cell lineages using dimensionality reduction combined with reversed graph embedding.                                & \cite{qiu_reversed_2017}         & \url{https://cole-trapnell-lab.github.io/monocle-release}           \\ \midrule
        \textbf{PAGA trees}      & Constructs a graph representing clusters of cells and abstracts it into a tree structure to capture hierarchical branching.                              & \cite{wolf_paga_2019}            & \url{https://dynverse.org/reference/dynmethods/other/ti_paga_tree/} \\ \midrule
        \textbf{PROSSTT}         & Simulates single-cell RNA-seq datasets for differentiation processes to generate lineage trees for benchmarking lineage inference methods.               & \cite{papadopoulos_prosstt_2019} & \url{https://github.com/soedinglab/prosstt}                         \\ \midrule
        \textbf{SoptSC}          & Builds a lineage tree by clustering and lineage inference using cell-to-cell similarity matrices.                                                        & \cite{wang_cell_2019}            & \url{https://github.com/WangShuxiong/SoptSC}                        \\ \midrule
        \textbf{CALISTA}         & Integrates clustering, lineage progression, transition gene identification, and pseudotime ordering into a unified framework to construct lineage trees. & \cite{papili_gao_calista_2020}   & \url{https://github.com/CABSEL/CALISTA}                             \\
        \bottomrule
    \end{tabular}
    \label{tab_classical_sc_lineage_tree}
\end{table*}

\begin{figure*}[t]
	\begin{center}
		\includegraphics[width=0.99\linewidth]{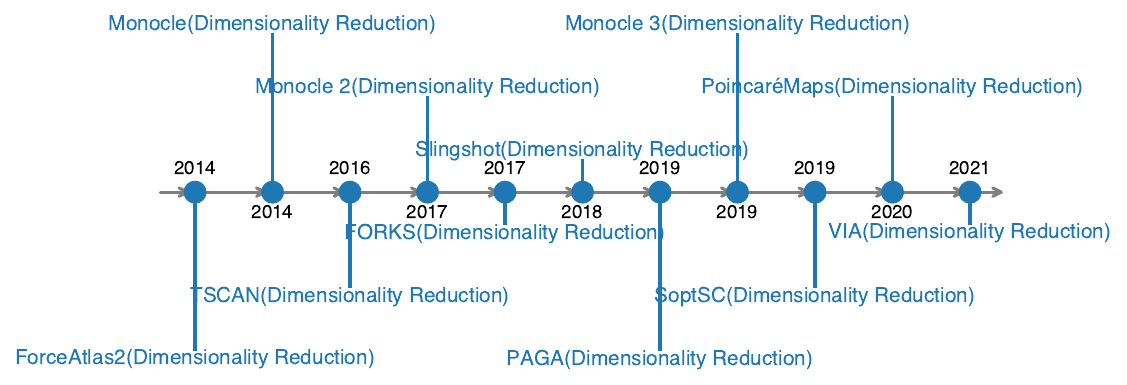}
	\end{center}
	\caption{\textit{The timeline of Dimensionality Reduction based Classical Single Cell Trajectory Inference Methods.} The figure shows the chronological development of trajectory inference methods based on single-cell RNA sequencing data. These methods have evolved by incorporating different types of prior knowledge to improve accuracy and computational efficiency in cell development analysis.}
	\label{fig_method_singlecell_dr}
\end{figure*}

\begin{figure*}[t]
	\begin{center}
		\includegraphics[width=0.99\linewidth]{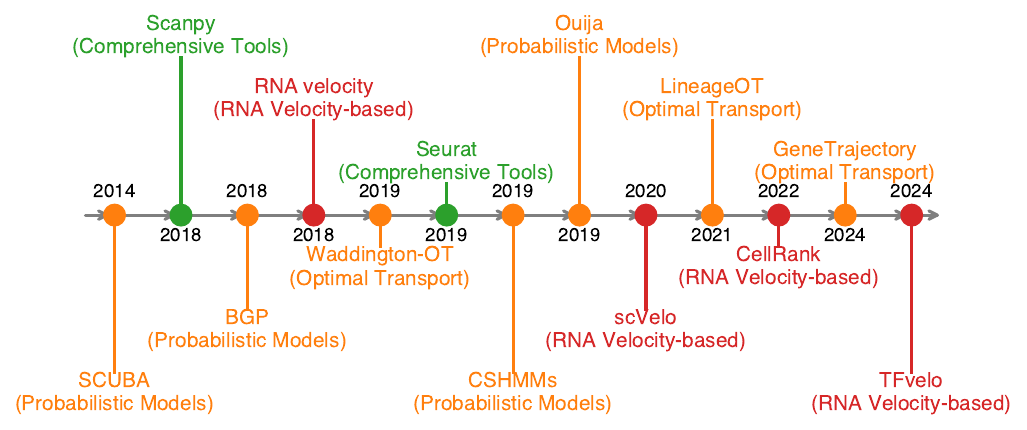}
	\end{center}
	\caption{\textit{The timeline of Classical Single Cell Trajectory Inference Methods.} The figure shows the chronological development of trajectory inference methods based on single-cell RNA sequencing data. These methods have evolved by incorporating different types of prior knowledge to improve accuracy and computational efficiency in cell development analysis.}
	\label{fig_method_singlecell_edr}
\end{figure*}

\begin{figure*}[t]
	\begin{center}
		\includegraphics[width=0.99\linewidth]{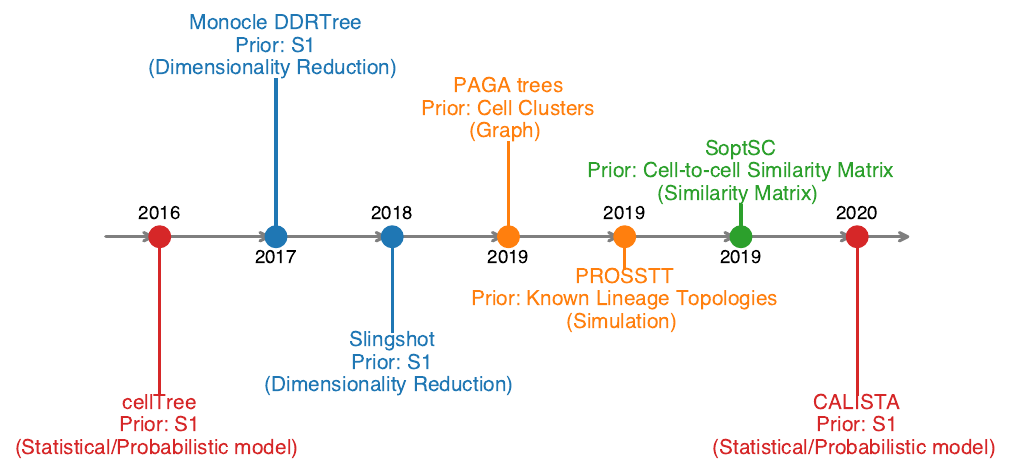}
	\end{center}
	\caption{\textit{The Timeline of Classical Single Cell Tree Construction Methods.} The figure shows the chronological development of tree-based methods for cell differentiation analysis based on single-cell RNA sequencing data. These methods have evolved by incorporating different types of prior knowledge to improve accuracy and computational efficiency in cell development analysis.}
	\label{fig_method_singlecell_tree}
\end{figure*}

In single-cell RNA sequencing (scRNA-seq) analysis, inferring developmental and differentiation trajectories is essential for unraveling complex biological processes. This involves three core tasks: trajectory, pseudo-time, and lineage inference. Various computational methods have been developed for these purposes, primarily falling into two categories: trajectory \& pseudo-time inference methods and lineage inference methods.

\subsubsection{Classical Single-cell Trajectory \& Pseudotime Inference Methods}

As shown in Table.~\ref{tab_classical_sc_lineage}, Figure.~\ref{fig_method_singlecell_dr} and Figure.~\ref{fig_method_singlecell_edr}, the trajectory inference methods aim to reconstruct the differentiation pathways of cells by organizing them along potential developmental trajectories. These methods use prior information \textit{Cell Type-Specific Marker Genes (S3)} to identify continuous progression and branching points that represent different lineage decisions. In contrast, pseudo-time inference, based on the prior assumption \textit{Pseudotime Ordering (S4)}, focuses on ordering cells along a temporal axis, estimating the relative progression of individual cells through a dynamic process. While pseudo-time methods do not necessarily infer explicit branching lineages, they capture the gradual changes in cell states over time. Both approaches are primarily grounded in prior knowledge high-dimension \textit{Cell Type-Specific Marker Genes (S3)}. The existing computational methods can be broadly categorized into three groups. The first two (dimensionality reduction and gene space-based probabilistic methods) link cells over time using gene expression, while the third (RNA velocity) relies on data from spliced and unspliced transcripts.

\paragraph{Dimensionality Reduction-based Methods for Trajectory \& Pseudotime Inference.}
Dimensionality reduction-based methods leverage lower-dimensional representations of cells to infer spanning trees or other graphical structures, which are then used to map cells and reconstruct trajectories. These methods allow for the simultaneous reconstruction of cellular trajectories and the visualization of cell distributions in an interpretable and accessible manner. The existing methods can generally be classified into three main categories: dimensionality reduction methods, dimensionality reduction combined with graph-based methods, and dimensionality reduction integrated with pseudo-time analysis.

For \textit{dimensionality reduction methods}, high-dimensional \textit{Cell Type-Specific Marker Genes (S3)} are reduced to a lower-dimensional space for trajectory inference directly. For instance, \textit{ForceAtlas2} \citep{jacomy_forceatlas2_2014} positions nodes in a graph by simulating a physical system where nodes repel each other like charged particles, while edges act like springs pulling connected nodes together, leading to a balanced and visually meaningful network structure for trajectory inference. The \textit{Monocle} \citep{trapnell_dynamics_2014} orders cells in pseudotime using independent component analysis (ICA) and constructs a spanning tree to infer linear trajectories. \textit{Monocle 2} \citep{qiu_reversed_2017} enhances Monocle with a reversed graph embedding technique to create a principal graph, enabling robust handling of both linear and branching trajectories. \textit{FORKS} \citep{sharma_forks_2017} infers bifurcating and linear trajectories using Steiner trees, enhancing robustness against noise and complexity. \textit{TSCAN} \citep{ji_tscan_2016} clusters cells based on gene expressions and constructs a minimum spanning tree (MST) for trajectory identification. \textit{Slingshot} \citep{street_slingshot_2018} fits smooth curves in the reduced-dimensional space for simultaneous pseudotime and lineage inference. \textit{PAGA} \citep{wolf_paga_2019} creates an abstracted graph of cellular relationships to capture both continuous and discrete transitions before refining the trajectories. \textit{Monocle 3} \citep{cao_single-cell_2019} combines the strengths of Monocle 2, UMAP, and PAGA to manage complex branching trajectories with improved accuracy and scalability. \textit{SoptSC} \citep{wang_cell_2019} constructs a cell similarity graph for pseudotime ordering and uses the shortest path for trajectory inference. \textit{PoincaréMaps} \citep{klimovskaia2020poincare} estimates pseudotime ordering using hyperbolic distances within hyperbolic space. \textit{Waddington-OT} \citep{schiebinger2019optimal} applies optimal transport to infer trajectories from scRNA-seq data. \textit{LineageOT} \citep{forrow2021lineageot} models lineage progression using optimal transport theory. \textit{GeneTrajectory} \citep{qu_gene_2024} employs optimal transport metrics to infer gene trajectories. \textit{Seurat} \citep{stuart2019integrative} and \textit{Scanpy} \citep{wolf2018scanpy} are comprehensive tools for single-cell RNA-seq trajectory inference. In addition, \textit{VIA} successfully identifies elusive lineages and rare cell fates across various prior knowledge, including \textit{Protein Expression Levels} and \textit{Epigenetic Modification}. It \citep{stassen_generalized_2021} employs random walks and MCMC simulations for trajectory reconstruction.

\paragraph{Probabilistic Models in Gene Space.}
Dimensionality reduction has the potential downside of inferring trajectories from only the most abundantly \textit{Cell Type-Specific Marker Genes (S3)}, which could hinder the ability to distinguish and accurately reconstruct cell state clusters that have fewer cells. Several methods have been proposed to overcome this limitation by inferring pseudotime and trajectories directly from the \textit{Gene Expression Profiles (S1)}. \textit{SCUBA} \citep{marco_bifurcation_2014} uses bifurcation analysis to model trajectories in gene space. \textit{CSHMMs} \citep{lin_continuous-state_2019} extend probabilistic methods to continuous trajectories, allowing cells to be assigned to any position along the trajectory graph. \textit{BGP} \citep{boukouvalas_bgp_2018} estimates branching times for individual genes, while \textit{Ouija} \citep{campbell_descriptive_2019} models gene expression along pseudotemporal trajectories.

\paragraph{RNA Velocity-based Methods.}
RNA velocity-based methods further utilize prior information \textit{RNA Velocity (S2)} to analyze spliced and unspliced transcripts, capturing transcriptional dynamics within cells. \textit{RNA velocity} \citep{la2018rna} provides insights into a cell's future trajectory by calculating the ratio of spliced and unspliced mRNAs. \textit{scVelo} \citep{bergen_generalizing_2020} generalizes RNA velocity analysis to diverse transcriptional kinetics. \textit{CellRank} \citep{lange_cellrank_2022} integrates RNA velocity with pseudotime inference to identify lineage drivers. \textit{TFvelo} \citep{li_tfvelo_2024} extends RNA velocity analysis by integrating gene regulatory data, enhancing the accuracy of cell dynamics and trajectory inference.

\subsubsection{Classical Single-cell Lineage Inference \& Tree Construction Methods}

As shown in Table.\ref{tab_classical_sc_lineage_tree} and Figure.\ref{fig_method_singlecell_tree}, single-cell lineage inference aims to reconstruct the hierarchical relationships between individual cells by analyzing their \textit{Gene Expression Profiles (S1)}. Its primary goal is to generate a lineage {tree} that represents the developmental paths cells take as they divide and differentiate. Each branch of the tree reflects how cells progress from a common progenitor to various specialized cell types.

\textit{Dimensionality reduction-based methods} map cell data into a low-dimensional space to reconstruct complex lineage {trees} with multiple branches, allowing for pseudotime inference and better noise handling during cell differentiation analysis. \textit{Slingshot} \citep{street_slingshot_2018} constructs lineage {trees} by embedding cells into a reduced dimensional space and connecting clusters through minimum spanning {trees}, thereby capturing the branching structure of cell lineages in the form of a {tree}. \textit{Monocle DDRTree} \citep{qiu_reversed_2017} explicitly builds a {tree} structure to represent cell developmental lineages by combining discriminative dimensionality reduction with reversed graph embedding, enabling the inference of cell trajectories from gene expression data within a {tree} framework.

\textit{Graph-based methods} utilize graph abstraction techniques to model relationships between cells and reconstruct lineage {trees}. \textit{PAGA trees} \citep{wolf_paga_2019} constructs a graph where nodes represent clusters of cells and edges represent the connectivity probabilities between them. By abstracting this graph into a simplified {tree} structure, PAGA enables the reconstruction of complex lineage topologies, capturing the hierarchical branching patterns inherent in cell differentiation processes.

\textit{Simulation-based methods} provide synthetic datasets with known lineage topologies to test and develop lineage reconstruction tools. \textit{PROSSTT} \citep{papadopoulos_prosstt_2019} simulates single-cell RNA-seq datasets for differentiation processes, generating lineage {trees} of any desired complexity, noise level, noise model, and size. By producing datasets with predefined {tree} structures, PROSSTT allows for benchmarking and evaluating the accuracy of lineage inference methods in reconstructing the true underlying {tree} topology.

\textit{Similarity matrix-based methods} utilize a cell-to-cell similarity matrix to analyze relationships between cells and construct lineage {trees} based on these similarities. \textit{SoptSC} \citep{wang_cell_2019} builds a lineage {tree} by performing clustering and lineage inference using cell-cell relationships derived from a similarity matrix, effectively capturing the hierarchical differentiation paths in a {tree} structure.

\textit{Statistical/Probabilistic model-based methods} rely on statistical or probabilistic models to account for noise and stochasticity in gene expression profiles while constructing lineage {trees}. \textit{cellTree} \citep{duverle_celltree_2016} models the gene expression data using a probabilistic framework to construct a {tree}-like structure that outlines hierarchical differentiation, explicitly representing cell lineages as branches of a {tree}. \textit{CALISTA} \citep{papili_gao_calista_2020} integrates clustering, lineage progression, transition gene identification, and pseudotime ordering into a unified framework, constructing lineage {trees} that represent the developmental trajectories of cells based on statistical modeling of gene expression patterns.

\subsection{Limitations of Traditional BioTree Construction Methods}
\paragraph{Computational Complexity.}
Traditional tree construction methods, such as \textit{Maximum Likelihood (ML)} and \textit{Bayesian Inference}, have been foundational in phylogenetics due to their robust statistical frameworks. However, as sequence numbers grow, the exponential increase in possible tree topologies renders exhaustive searches infeasible. While heuristic approaches like \textit{RAxML} and \textit{MrBayes} mitigate these challenges, they remain computationally demanding, requiring significant resources for high-throughput sequencing datasets, potentially limiting scalability.

\paragraph{Scalability Challenges.}
The rise of multi-omics approaches introduces complex data integration demands that traditional methods struggle to address. These methods, often tailored for single sequence types, face difficulties in capturing the biological context of genomic, transcriptomic, and proteomic interrelationships. Advances in statistical models are gradually improving adaptability, but the challenges of scalability and dimensionality remain significant.

\paragraph{Model Dependency.}
Predefined evolutionary models, such as \textit{substitution models}, simplify phylogenetic analysis but may not fully reflect real evolutionary dynamics, where rates vary across lineages and selective pressures differ among genes. This dependency introduces biases that modern flexible models aim to address, allowing for more accurate evolutionary representations.

\paragraph{Handling Uncertain and Noisy Data.}
Sequencing errors, gene loss, and missing data are common in real-world datasets and pose challenges for robust tree construction. Traditional methods are sensitive to these uncertainties, often yielding less reliable topologies. Advances in preprocessing and uncertainty-aware frameworks are enhancing resilience, enabling these methods to better accommodate noisy data while maintaining accuracy.

\section{Deep Learning-Based BioTree Construction Methods} \label{sec_deep}
The rapid development of deep learning has revolutionized BioTree construction by introducing methods capable of capturing complex biological relationships from diverse and high-dimensional data. Unlike traditional approaches, which often rely on single data modalities, deep learning excels in **information fusion**, seamlessly integrating data from genomic sequences, protein structures, transcriptomics, and single-cell omics. This ability to combine heterogeneous data types not only enhances the accuracy of tree inference but also uncovers hidden patterns across biological systems. By leveraging advanced neural network architectures and embedding prior biological knowledge, these methods address critical challenges such as scalability, noise robustness, and interpretability. This section provides an overview of deep learning frameworks for BioTree construction, categorized by their applications to general datasets, gene-based trees, protein-based trees, and single-cell lineage trees, highlighting the transformative potential of information fusion in phylogenetics.

\subsection{Deep General BioTree Construction Methods}

\begin{figure*}[t]
	\begin{center}
		\includegraphics[width=0.8\linewidth]{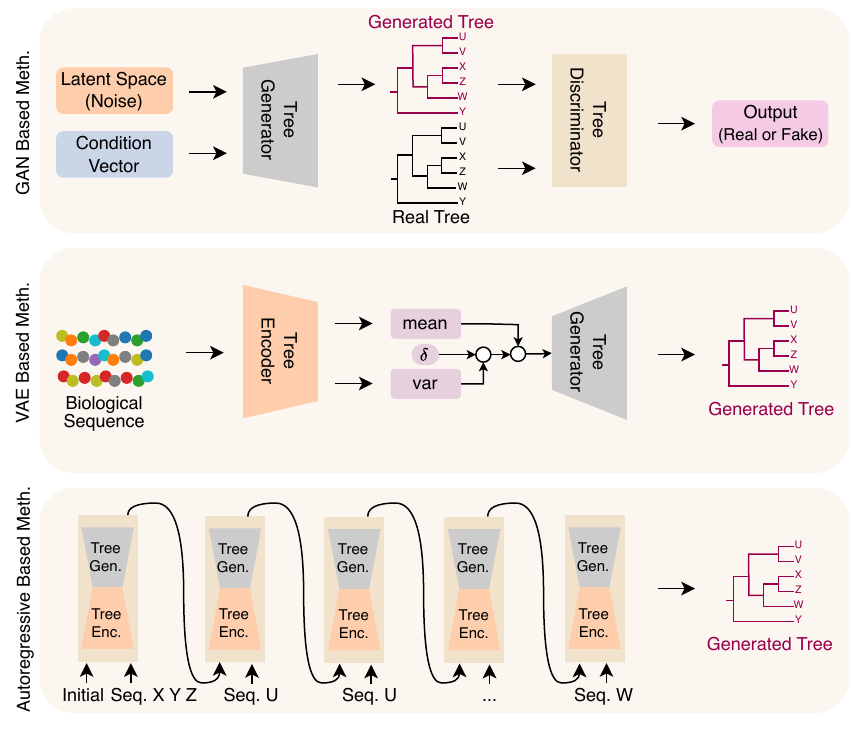}
	\end{center}
	\caption{
		\textit{The Deep Learning-Based BioTree Construction Methods.} This figure summarizes three common tree generation methods for biological sequence analysis: \textit{GAN-Based Method}, which uses a latent space and a condition vector to generate trees, with a discriminator distinguishing real from generated trees; \textit{VAE-Based Method}, which encodes sequences into a latent space and generates trees by sampling from it; and \textit{Autoregressive-Based Method}, which iteratively generates trees from an initial sequence and subsequent sequences using an autoregressive model.
		}
	\label{fig_deep}
\end{figure*}

Tree generation is a critical research problem with diverse applications, including biological evolution analysis, lineage tracing, and the construction of hierarchical classification systems. Unlike general graph generation tasks, tree generation must adhere to strict structural constraints, such as acyclicity, single-root properties, and hierarchical relationships, which reflect the clear evolutionary directionality inherent in many biological systems. These requirements introduce unique challenges, as tree generation must not only capture complex structural features but also ensure biologically meaningful outputs. 

In this section, we review recent advances in deep learning-based tree generation methods, including \textit{Generative Adversarial Networks (GANs)}, \textit{Variational Autoencoders (VAEs)}, and \textit{autoregressive models}. These methods leverage data-driven approaches to model tree structures while addressing challenges such as scalability, uncertainty, and multimodal data integration. We discuss each method's key characteristics, applications, and limitations, highlighting their potential for advancing tree generation in diverse biological contexts. Figure~\ref{fig_deep} provides an overview of the three common tree generation frameworks explored in this section.

\paragraph{GAN-Based BioTree Construction Methods.}
Generative Adversarial Networks (GANs) employ adversarial training between a generator that produces graph structures and a discriminator that evaluates their realism, playing a significant role in graph generation tasks. Classical models like \textit{NetGAN} generate graphs by learning random walk sequences on existing graphs, showcasing effectiveness in network reconstruction tasks \citep{bojchevski2018netgan}. Building on this, {MolGAN} extends the GAN framework to molecular graphs, focusing on chemical properties, which has significant applications in drug design \citep{de2018molgan}.

More sophisticated GANs, such as \textit{Hierarchical GANs}, introduce complex generative structures, including \textit{GAN-Tree} and \textit{Hierarchical GAN-Tree}, to handle multimodal data distributions and multi-label classification tasks \citep{kundu_gan-tree_2019, wang_hierarchical_2022}. The GAN-Tree model incrementally learns a hierarchical generative structure for multimodal data, offering a versatile framework for multimodal data generation. This incremental learning of tree-like structures enables it to effectively handle image generation and multi-label classification tasks, outperforming traditional GAN models in these scenarios.

Further advancing the GAN-based approach, \textit{HC-MGAN} introduces a hierarchical generation strategy using multi-generator GANs (MGANs) for deep clustering \citep{mello_top-down_2022}. It achieves hierarchical data organization through top-down clustering trees, offering meaningful clustering of real data distributions and a novel method for tree structure generation tasks. Additionally, the \textit{Hierarchical GAN-Tree (HGT)} model combines bidirectional capsule networks to enhance feature generation through unsupervised divisive clustering, addressing mode collapse issues commonly found in traditional GANs \citep{wang_hierarchical_2022}.

These GAN-based tree generation methods excel in managing complex data distributions and hierarchical structures. However, they still face challenges under strict tree structure constraints, such as acyclicity. Their performance can potentially be enhanced by integrating other generative strategies like VAEs or autoregressive models, especially for generating larger and more intricate tree structures.

\paragraph{VAE-Based BioTree Construction Methods.}
Variational Autoencoders (VAEs) offer a probabilistic approach to learning latent representations of graph structures, providing potential solutions for generating specific tree structures. Although traditional VAEs, like \textit{VGAE}, have shown great performance in graph representation and link prediction tasks \citep{kipf2016variational}, their unconstrained generation process can result in structures that do not adhere to the hierarchy and acyclicity requirements of trees.

To address these constraints, the \textit{Tree Variational Autoencoder (TreeVAE)} introduces a generative hierarchical clustering model that learns a flexible tree-based posterior distribution over latent variables \citep{manduchi_tree_2023}. This model enables the generation of samples while preserving the hierarchical structure, proving effective in data clustering and generation tasks. Similarly, the \textit{Junction Tree Variational Autoencoder (JTVAE)} tackles the challenge of chemical graph generation by converting the problem into tree generation \citep{jin_junction_2018}. It first generates a tree-structured scaffold, followed by a message-passing network that reconstructs the molecular graph. This two-step method ensures chemical validity and has demonstrated superiority over previous state-of-the-art methods in various molecular design tasks.

\textit{Diffuse-TreeVAE} further enhances VAE-based tree generation by integrating it into the framework of Denoising Diffusion Probabilistic Models (DDPMs) for image generation \citep{goncalves_structured_2024}. This approach generates root embeddings for a learned latent tree structure, propagating through hierarchical paths, and uses a second-stage DDPM to refine and produce high-quality images. It overcomes the limitations of traditional VAE models, contributing to advancements in clustering-based generative modeling. Additionally, researchers have emphasized uncertainty quantification (UQ) in generative models. For instance, \textit{Leveraging Active Subspaces for Epistemic Model Uncertainty} captures model uncertainty in the JT-VAE model by leveraging low-dimensional active subspaces without altering the model architecture \citep{abeer_leveraging_2024}. This method has shown effectiveness in molecular optimization tasks.

Overall, VAE-based methods, particularly those employing hierarchical structures like TreeVAE and JTVAE, address the constraints required for tree generation. However, they still need refinement in scaling to larger and more complex tree structures.

\paragraph{Autoregressive BioTree Construction Methods.}
Autoregressive models, such as \textit{GraphRNN}\citep{you2018graphrnn}, treat graph generation as a sequential process, where nodes and edges are generated step-by-step. This sequential nature allows for fine-grained control over hierarchical relationships and dependencies inherent in tree structures. By explicitly modeling the generation order, GraphRNN ensures the preservation of acyclicity and hierarchical properties, making it particularly suited for generating trees.

Applications of GraphRNN to tree generation include the construction of biological family trees and evolutionary trees, where maintaining hierarchical information is crucial. The stepwise approach of autoregressive models offers advantages in controlling the generated structure's complexity and depth, providing flexibility in the creation of diverse tree structures. However, the inherent sequential process can be computationally intensive, particularly as the tree size increases.

In summary, deep learning-based tree generation methods offer diverse approaches, each with its own set of strengths and limitations. GAN-based models are powerful in handling complex data distributions but face challenges in strictly adhering to tree constraints. VAE-based methods provide a probabilistic framework suitable for hierarchical clustering and molecular design but require further enhancement to scale to larger tree structures. Autoregressive models, while maintaining strict control over hierarchical generation, may encounter computational limitations as tree complexity grows. Future research may benefit from combining these methods to leverage their individual strengths, creating more robust and scalable solutions for tree generation tasks.

\subsection{Deep Gene-Based Phylogenetic BioTree Construction Methods}
\begin{figure*}[t]
	\begin{center}
		\includegraphics[width=0.99\linewidth]{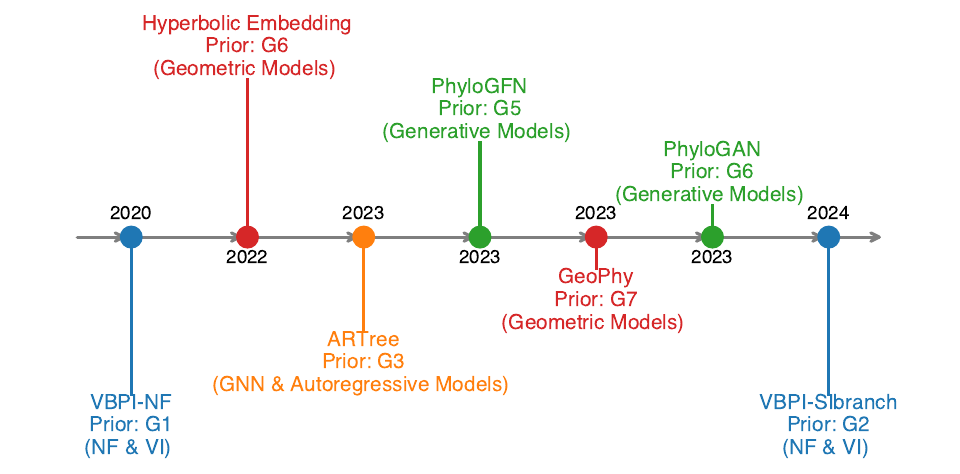}
	\end{center}
	\caption{\textbf{The Timeline of Deep Gene BioTree Construction Methods.} The figure shows the development of deep learning-based gene tree construction methods in phylogenetics from 2020 to 2024, categorized into normalizing flows and variational inference methods, graph neural network (GNN) and autoregressive models, and geometric and generative models. Different colors indicate different categories.}
	\label{fig_method_deep_gene}
\end{figure*}

\begin{table*}[t]
    \centering
    \scriptsize
    \caption{Overview of the Classical Gene-based Tree Construction Methods.}
    \begin{tabular}{p{1.5cm}|p{5cm} p{1cm} p{4.5cm}}
        \toprule
        \textbf{Method Name}          & \textbf{Description}                                                                                                                                & \textbf{Ref.}             & \textbf{URL}                                   \\
        \midrule
        \textbf{VBPI-NF}              & Uses normalizing flows to model branch length distributions across tree topologies, improving flexibility in non-Euclidean tree space.              & \cite{zhang_improved_2020}     & \url{https://github.com/zcrabbit/vbpi-nf}      \\ \midrule
        \textbf{Hyperbolic Embedding} & Embeds gene sequences into hyperbolic spaces to reduce distance distortion, improving species tree distance modeling.                               & \cite{jiang_learning_2022}     & \url{https://github.com/yueyujiang/hdepp}      \\ \midrule
        \textbf{ARTree}               & Autoregressive model that decomposes tree topology into sequences of leaf node additions, using GNNs for tree topology estimation.                  & \cite{xie_artree_2023}         & \url{https://github.com/tyuxie/ARTree}         \\ \midrule
        \textbf{PhyloGFN}             & Utilizes generative flow networks (GFlowNets) to sample from the multimodal posterior distribution over tree topologies and evolutionary distances. & \cite{zhou_phylogfn_2023}      & \url{https://github.com/zmy1116/phylogfn}      \\ \midrule
        \textbf{Geophy}               & Fully differentiable method for phylogenetic inference in continuous geometric spaces, incorporating chromatin accessibility data.                  & \cite{mimori_geophy_2023}      & \url{https://github.com/m1m0r1/geophy}         \\ \midrule
        \textbf{PhyloGAN}             & Generative adversarial network (GAN) model for inferring phylogenetic relationships by generating data similar to real evolutionary data.           & \cite{smith_phylogenetic_2023} & \url{https://github.com/meganlsmith/phyloGAN/} \\ \midrule
        \textbf{VBPI-SIbranch}        & Applies graph neural networks (GNNs) to handle non-Euclidean branch length space with improved computational efficiency.                            & \cite{xie_variational_2024}    & \url{https://github.com/tyuxie/vbpi-sibranch}  \\
        \bottomrule
    \end{tabular}
    \label{tab_deep_gene}
\end{table*}

As shown in Table~\ref{tab_deep_gene} and Figure~\ref{fig_method_deep_gene}, recent advances in deep learning have significantly advanced the field of phylogenetics, leading to the development of novel algorithms and techniques that improve the accuracy, efficiency, and scalability of phylogenetic inference. These methods leverage deep learning architectures and the concept of information fusion to combine prior biological knowledge, such as conserved genomic regions, evolutionary substitution models, and gene duplication events, with data-driven approaches to address challenges faced by traditional methods. Based on the prior knowledge they utilize and the problems they tackle, existing deep learning methods can be categorized into three main groups: normalizing flows and variational inference methods, graph neural network (GNN) and autoregressive models, and geometric and generative models.

Normalizing flows and variational inference methods excel in managing the complex, non-Euclidean tree space required for phylogenetic inference. By integrating information fusion, methods such as \textit{VBPI-NF} \citep{zhang_improved_2020} utilize conserved genomic regions to guide the modeling of branch length distributions across tree topologies, while combining this prior knowledge with data-driven variational frameworks for improved uncertainty management. Similarly, \textit{VBPI-SIbranch} \citep{xie_variational_2024} enhances efficiency by incorporating evolutionary substitution models to model nucleotide sequence changes, demonstrating how information fusion bridges theoretical models and empirical data.

Graph neural network (GNN) and autoregressive models adopt flexible probabilistic frameworks, leveraging information fusion to combine heuristic-free data-driven learning with biological priors. For instance, \textit{ARTree} \citep{xie_artree_2023} decomposes tree topologies into node addition operations, effectively utilizing evolutionary substitution models alongside learned conditional distributions to enhance phylogenetic tree generation.

Geometric and generative models take a distinct approach by embedding tree topologies in continuous geometric spaces. These methods emphasize information fusion by integrating multimodal data sources and biological priors. For example, \textit{PhyloGFN} \citep{zhou_phylogfn_2023} utilizes sequence homology and multimodal evolutionary data to sample tree topologies, addressing challenges in parsimony and Bayesian inference. The \textit{hyperbolic embedding method} \citep{jiang_learning_2022} demonstrates how hyperbolic geometry, enriched by genomic linear order and gene duplication events, reduces distortion compared to Euclidean spaces. Similarly, \textit{GeoPhy} \citep{mimori_geophy_2023} combines biological priors with end-to-end geometric transformations, optimizing tree generation.

Generative adversarial networks (GANs) push the boundaries of phylogenetic inference by introducing information fusion into evolutionary data generation. Methods like \textit{PhyloGAN} \citep{smith_phylogenetic_2023} leverage gene duplication and loss events as prior information, improving data-driven heuristic searches and enabling exploration of complex model spaces beyond the reach of traditional methods.

\subsection{Deep Protein-Based Phylogenetic BioTree Construction Methods}
\begin{figure*}[t]
	\begin{center}
		\includegraphics[width=0.99\linewidth]{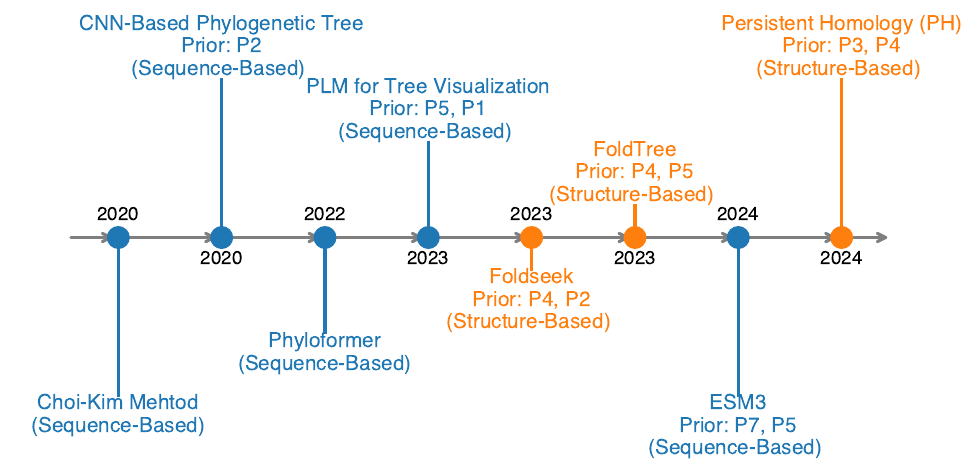}
	\end{center}
	\caption{\textbf{The Timeline of Deep Protein BioTree Construction Methods.} The figure shows the chronological development of phylogenetic tree construction methods based on protein sequence and structural information. These methods have evolved by incorporating different types of prior knowledge to improve accuracy and computational efficiency in evolutionary analysis.}
	\label{fig_method_deep_protein}
\end{figure*}

\begin{table*}[t]
    \centering
    \scriptsize
    \caption{Overview of the Classical Protein-based Tree Construction Methods.}
    \begin{tabular}{p{1.5cm}|p{5cm} p{1cm} p{4.5cm}}
        \toprule
        \textbf{Method Name}                 & \textbf{Description}                                                                                                                            & \textbf{Ref.}                 & \textbf{URL}                                             \\
        \midrule
        \textbf{Choi-Kim Mehtod}             & Sequence-based method using whole-proteome data and evolutionary substitution models to infer phylogenetic relationships.                       & \cite{choi_whole-proteome_2020}    & \url{https://github.com/jaejinchoi/FFP}                  \\ \midrule
        \textbf{CNN-Based Phylogenetic Tree} & CNN-based method for inferring tree topologies from multiple sequence alignments, improving accuracy and speed.                                 & \cite{suvorov_accurate_2020}       & \url{https://github.com/SchriderLab/Tree_learning}       \\ \midrule
        \textbf{Phyloformer}                 & Transformer-based network architecture that predicts evolutionary distances between sequences, allowing for rapid tree topology reconstruction. & \cite{nesterenko_phyloformer_2022} & \url{https://github.com/lucanest/Phyloformer}            \\ \midrule
        \textbf{PLM for Tree Visualization}  & Embedding-based tree visualization to enhance functional clustering of protein sequences.                                                       & \cite{yeung_tree_2023}             & \url{github.com/esbgkannan/chumby}                       \\ \midrule
        \textbf{Foldseek}                    & Converts protein structures into structural alphabets for fast search and alignment.                                                            & \cite{van_kempen_fast_2023}        & \url{https://github.com/steineggerlab/foldseek}          \\ \midrule
        \textbf{FoldTree}                    & Infers relationships using tertiary structure and functional site conservation.                                                                 & \cite{moi_structural_2023}         & \url{https://github.com/DessimozLab/fold_tree}           \\ \midrule
        \textbf{ESM3}                        & Language model for simulating protein evolution using co-evolutionary relationships.                                                            & \cite{hayes_simulating_2024}       & \url{https://www.evolutionaryscale.ai/blog/esm3-release} \\ \midrule
        \textbf{Persistent Homology (PH)}    & Applies topological data analysis to capture structural phylogenetic signals.                                                                   & \cite{bou_dagher_persistent_2024}  & N/A                                                      \\
        \bottomrule
    \end{tabular}
    \label{tab_deep_protein}
\end{table*}

As shown in Table~\ref{tab_deep_protein} and Figure~\ref{fig_method_deep_protein}, phylogenetic inference methods based on protein sequence and structure have made significant advances, particularly in improving efficiency and accuracy when handling large-scale datasets. These methods can be broadly categorized into two main types: sequence-based and structure-based inference methods. As data volume continues to grow, traditional methods have encountered challenges related to computational complexity, which have prompted the introduction of novel algorithms, prior knowledge, and deep learning techniques to drive further innovation in the field of phylogenetic inference.

\subsubsection{Deep Protein Sequence-Based Phylogenetic Tree Methods.}

The \textit{Choi-Kim Method} \cite{choi_whole-proteome_2020} utilized whole-proteome data to construct a tree of life, revealing the evolutionary relationships among extant organisms. This approach applied information-theoretic methods to construct a topologically stable tree and proposed the concept of a deep burst of organismal diversity near the root of the evolutionary tree. It incorporated \textit{Conserved Protein Domains (P1)} as prior knowledge, employing the indicator function \( I(d_i^p, d_j^p) \) to identify conserved regions within protein sequences, reflecting their functional importance \cite{Murzin1995, Marchler2011}. This effectively grounded the method in biological priors while addressing large-scale evolutionary studies.

To handle the challenges of large datasets, \cite{suvorov_accurate_2020} proposed a convolutional neural network (CNN)-based approach to infer phylogenetic tree topologies from multiple sequence alignments (\textit{CNN-Based Phylogenetic Tree}). This method extracted features from sequence alignments and optimized the inference process by utilizing \textit{Evolutionary Models for Amino Acid Substitution (P2)}, described by the substitution matrix \( Q \), to account for the rate of amino acid substitutions over evolutionary time \cite{Jones1992, Dayhoff1978}. The integration of substitution models improved phylogenetic accuracy without adding significant computational overhead.

Deep learning frameworks have also enabled innovative approaches by combining various predictive models. For instance, \textit{Phyloformer} \cite{nesterenko_phyloformer_2022} employed a transformer-based architecture to predict evolutionary distances and reconstruct tree topologies. Meanwhile, \cite{yeung_tree_2023} developed a sequence embedding tree visualization method (\textit{PLM for Tree Visualization}), leveraging protein language models to generate tree-like structures that effectively capture global topological relationships and local functional clustering. These methods utilized \textit{Protein Family Classification (P6)} as prior knowledge to group proteins based on sequence and structural similarity, enhancing their interpretative power in high-dimensional datasets \cite{Bateman2002, Finn2016}.

In addition, \cite{hayes_simulating_2024} introduced \textit{ESM3}, a multimodal generative language model capable of simulating evolutionary processes over hundreds of millions of years. This model generated highly divergent functional proteins while incorporating \textit{Functional Site Conservation (P5)} as prior knowledge, represented by the function \( F(x_i^p, x_j^p) \), to identify and prioritize critical functional sites within proteins \cite{Bartlett2002, Thornton2000}. This approach demonstrated its potential for tackling complex evolutionary tasks and generating novel functional proteins efficiently.

\subsubsection{Deep Structure-Based Phylogenetic Tree Methods.}

In structure-based methods, protein structure information has provided deeper insights into evolutionary relationships. \cite{van_kempen_fast_2023} proposed \textit{Foldseek}, a method that converts protein tertiary structure into structural alphabets to significantly improve structure search speed. Foldseek relied on structural alignment to enable fast inference across large protein structure datasets. In these methods, \textit{Tertiary Structure Conservation (P4)} serves as crucial prior knowledge, with the root-mean-square deviation (RMSD) used to measure the conservation of protein 3D structure, which is often more conserved than the primary sequence \cite{Sali1994}.

Building on structural analysis, \cite{bou_dagher_persistent_2024} introduced \textit{Persistent Homology (PH)} for phylogenetic inference, marking the first application of topological data analysis in this field. PH calculated the topological features of protein tertiary structures to measure evolutionary distances. This method captured strong phylogenetic signals within protein structures, offering a novel approach for analyzing evolutionary relationships at both small and large evolutionary scales. Here, \textit{Protein Secondary Structure Information (P3)} was utilized as prior knowledge, employing the similarity matrix \( S \) to identify conserved secondary structures such as alpha-helices and beta-sheets, reflecting important evolutionary features \cite{Kabsch1983, Chothia1984}.

\cite{moi_structural_2023} extended structure-based methods with \textit{FoldTree}, a method designed to infer evolutionary relationships between proteins with large evolutionary distances. The application of FoldTree in studying the evolutionary diversification of protein families demonstrated its strength in handling complex evolutionary histories by combining structural conservation and functional site information. In this context, \textit{Functional Site Conservation (P3)} was again used as prior knowledge, leveraging the function \( F(x_i^p, x_j^p) \) to identify critical functional sites within proteins \cite{Bartlett2002, Thornton2000}, thus improving the accuracy of phylogenetic tree construction.

\subsection{Deep Single-Cell-Based Lineage BioTree Construction Methods}
\begin{figure*}[t]
	\begin{center}
		\includegraphics[width=0.99\linewidth]{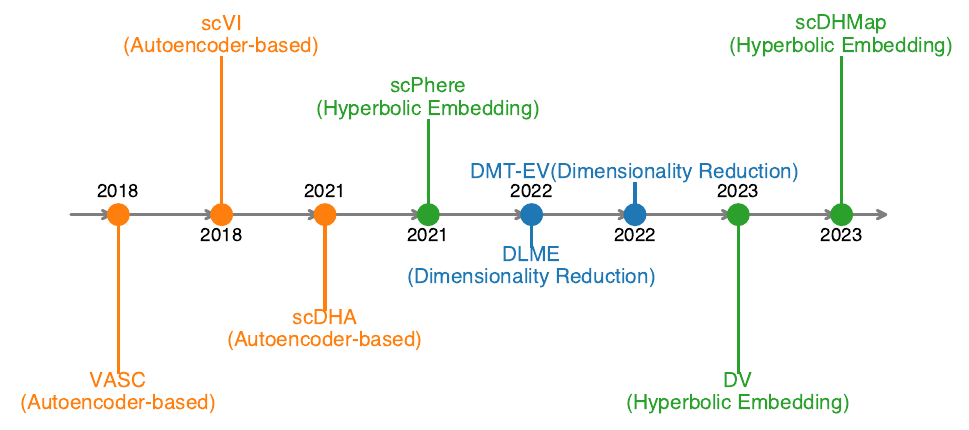}
	\end{center}
	\caption{\textbf{The Timeline of Deep Single Cell BioTree Construction Methods.} The figure shows the chronological development of trajectory inference methods based on single-cell RNA sequencing data. These methods have evolved by incorporating different types of prior knowledge to improve accuracy and computational efficiency in cell development analysis.}
	\label{fig_method_deep_singlecell}
\end{figure*}

\begin{figure*}[t]
	\begin{center}
		\includegraphics[width=0.99\linewidth]{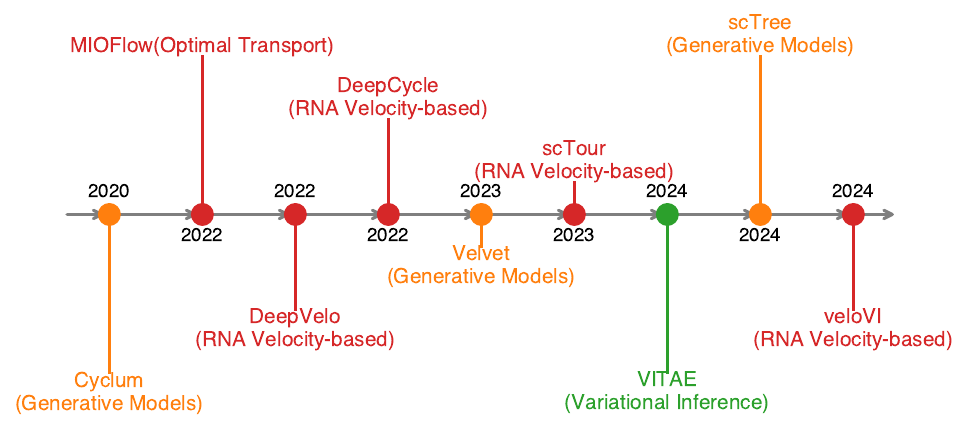}
	\end{center}
	\caption{\textbf{The Timeline of Deep Single Cell BioTree Construction Methods.} The figure shows the chronological development of trajectory inference methods based on single-cell RNA sequencing data. These methods have evolved by incorporating different types of prior knowledge to improve accuracy and computational efficiency in cell development analysis.}
	\label{fig_method_deep_singlecell_edr}
\end{figure*}

\clearpage
\begin{table}[H]
    \centering
    \scriptsize
    \caption{Overview of Deep Learning Methods in Single-Cell Trajectory Inference.}
    \begin{tabular}{p{1.5cm}|p{5cm} p{1cm} p{4.5cm}}
        \toprule
        \textbf{Method Name} & \textbf{Description}                                                                                        & \textbf{Ref.}             & \textbf{URL}                                              \\
        \midrule
        \multicolumn{4}{c}{\textbf{Dimensionality Reduction-based Methods}}                                                                                                                                                             \\ \midrule
        \textbf{VASC}        & Models scRNA-seq data distribution and clusters latent space for improved dimensionality reduction.         & \cite{wang2018vasc}            & \url{https://github.com/wang-research/VASC}               \\
        \textbf{scVI}        & Applies VAE to single-cell transcriptomic data, addressing noise and dropout events.                        & \cite{lopez2018deep}           & \url{https://github.com/YosefLab/scVI}                    \\
        \textbf{scDHA}       & Uses a non-negative kernel autoencoder for filtering insignificant genes in scRNA-seq data.                 & \cite{tran2021fast}            & \url{https://github.com/duct317/scDHA}                    \\
        \textbf{scPhere}     & Uses deep hyperbolic embedding to compute pseudotime in hyperbolic space.                                   & \cite{ding2021deep}            & \url{https://github.com/klarman-cell-observatory/scPhere} \\
        \textbf{DLME}        & Addresses under-sampled data through data augmentation and local flatness constraints.                      & \cite{zang_dlme_2022}          & \url{https://github.com/zangzelin/code_ECCV2022_DLME}     \\
        \textbf{DMT-EV}      & Enhances dimensionality reduction performance and explainability using manifold-based loss functions.       & \cite{zang2022dmt}             & \url{https://github.com/zangzelin/code_EVNet_DMTEV}       \\
        \textbf{MIOFlow}     & Aligns geodesic distances on the data manifold to accurately reconstruct trajectories.                      & \cite{huguet_manifold_2022}    & \url{https://github.com/KrishnaswamyLab/MIOFlow}          \\
        \textbf{VITAE}       & Combines hierarchical models with VAEs to map the latent space of single-cell data.                         & \cite{du_joint_2024}           & \url{https://github.com/jaydu1/VITAE}                     \\ \midrule

        \multicolumn{4}{c}{\textbf{Deep Generative Models}}                                                                                                                                                                             \\ \midrule
        \textbf{Cyclum}      & Uses autoencoders to identify cyclic trajectories in gene expression data.                                  & \cite{liang_latent_2020}       & \url{https://github.com/KChen-lab/cyclum}                 \\
        \textbf{scTree}      & VAE-based method integrating hierarchical clustering with batch correction.                                 & \cite{vandenhirtz_sctree_2024} & \url{https://github.com/mvandenhi/sctree-public}          \\
        \textbf{Velvet}      & Models gene expression dynamics using a VAE and neural stochastic differential equation system.             & \cite{maizels_deep_2023}       & \url{https://github.com/rorymaizels/velvet}               \\\midrule

        \multicolumn{4}{c}{\textbf{RNA Velocity-based Methods}}                                                                                                                                                                         \\\midrule
        \textbf{DeepVelo}    & Uses neural network-based ODE framework to model transcriptional dynamics and RNA velocity.                 & \cite{chen_deepvelo_2022}      & \url{https://github.com/bowang-lab/DeepVelo}              \\
        \textbf{DeepCycle}   & Analyzes cell cycle gene regulation dynamics in scRNA-seq data using deep learning.                         & \cite{riba_cell_2022}          & \url{https://github.com/andreariba/DeepCycle}             \\
        \textbf{scTour}      & Infers cellular dynamics using a VAE and neural ODE framework, minimizing batch effects.                    & \cite{li_sctour_2023}          & \url{https://github.com/LiQian-XC/sctour}                 \\
        \textbf{veloVI}      & Shares information across all cells to learn kinetic parameters and latent time for RNA velocity inference. & \cite{gayoso_deep_2024}        & \url{https://github.com/YosefLab/velovi}                  \\

        \bottomrule
    \end{tabular}
    \label{tab_deep_single_cell}
\end{table}

As shown in Table~\ref{tab_deep_single_cell} and Figure~\ref{fig_method_deep_singlecell}, in the field of single-cell RNA sequencing (scRNA-seq), deep learning techniques have emerged as powerful tools for handling high-dimensional and sparse data, particularly in inferring cellular differentiation pathways and generating differentiation trees. These methods incorporate advanced techniques such as dimensionality reduction and pseudo-time analysis, enabling the modeling of complex biological processes. In some cases, they also benefit from information fusion, which facilitates the integration of diverse biological data sources, such as gene expression profiles, RNA velocity, and lineage-specific markers, to enhance the interpretability of results. This section focuses on various approaches, including dimensionality reduction-based methods, deep generative models, and RNA velocity-based methods. By leveraging the strengths of deep learning, these techniques significantly improve the accuracy and scalability of differentiation tree construction while offering new tools for understanding the dynamic nature of cell development.

\paragraph{Dimensionality Reduction-based Methods.}
The existing methods can generally be classified into two main categories: dimensionality reduction methods and dimensionality reduction integrated with pseudo-time analysis, both contributing to the generation of differentiation trees by capturing the hierarchical structure of cell states.

For \textit{dimensionality reduction methods}, high-dimensional \textit{Cell Type-Specific Marker Genes (S3)} are projected into a lower-dimensional space, which serves as a foundation for constructing the differentiation tree by identifying distinct cellular states. Deep manifold learning methods have been increasingly utilized for dimensionality reduction in single-cell data analysis, thereby aiding in the generation of differentiation trees. \textit{DMAGE (deep manifold attributed graph embedding)} \citep{zang_unsupervised_2021} effectively captures both structural and feature information in latent spaces by leveraging node-to-node geodesic similarities. This allows for a more accurate reconstruction of cellular relationships, which is crucial for inferring cell differentiation pathways. Their subsequent works, \textit{DLME (deep local-flatness manifold embedding)} \citep{zang_dlme_2022}, address the challenges posed by under-sampled data through data augmentation \citep{zang_diffaug_2024} and local flatness constraints, further enhancing the accuracy of cell state embeddings and thus improving differentiation tree construction. Similarly, \textit{UDRN (unified dimensional reduction neural-network)} \citep{zang2023udrn} integrates feature selection and feature projection, ensuring that the essential cellular features are preserved in the reduced space, facilitating the differentiation tree generation process. \textit{DMT-EV} \citep{zang2022dmt} enhances both performance and explainability by using manifold-based loss functions to maintain cellular hierarchical structures in the latent space, which directly benefits the generation of differentiation trees.

Autoencoder-based methods, such as \textit{VASC} \citep{wang2018vasc} and \textit{scVI} \citep{lopez2018deep}, encode high-dimensional \textit{Gene Expression Profiles (S1)} into lower-dimensional latent spaces, capturing key information about cellular states. These methods not only improve the visualization and clustering of cells but also support the construction of differentiation trees by revealing the underlying branching patterns of cell lineages. \textit{scDHA (single-cell decomposition using hierarchical autoencoder)} \citep{tran2021fast, zhang_robust_2023} filters insignificant genes and projects data into a lower-dimensional space, providing a more focused view of the essential differentiation trajectories.

\textit{Dimensionality reduction integrated with pseudo-time analysis} incorporates prior information on \textit{Pseudotime Ordering (S4)}, facilitating differentiation tree generation by tracking the transitions between cell states over time. Deep hyperbolic embedding methods, such as \textit{scPhere} \citep{ding2021deep} and \textit{scDHMap} \citep{tian2023complex}, compute hyperbolic distances in latent space to infer pseudotime, effectively reconstructing differentiation pathways. By integrating pseudo-time and cell embeddings, these methods generate more accurate differentiation trees that represent the temporal progression and branching of cellular differentiation processes. Additionally, \textit{VITAE (variational inference for trajectory by autoEncoder)} \citep{du_joint_2024} provides a hierarchical model that assigns edge scores to cell transitions, directly informing the construction of the differentiation tree's backbone.

\paragraph{Deep Generative Models.}
Deep generative models, such as autoencoders and VAEs, focus on capturing the latent distribution of \textit{Gene Expression Profiles (S1)} to simulate cell state transitions, thereby serving as critical tools in differentiation tree generation. For instance, \textit{Cyclum} \citep{liang_latent_2020} uses autoencoders to identify cyclic trajectories in gene expression, helping to elucidate differentiation cycles within the differentiation tree. \textit{scTree} \citep{vandenhirtz_sctree_2024} integrates hierarchical clustering with batch correction to enhance the identification of cellular hierarchies, using a tree-structured approach to represent differentiation paths. Similarly, \textit{Velvet} \citep{maizels_deep_2023} models global gene expression dynamics in latent space, providing a comprehensive view of the differentiation landscape.

\paragraph{RNA Velocity-based Methods.}
Several methods estimate \textit{RNA Velocity (S2)} to model cellular trajectories and generate differentiation trees. \textit{veloVI} \citep{gayoso_deep_2024} shares information across cells and genes to learn latent time and kinetic parameters, improving the accuracy of inferred differentiation paths. \textit{scTour} \citep{li_sctour_2023} uses a deep learning architecture built on VAE and neural ODEs to estimate pseudotime and map cells into a latent space, facilitating differentiation tree generation. By modeling continuous transcriptional dynamics, \textit{DeepVelo} \citep{chen_deepvelo_2022} provides a refined view of gene expression changes, directly contributing to the construction of high-resolution differentiation trees. \textit{DeepCycle} \citep{riba_cell_2022} fits cycling patterns observed in the unspliced-spliced RNA space, offering a detailed map of differentiation processes during the cell cycle.

\section{Applications of BioTree} \label{sec_applications}

BioTree Construction, also known as evolutionary trees or phylogeny, have widespread applications in biology, spanning from species evolution analysis to molecular phylogenetics. This section provides a detailed overview of these applications along with specific examples.

\subsection{BioTree for Infectious Diseases}
Phylogenetic trees play a pivotal role in infectious disease research, serving as essential tools for tracing the origins, transmission, and evolutionary dynamics of pathogens across various biological scales. By integrating molecular data with evolutionary models, these analyses offer insights into the complex processes underlying the emergence and spread of infectious agents, with significant implications for public health interventions.

At the molecular and evolutionary level, phylogenetic analyses are indispensable for identifying the origins and reconstructing the evolutionary trajectories of viral pathogens. One prominent example is the classification of SARS-CoV-2 as a novel coronavirus, achieved through comprehensive phylogenetic analyses that revealed its close genetic relationship to bat coronaviruses. This classification provided the foundation for understanding SARS-CoV-2 as the causative agent of the COVID-19 pandemic \citep{gorbalenya_species_2020}. Furthermore, phylogenetic methods have been critical in tracking the evolutionary divergence of SARS-CoV-2 variants, including the Omicron subvariants BA.4, BA.5, and XBB. These analyses not only traced the lineage-specific mutations that differentiated these variants but also shed light on their global spread and potential public health impacts, aiding in the timely identification of new threats \citep{tegally_emergence_2022, tamura_virological_2023}.

Beyond the molecular level, phylogenetic tools have been extensively applied to monitor virus transmission dynamics within and between populations. These analyses provide critical insights into how pathogens adapt and evolve over time, often revealing the complex interplay between viral evolution and transmission patterns. For example, research on the spread of highly pathogenic avian influenza A (H5N1) among marine mammals and seabirds in Peru utilized phylogenetic trees to trace genetic reassortments that facilitated cross-species transmission, highlighting the zoonotic potential of these viruses and underscoring the importance of phylogenetic analysis in predicting future spillover events \citep{leguia_highly_2023}. Similarly, studies on SARS-CoV-2 transmission within immunocompromised individuals have demonstrated how intrahost viral evolution can contribute to the emergence of new variants, further complicating efforts to control the pandemic and emphasizing the role of phylogenetics in understanding viral persistence and adaptation in specific host populations \citep{gonzalez-reiche_sequential_2023}.

In addition to its application in pandemic contexts, phylogenetic analysis has been employed to explore co-infections involving non-pandemic viruses, broadening its utility in virology. A notable case is the investigation of Adeno-associated virus type 2 (AAV2) in U.S. children with acute severe hepatitis, where phylogenetic methods were used to assess viral relationships and explore the role of co-infections in disease severity. This example demonstrates the versatility of phylogenetic tools beyond pandemic viruses, showcasing their broader applicability in elucidating complex viral interactions \citep{servellita_adeno-associated_2023}.

Overall, phylogenetic trees are invaluable in infectious disease research, providing detailed insights into pathogen evolution, transmission dynamics, and cross-species interactions. By tracing the evolutionary pathways of pathogens and predicting future outbreaks, phylogenetic analyses are instrumental in informing public health strategies and shaping global responses to emerging infectious diseases.

\subsection{BioTree for Biomarker Discovery}
The integration of phylogenetic trees in biomarker discovery has emerged as a powerful analytical approach across various biological levels, offering insights into evolutionary relationships that guide the identification and validation of biomarkers. Spanning scales from microbial communities to gene family diversification, population genetics, and species-level comparative genomics, phylogenetic analysis enriches our biological understanding while presenting new opportunities for applications in precision medicine, agriculture, and environmental conservation.

At the microbial and environmental level, phylogenetic trees have become indispensable tools in metagenomics and environmental microbiology. By reconstructing evolutionary relationships within microbial communities, these trees help elucidate the functional roles of microbes in ecosystems and their potential as disease biomarkers. For instance, phylogenetic analysis has been applied to study sulfur metabolic genes in the human gut microbiome, where specific microbial genes were identified as potential biomarkers for colorectal cancer \citep{wolf_diversity_2022, zang_udrn_2023}. This approach demonstrates how the evolutionary study of microbial genes can provide actionable insights for disease diagnosis and treatment. Similarly, the discovery of novel circular DNA viruses through phylogenetic analyses highlights the method's capacity to uncover viral diversity in previously uncharacterized environments, broadening our understanding of virology \citep{tisza_discovery_2020}. Such findings underscore the crucial role of phylogenetic trees in expanding our knowledge of microbial evolution and their application in biomarker discovery within environmental and health-related contexts.

As research transitions from microbial ecosystems to gene-level analyses, phylogenetic trees continue to play a crucial role in exploring the evolutionary history and diversification of gene families. This line of research has significant implications for identifying biomarkers related to disease resistance and functional gene evolution. For example, the structural evolution of the LRR-RLK gene family, which drives diversification in plant defense mechanisms, was explored through phylogenetic methods, offering insights into the genetic underpinnings of disease resistance \citep{man_structural_2020}. Similarly, the evolutionary expansion of the CHS-L gene family in \textit{Senna tora} was linked to the biosynthesis of anthraquinones, a class of compounds with pharmaceutical relevance \citep{kang_genome-enabled_2020}. These studies demonstrate how phylogenetic analysis of gene family diversity and structural evolution can inform functional genomics and facilitate the discovery of potential biomarkers.

At the population genetics level, phylogenetic trees provide a framework for uncovering genetic diversity and structural variations associated with disease susceptibility. By integrating phylogenetic analyses with genomic data, researchers can identify population-specific biomarkers and uncover the genetic bases for gene-environment interactions. For instance, the combination of phylogenetic and structural variation analysis in diverse human populations has led to the identification of population-specific biomarkers, revealing how genetic diversity impacts disease susceptibility \citep{ebert_haplotype-resolved_2021}. Furthermore, stress-responsive genes in \textit{Nitraria tangutorum} were identified through genome-wide analysis, shedding light on the genetic mechanisms underlying adaptation to environmental stressors \citep{zhu_genome-wide_2023}. These studies highlight how phylogenetic trees can reveal complex genetic structures and their implications for population health and adaptation.

On a broader, species-level scale, phylogenetic trees play a fundamental role in comparative genomics, enabling the identification of species-specific biomarkers related to adaptive traits. Through cross-species comparisons, researchers can trace the evolutionary conservation and divergence of genes across species, which is crucial for understanding trait evolution and adaptation. For example, phylogenetic mapping of resistance genes in winter wheat provided valuable insights into gene conservation at the species level, with direct implications for crop improvement and disease resistance \citep{kale_catalogue_2022}. In a similar vein, studies exploring gene transfer mechanisms across domains revealed evolutionary connections between archaea and eukaryotes, emphasizing the utility of phylogenetic trees in tracing gene function evolution and speciation events \citep{ghaly_discovery_2022, moi_discovery_2022}. These investigations demonstrate the power of phylogenetic analysis in revealing the evolutionary forces shaping species and their potential for informing biomarker discovery related to environmental adaptation.

In summary, phylogenetic trees serve as critical tools across multiple biological scales, offering a comprehensive approach to biomarker discovery that integrates evolutionary insights from microbial ecosystems to species-wide genomic comparisons. Whether analyzing microbial community dynamics, gene family diversification, population genetics, or species-level evolution, phylogenetic analysis provides a robust framework for understanding the complex biological processes underlying biomarker discovery. These applications not only expand our understanding of biodiversity and evolutionary mechanisms but also offer practical strategies for advancing fields such as precision medicine, agricultural enhancement, and environmental conservation.

\subsection{BioTree for Cancer Evolution and Tumor Classification}
The application of evolutionary approaches in cancer research has significantly enhanced our understanding of the onset, progression, and therapeutic resistance of tumors. Phylogenetic trees, in particular, have proven to be indispensable tools, providing deeper insights into cancer resistance mechanisms, tumor evolution under selective pressures, and the functional genomics of cancer driver genes. This section categorizes the applications of evolutionary trees in cancer research into three major areas: understanding cancer resistance mechanisms, analyzing tumor evolution and therapeutic resistance, and exploring cancer driver mechanisms through functional genomics.

\paragraph{Understanding Cancer Resistance Mechanisms through Evolutionary Trees.}
Phylogenetic trees have been instrumental in investigating natural cancer resistance mechanisms in various species. These studies aim to uncover how evolutionary adaptations, such as duplications in tumor suppressor genes, contribute to reduced cancer risk in certain species. By tracing the evolutionary pathways of these adaptations, researchers can better understand the genetic foundations of cancer resistance and potentially apply these findings to human cancer therapies.

One such study by \cite{vazquez_parallel_2022} explored the parallel evolution of reduced cancer risk in Xenarthran lineages, such as sloths and armadillos, through phylogenetic analyses. The research found that bursts of tumor suppressor gene duplications coincided with reduced cancer risk, suggesting that these genetic duplications play a pivotal role in enhancing natural cancer resistance. Similarly, \cite{kolora_origins_2021} examined Pacific Ocean rockfish species, identifying genetic determinants associated with longevity and cancer resistance. Their findings highlighted the role of positive selection in DNA repair pathways, illustrating how evolutionary innovations contribute to cancer resistance. In another study, \cite{wang_phylovelo_2024} introduced PhyloVelo, a computational tool that integrates phylogenetic analysis to infer cell differentiation trajectories. This tool tracks lineage-specific adaptations and evolutionary dynamics, advancing our understanding of the molecular mechanisms underlying cancer resistance.

Collectively, these studies demonstrate how evolutionary trees can elucidate the genetic basis of natural cancer resistance, offering a foundation for developing new cancer therapies based on these insights.

\paragraph{Uncovering Tumor Evolution and Therapeutic Resistance through Phylogenetic Analysis.}
Phylogenetic trees are also employed to study tumor evolution, particularly in the context of therapeutic resistance. By reconstructing the evolutionary trajectories of tumors, researchers gain a deeper understanding of how tumors adapt to therapeutic interventions and develop resistance over time. This knowledge is crucial for designing more effective treatment strategies that target the evolutionary dynamics of cancer cells.

For example, \cite{fisk_premetastatic_2022} used phylogenetic analysis to study mutational processes in EGFR-driven lung adenocarcinoma. The research revealed that both endogenous factors, such as mutator gene mutations, and exogenous factors, such as mutagenic therapies, contribute to the emergence of therapeutic resistance. The study underscored the importance of considering the evolutionary pressures exerted on cancer cells when designing treatment strategies. Similarly, \cite{kwon_evolution_2020} traced the lineage dynamics of transmissible cancer in Tasmanian devils, uncovering how cancer cells adapt to different environmental and parasitic niches. This research highlighted the significance of understanding tumor evolution to combat the persistence and spread of cancer. In another example, \cite{schmidt_zero-agnostic_2023} developed the zero-agnostic copy number transformation (ZCNT) model, which optimizes tumor phylogeny inference and reveals gene changes associated with therapeutic resistance. The model represents a computational advancement in accurately modeling the evolutionary processes that lead to resistance.

These studies highlight the critical role of phylogenetic analysis in understanding the complex evolutionary processes that tumors undergo, particularly in the face of therapeutic pressures. By uncovering these dynamics, researchers can better predict resistance patterns and develop targeted treatment strategies.

\paragraph{Exploring Cancer Driver Mechanisms through Functional Genomics Based on Evolutionary Trees.}
In addition to studying cancer resistance and tumor evolution, phylogenetic trees are used to explore the functional genomics of cancer driver genes. By analyzing the evolutionary conservation and divergence of key genes, researchers can identify potential therapeutic targets and gain insight into the molecular mechanisms driving tumor progression.

For instance, \cite{jonsson_clec18a_2024} investigated the role of the gene CLEC18A in clear cell renal cell carcinoma (ccRCC), utilizing phylogenetic analysis to trace its evolutionary conservation and functional divergence in cancer. This study provided insights into how CLEC18A is regulated within the tumor microenvironment and its role in tumor progression. Similarly, \cite{zhang_heterologous_2024} explored the evolutionary dynamics of DNA transposable elements (TEs) in cancer cells, offering insights into genome engineering for cancer therapy. These studies underscore the value of evolutionary trees in understanding gene function evolution in the context of cancer. Furthermore, \cite{edogbanya_evolution_2021} examined the evolutionary history of the gene C1ORF112, revealing its role in DNA replication and DNA damage response, key processes implicated in cancer development. The study by \cite{julca_genomic_2023} provided a comprehensive genomic and metabolomic analysis of the medicinal plant \textit{Oldenlandia corymbosa}, revealing biosynthetic pathways with anticancer properties, which offers a unique perspective on the evolutionary basis of therapeutic compounds. Lastly, \cite{schmidt_zero-agnostic_2023} applied the ZCNT model in functional genomics to better understand cancer driver mechanisms within complex genomic datasets.

These studies demonstrate how phylogenetic trees can be applied to uncover the evolutionary dynamics of cancer driver genes, shedding light on their roles in tumor progression and offering new avenues for therapeutic development.

In summary, phylogenetic trees have become essential tools in cancer research, enabling scientists to investigate the evolution of cancer resistance, the mechanisms underlying tumor progression and therapeutic resistance, and the functional genomics of cancer driver genes. By integrating evolutionary insights with modern computational tools, researchers can develop more effective strategies for cancer diagnosis, treatment, and prevention, paving the way for improved outcomes in cancer therapy.

\subsection{BioTree for Agriculture and Crop Improvement}
Evolutionary trees are integral to plant science research, serving as a foundational tool for evolutionary analysis across a broad spectrum of applications. They are widely used to study genomic diversity, pathogen evolution, ecosystem management, and the functional evolution of plant genes. By constructing and analyzing phylogenetic trees, researchers can uncover the evolutionary relationships among species, the patterns of genome evolution, and the adaptive strategies plants employ in diverse ecological environments. This section reviews the methodologies and applications of evolutionary trees in plant science, underscoring their essential role in advancing the field.

\paragraph{Application of Evolutionary Trees in Plant Genomic Diversity and Domestication Traits.}
In studying plant genomic diversity and domestication traits, evolutionary trees are extensively employed to analyze structural variations in genomes and to trace the evolutionary relationships of specific genes. Pangenome analysis, for example, constructs a composite genome from multiple species or varieties and integrates evolutionary trees to reveal how selective pressures and adaptive changes have shaped different genes during evolution. \cite{chen_pangenome_2023} utilized this approach to identify genetic variations associated with domestication traits in broomcorn millet, providing key insights into the genomic changes that occurred during the domestication process. Similarly, phylogenomic methods apply large-scale genomic data to build evolutionary trees that unravel the complexity of species diversity and phylogenetic relationships, offering a deeper understanding of plant evolutionary history. \cite{guo_phylogenomics_2023} demonstrated how these phylogenetic analyses could support plant taxonomy and agricultural enhancement by identifying genetic diversity critical to adaptation and crop improvement. Additionally, co-expression network analysis, in conjunction with evolutionary trees, has been used to investigate the co-evolution and functional clustering of genes, offering molecular insights into plants' environmental adaptability and multicellular development \citep{feng_genomes_2024}. These examples underscore the utility of evolutionary trees in providing a comprehensive picture of plant genome evolution and their role in improving domestication practices.

\paragraph{Application of Evolutionary Trees in Plant Pathogen Evolution and Ecosystem Management.}
In the realm of plant pathogen evolution and ecosystem management, evolutionary trees serve as crucial tools for understanding pathogen diversity and tracing ecological dissemination pathways. Phylogenetic meta-analysis, which integrates molecular sequence data from plant pathogens, uses evolutionary trees to reveal the distribution patterns and evolutionary relationships of different pathogens. For example, \cite{bourret_cataloging_2023} employed evolutionary tree analysis to study the distribution and ecological risks of plant pathogens in California, offering vital data to inform plant protection strategies. The use of evolutionary models in combination with ecological management approaches provides insights into pest evolution and resistance patterns, helping optimize management strategies in agricultural ecosystems. \cite{thrall_evolution_2011} used evolutionary tree-based models to study the mechanisms of pathogen evolution, which enabled the development of proactive management tools aimed at mitigating pest threats in agro-ecosystems. Further research, such as the work by \cite{kan_cytonuclear_2024}, explored the co-evolution of plant genomes and their interactions with pathogens, emphasizing how evolutionary trees can elucidate the molecular mechanisms behind ecological adaptation and pathogen resistance in plants.

\paragraph{Application of Evolutionary Trees in Plant Genomic Evolution and Functional Studies.}
Evolutionary trees are also pivotal in investigating plant genomic evolution and functional studies, particularly in revealing the adaptive mechanisms that underpin plant survival across diverse environments. Cytonuclear interaction analyses, which focus on the co-evolution of nuclear and organellar genomes, rely on evolutionary trees to trace how these genetic systems evolve in coordination. By analyzing whole-genome data, \cite{kan_cytonuclear_2024} demonstrated that the co-evolution of nuclear and organellar genes plays a critical role in maintaining genomic stability during polyploidization, a process that has significantly influenced the diversification of Brassica species. Multi-omics approaches, which integrate genomic, transcriptomic, and proteomic data, further utilize evolutionary trees to explore the functional evolution of genes, shedding light on how plants adapt to environmental stresses \citep{jia_origin_2023}. For instance, evolutionary analysis combined with chromosome-level genome assembly has been employed to study gene family expansion and evolutionary patterns, revealing the molecular underpinnings of plant ecological adaptations and behaviors, such as predation, as shown by \cite{yuan_chromosome-level_2023}. These applications demonstrate the versatility of evolutionary trees in studying plant genomic evolution and function, providing critical insights into both basic plant biology and applied agricultural science.

In summary, evolutionary trees are indispensable tools in plant research, offering profound insights into the mechanisms underlying genomic diversity, pathogen evolution, and functional gene adaptation. Their application spans multiple biological scales, from studying individual gene evolution to managing large-scale ecological systems. Through the construction and interpretation of evolutionary trees, researchers can uncover the intricate relationships that drive plant evolution, enabling advancements in agricultural improvement, ecosystem management, and the broader understanding of plant sciences. As plant science continues to evolve, the role of phylogenetic trees in uncovering the molecular mechanisms of plant adaptation and survival will remain essential, contributing to both theoretical research and practical applications in the field.

\subsection{BioTree for Ecology and Environmental Studies}
Evolutionary biology seeks to uncover the origins of species, their relationships, and the adaptive changes they undergo. Recent advancements in molecular phylogenetics, genomics, and ecology have enabled researchers to probe the complexity of species evolution and their responses to ecological and environmental contexts more deeply. This review focuses on three central themes in current research: phylogenetic reconstruction and evolutionary relationships, genomic evolution and adaptive studies, and species diversity and biogeography. These themes help elucidate the mechanisms behind biodiversity, ecological adaptation strategies, and the role of environmental factors in shaping species evolution.

\paragraph{Phylogenetics and Evolutionary Relationship Reconstruction.}
Phylogenetic reconstruction is essential for understanding the evolutionary history of species and their adaptations to ecological pressures. By analyzing molecular data and constructing evolutionary trees, researchers can infer species relationships and divergence patterns, providing insights into how species respond to environmental challenges.

Recent studies highlight the importance of taxon sampling in evolutionary inference, as small changes in sampling can significantly alter phylogenetic outcomes. For instance, \cite{bernot_major_2023} revised the phylogeny of crustaceans and hexapods, showing that variations in sampling influence tree topologies and, consequently, our understanding of species' ecological adaptations. This study challenges existing phylogenetic hypotheses and underscores the significance of environmental diversity in evolutionary relationship studies. Similarly, \cite{eme_inference_2023} reconstructed the evolutionary relationships between Asgard archaea and eukaryotes, shedding light on gene duplication and loss during early life evolution, providing insights into species' adaptations to different ecological niches.

Phylogenetic analyses have also been applied to clarify the evolutionary positions of rare species. For example, \cite{lax_molecular_2023} employed single-cell transcriptomics and phylogenetic tools to study \textit{Dolium sedentarium}, confirming its unique evolutionary position in specific ecological contexts. These studies demonstrate how molecular phylogenetic methods can resolve uncertainties in evolutionary histories, offering a pathway for more precise species classification. Furthermore, studies like \cite{maurya_molecular_2023}, which examined the phytogeographic history of \textit{Capparis}, reveal how species differentiation and migration are influenced by environmental factors, further contributing to our understanding of species evolution and reclassification.

\paragraph{Genomic Evolution and Adaptive Studies.}
Research on genomic evolution investigates how structural and functional changes in genomes drive species' adaptations to diverse environments. Trait innovations, gene expansions, and genome rearrangements are key processes in ecological adaptation and diversification.

For example, the comparative genomics of multicellular algae and land plants studied by \cite{feng_genomes_2024} revealed that specific gene expansions and signaling network modifications were crucial for plant adaptation to terrestrial environments. These findings provide a theoretical foundation for understanding how genomic changes facilitate ecological adaptation. Similarly, research by \cite{blaimer_key_2023} on the phylogeny of Hymenoptera insects demonstrated how trait innovations like parasitism and phytophagy drive species diversification in response to environmental conditions.

In addition, studies of genome rearrangements have revealed how structural changes enable the evolution of new phenotypic traits. For instance, \cite{marletaz_little_2023} analyzed the genome of the little skate, uncovering how regulatory networks and genome rearrangements facilitated the evolution of its wing-like fins. These studies suggest that environmental changes are key drivers of genomic evolution and highlight the importance of understanding these dynamics for evolutionary biology.

\paragraph{Species Diversity and Evolutionary Biogeography.}
Research in species diversity and evolutionary biogeography integrates ecological and environmental data to understand how historical processes and environmental changes shape species adaptation and diversification. This approach reveals how geographical environments influence evolutionary pathways and species distributions.

The impact of human activities on species diversity and evolution has been a major focus of recent studies. \cite{chen_evidence_2023} explored the domestication history of yaks, taurine cattle, and their hybrids on the Tibetan Plateau, showing how human activities and natural selection have jointly shaped these species' ecological adaptations. Similarly, \cite{guo_phylogenomics_2023} analyzed the phylogeny of flowering plants, revealing the influence of whole-genome duplication and hybridization on species biogeography, further illustrating how evolutionary processes differ across ecological environments.

Genomic studies on plant domestication have also contributed to our understanding of species adaptation to environmental changes. For instance, \cite{chen_pangenome_2023} conducted a pangenome analysis of broomcorn millet, linking genomic variations to domestication traits and offering critical data for crop improvement and ecological adaptation research. These studies emphasize how environmental conditions and genomic changes interact to influence species' evolutionary trajectories, demonstrating the importance of evolutionary biogeography in understanding species diversity.

Research in phylogenetics, genomic evolution, and species diversity plays a pivotal role in modern evolutionary biology, offering a comprehensive view of biodiversity formation and species adaptation. By integrating phylogenetic reconstruction, genomic analysis, and biogeographical methods, researchers can reveal the mechanisms underlying evolutionary processes, particularly in response to changing ecological environments. These studies not only advance evolutionary biology theories but also provide essential insights for ecological conservation, biodiversity management, environmental monitoring, and agricultural development. Future research will benefit from further integration of ecological and molecular data, offering an increasingly dynamic understanding of biological evolution.

\section{Current Limitations of BioTree Construction} \label{sec_limitations}

\subsection{Limitations of Classical BioTree Construction Methods}
The limitations of classical BioTree construction methods in phylogenetic analysis stem from the intrinsic characteristics of their algorithms, theoretical assumptions, and the disparity between the complexity of biological data and the evolving demands of modern bioinformatics. Recognizing these limitations is essential for refining existing methods and designing innovative tools that address the unique challenges posed by contemporary biological research.

A fundamental challenge lies in scalability and computational complexity, which restricts the utility of classical methods for large-scale datasets. Techniques like Maximum Likelihood (ML) and Bayesian Inference, though effective for small datasets, rely on exhaustive searches through possible tree structures. As dataset sizes grow and taxa numbers increase, the combinatorial explosion drastically escalates computational time and resource requirements. This computational bottleneck hinders large-scale phylogenetic analysis, slowing biological discovery and constraining the practical use of evolutionary trees in applications such as ecosystem conservation and drug target identification \citep{Stamatakis2014}. In metagenomics and environmental genomics, where massive volumes of sequence data demand rapid analysis, classical methods struggle to meet the efficiency required for actionable insights. While computational optimizations have been explored, the absence of mechanisms to integrate prior knowledge or data-driven strategies further limits their scalability.

Another critical issue is the inadequate handling of uncertainty and missing data, reflecting classical methods' dependence on complete, high-quality datasets. Biological data, particularly from field samples or historical specimens, often contain gaps or noise. Classical approaches like ML and Bayesian Inference are not equipped to robustly handle such uncertainties, leading to phylogenetic inferences that may diverge significantly from true evolutionary histories \citep{Guindon2003}. This limitation is especially apparent in contexts such as viral evolution studies, where high mutation rates and incomplete genomic sequences prevail. In such scenarios, the inability to incorporate incomplete data and appropriately model uncertainty can result in substantial misinterpretations of key evolutionary pathways. Though modern approaches increasingly emphasize the fusion of incomplete datasets to enhance reliability, this remains underexplored in classical frameworks.

The dependence on rigid model assumptions further constrains the applicability of classical methods. These methods often rely on fixed evolutionary models, such as the molecular clock hypothesis or constant substitution rates, which do not align with the complexities of real biological processes. Factors like rate heterogeneity, lineage-specific substitution patterns, and events such as horizontal gene transfer or genome duplications are challenging to capture within traditional frameworks \citep{Ronquist2012}. Bayesian approaches, despite offering flexibility through priors, are highly sensitive to model selection, where incorrect assumptions can lead to biased or erroneous results. For example, in polyploid plants or recombinant pathogens with intricate evolutionary histories, classical models often fail to provide biologically plausible insights. Incorporating information fusion techniques that blend empirical data with adaptive model selection may offer a promising avenue to address this gap.

Lastly, classical methods exhibit limited capability in managing data complexity and diversity, particularly in the context of modern multi-omics studies. The integration of genomic, transcriptomic, epigenomic, and metabolomic data is increasingly critical for capturing organismal function and evolutionary trajectories. However, classical BioTree construction methods are predominantly designed for single-data-type analysis and lack robust mechanisms for combining multiple data sources \citep{Nguyen2015}. When evolutionary signals conflict across omics layers, these methods fail to produce reliable integrated results. This deficiency hampers the holistic understanding of evolutionary processes and multi-level biological systems. While deep learning approaches have begun to leverage data-driven strategies for fusion, classical methods remain inadequate in addressing this integration challenge.

\subsection{Challenges of Deep Learning-Based BioTree Construction Methods}

Deep learning-based methods have become powerful tools for constructing phylogenetic trees due to their ability to model complex patterns from high-dimensional data. However, these methods face several critical challenges in their effective application.

One prominent challenge is the interpretability of deep learning models. Unlike classical methods, deep learning approaches such as deep neural networks, generative adversarial networks (\textit{GANs}), and variational autoencoders (\textit{VAEs}) are often treated as "black boxes." These models capture intricate patterns in the data through their multi-layered architectures, but this complexity makes it difficult to intuitively explain the results or connect them to underlying biological phenomena \citep{you2018graph, bojchevski2018netgan}. The lack of interpretability can obscure evolutionary relationships, particularly in cases where precise pathways or mechanisms must be identified\cite{jin2018junction}. Although efforts have been made to improve interpretability by incorporating visualization techniques or simplifying model architectures, these approaches often come at the cost of reduced performance. Information fusion, where prior biological knowledge is combined with model outputs, has the potential to enhance interpretability by aligning learned representations with domain-specific insights, though its integration into deep learning frameworks remains a challenge.

Another key challenge lies in the data requirements and generalization capabilities of deep learning models \cite{li2018learning, you2018graphrnn}. These methods typically rely on large, labeled datasets to achieve robust performance, yet biological datasets are often sparse, incomplete, or biased. This can lead to overfitting, where models perform well on training data but fail to generalize to unseen or diverse datasets \citep{you2018graph, jin2018junction}. This limitation hinders the practical utility of deep learning models, as errors in phylogenetic tree construction can misrepresent evolutionary pathways and compromise subsequent biological analyses. Data augmentation techniques and unsupervised learning strategies have been proposed to mitigate these challenges, but they often require careful tuning and significant computational resources.

The integration of biological prior knowledge into deep learning models presents another significant hurdle. While deep learning excels at data-driven learning, its frameworks often lack mechanisms to incorporate domain-specific knowledge, such as evolutionary constraints or known phylogenetic priors. This shortcoming can result in tree structures that, while computationally optimized, fail to reflect biologically plausible evolutionary relationships \citep{de2018molgan, jin2018junction, manduchi_tree_2023}. For example, tree nodes inferred without considering known mutation rates or lineage-specific traits may diverge from actual evolutionary histories. Approaches that fuse data-driven methods with explicit prior knowledge have shown promise but are not yet widely adopted in phylogenetic applications.

Finally, the computational costs and resource limitations of deep learning methods represent a substantial barrier \cite{agapito2023overview}. Training deep learning models demands high-performance computing resources, including GPUs and large memory capacities. This requirement becomes especially pronounced when handling large-scale biological datasets \citep{beaini2023towards,wang_hierarchical_2022}. The computational burden can slow research progress, limit accessibility to resource-constrained teams, and impede the development and validation of novel algorithms. Although innovations in distributed computing and model optimization have alleviated some of these concerns, achieving a balance between computational efficiency and model performance remains an ongoing challenge.

While deep learning-based methods hold great promise for advancing phylogenetic analysis, these challenges underscore the need for improvements in interpretability, data integration, and computational efficiency. Developing hybrid approaches that combine classical and deep learning techniques may offer a way forward by leveraging the strengths of both paradigms, particularly in the context of information fusion to align computational outputs with biological realities.

\section{Opportunities in BioTree Construction}

\subsection{Fusion of multimodal information for co-modeling}
\begin{figure*}[t]
	\begin{center}
		\includegraphics[width=0.99\linewidth]{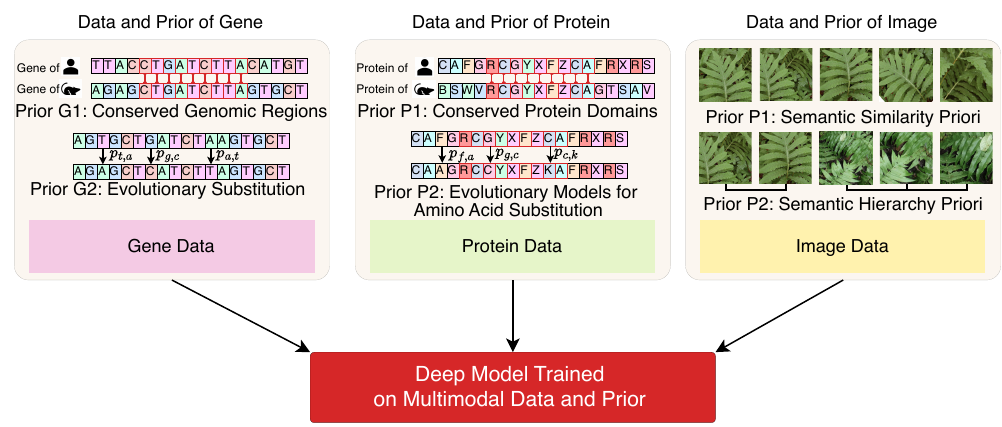}
	\end{center}
	\caption{\textbf{The futurework for Fusion of multimodal information in biological research.} The figure illustrates the integration of multi-modal data in deep learning models for biological research, combining genomic, proteomic, transcriptomic, metabolomic, and epigenetic data to enhance model performance and uncover comprehensive biological information.}
	\label{fig_method_multimodal}
\end{figure*}

Single-modal studies dominate current research in evolutionary and differentiation tree construction, focusing primarily on genomic sequences, protein sequences, or single-cell transcriptomics RNA sequencing data \citep{rao2021deep}. However, single-modal approaches have significant limitations in capturing the complex, multi-layered nature of biological systems. These systems involve dynamic interactions among genes, proteins, metabolites, cells, and tissues, which cannot be fully understood through isolated analysis. The integration of multimodal data addresses these limitations by leveraging diverse datasets to uncover comprehensive biological insights, offering a powerful solution for complex biological questions.

Each modality contributes unique prior knowledge that complements the others. For instance, \textit{genomic data} reveal genetic variations and structural rearrangements, while \textit{proteomic data} highlight protein interactions and modifications \citep{hasin2017multi, muller2023expanding}. \textit{Transcriptomic data} elucidate regulatory relationships, \textit{metabolomic data} reflect cellular metabolic states, and \textit{epigenetic data} provide insights into gene regulation. These complementary layers of information enhance the robustness and predictive accuracy of deep learning models, allowing for a more holistic understanding of biological evolution and differentiation processes.

Recent advancements in deep learning have accelerated the development of multimodal models capable of integrating such diverse data. For example, models like \textit{BLIP} (Bootstrapping Language-Image Pre-training) \citep{li2022blip, li2023blip}, \textit{CLIP} (Contrastive Language-Image Pre-training) \citep{radford2021learning}, and \textit{HuggingGPT} \citep{shen2024hugginggpt} demonstrate how unified frameworks can align features across modalities. These models effectively capture cross-modal relationships, as evidenced by \textit{BLIP}'s success in tasks like image captioning and \textit{CLIP}'s performance on large-scale image-text datasets. Similarly, \textit{Graph Neural Networks} (GNNs) have been employed to integrate multi-omics data for tasks like cancer type prediction, outperforming single-modal approaches \citep{huang2020fusion}. Generative models such as \textit{Variational Autoencoders} (VAEs) \citep{kingma2013auto} and \textit{Generative Adversarial Networks} (GANs) \citep{goodfellow2014generative} also facilitate the fusion of multimodal data by creating shared feature spaces for diverse datasets.

Despite these advancements, integrating multimodal data remains challenging due to differences in data scales, noise levels, and missing values. Effective alignment and integration require robust algorithms. Strategies to address these issues include using \textit{aligned embeddings} to map modalities into a common feature space \citep{zhang2018multi}, applying \textit{cross-modal attention mechanisms} to dynamically weigh and fuse information \citep{tsai2019multimodal}, and incorporating biological priors like protein-protein interaction networks to guide model training \citep{meng2019gene}. For example, in cancer research, the integration of genomic and proteomic data can reveal gene-protein interaction patterns crucial for identifying biomarkers, even when data from certain modalities are incomplete or noisy.

\subsection{Enhancing Interpretability of Deep Learning Models}
\begin{figure*}[t]
	\begin{center}
		\includegraphics[width=0.99\linewidth]{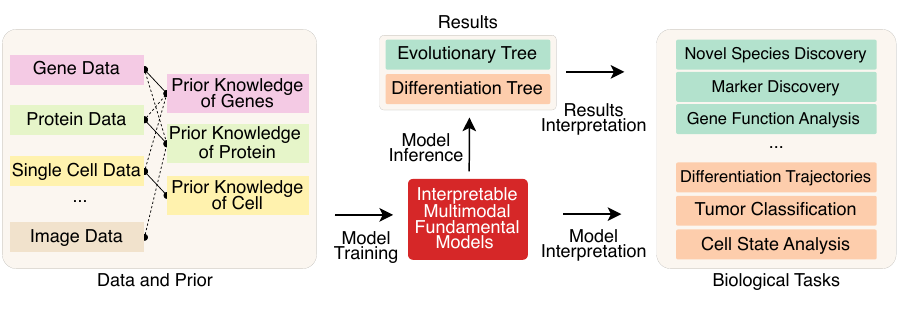}
	\end{center}
	\caption{\textbf{Integrative Framework for Interpretable Multimodal Deep Learning in Biological Research.} The figure illustrates the integration of multimodal biological data and prior knowledge in deep learning models to enhance model interpretability and transparency. By combining multimodal data and prior knowledge, deep learning models can provide accurate predictions while uncovering biological knowledge through interpretable results.}
	\label{fig_method_Interpretability}
\end{figure*}

While deep learning models excel at learning complex patterns from high-dimensional data, their "black-box" nature limits their acceptance in evolutionary biology research. Therefore, improving the interpretability and transparency of these models is a crucial direction for future research (see Figure~\ref{fig_method_Interpretability}). By training deep learning models with multimodal biological data (e.g., gene, protein, single-cell, and image data) and their prior knowledge, we can not only improve the prediction accuracy of these models but also perform various downstream tasks (e.g., evolutionary tree and differentiation tree construction, species discovery, gene function analysis) based on the model outputs and their interpretable results. This approach enables us to achieve high-accuracy predictions while uncovering biological knowledge through deep models.

New neural network architectures, such as attention-based models and self-explainable neural networks\citep{hu2024self,zang_dmt-ev_2024,wei2024towards}, provide methods for automatically explaining or visualizing important features, thereby enhancing model interpretability. Techniques like SHAP (Shapley Additive Explanations)\cite{antonini2024machine} and LIME (Local Interpretable Model-Agnostic Explanations)\cite{tan2024glime} can quantify the contribution of each input feature to the final prediction outcome. These methods help uncover the biological patterns learned by deep learning models and verify whether these results are consistent with existing biological knowledge, thus avoiding potential misunderstandings. For example, in applications such as the Junction Tree Variational Autoencoder (JT-VAE) \cite{jin2018junction} for molecular graph generation, interpretability can provide insights into how the model captures chemical substructures and their contributions to molecular biological functions.

In the field of life sciences, attempts to achieve reliable interpretable analyses and explore new biological knowledge remain limited \cite{chen2024applying}. Post-hoc interpretability methods like SHAP and LIME, while useful to some extent, often fall short in terms of stability and effectiveness for practical biological discovery. These methods \cite{esser2023reliable} rely on the relationships between perturbations in input data and model outputs, making their results highly sensitive to data distribution and model changes. Consequently, post-hoc interpretability methods may exhibit inconsistencies across different datasets or model architectures, limiting their application in complex biological problems. Therefore, to better meet the needs of biological discovery, it is essential to design more interpretable and robust deep learning models that can provide stable and reliable interpretative results while handling high-dimensional and diverse biological data.

To further enhance the interpretability of deep learning models, integrating biological prior knowledge into the model architecture design and training processes could be considered. For example, introducing domain-specific evolutionary constraints or priors in biological tree construction, combined with a hierarchical interpretation framework, can provide a clearer explanation path for complex biological evolutionary processes. This combination can significantly improve the credibility and application value of deep learning models. Thus, by adopting diverse interpretability techniques and leveraging the outputs of deep models for various downstream biological tasks (see Figure~\ref{fig_method_Interpretability}), future deep learning models can provide strong biological explanations while improving prediction accuracy, thereby promoting their widespread application in bioinformatics, phylogenetics, and other related fields.

\subsection{Fusion of Cellular and Species-Level Information for Downstream Tasks}
In biological research, evolutionary trees (phylogenetic trees) and differentiation trees (developmental pathways) are fundamental tools for understanding the evolutionary relationships among species and the developmental differentiation pathways of cells. These two tools are often studied independently, focusing either on macro-level species evolution or micro-level cell differentiation. However, by integrating evolutionary and differentiation trees, researchers can uncover deeper insights into biological processes, bridging the gap between cellular and species-level information.

For instance, in \textit{species discovery}, the combination of genetic information and cellular differentiation patterns enhances the identification of new species and subspecies, while simultaneously revealing their evolutionary pathways \citep{bock1959preadaptation, spain2023late}. Similarly, in \textit{gene function analysis}, linking evolutionary conservation patterns with cellular differentiation processes illuminates gene regulatory networks and their roles in development.

From a technical perspective, mutual validation between evolutionary and differentiation trees not only improves the reliability of existing models but also highlights potential areas for refinement. For example, in \textit{disease progression modeling}, understanding abnormal cancer cell evolution and differentiation pathways can lead to new biomarkers and therapeutic strategies \citep{zhao2023understanding}.

Moreover, leveraging deep learning models with advanced techniques like \textit{hierarchical attention networks} and \textit{multi-task learning} enables effective integration of evolutionary and differentiation data. Incorporating \textit{biological prior knowledge} into these models further enhances interpretability and alignment with biological principles. Despite these advancements, challenges remain, such as handling noise, missing data, and the alignment of multi-modal datasets. Methods like \textit{canonical correlation analysis} and \textit{manifold alignment} can play a pivotal role in addressing these issues.

By combining the strengths of evolutionary and differentiation trees, future research is expected to achieve more accurate predictions and provide richer biological insights. This integrated approach will drive progress in phylogenetics, developmental biology, and personalized medicine.

\section{Data Availability}
No data was generated or used in this review.

\section{Code Availability}
In Figure.~2 and Figure.~3, the code for DeepSeek-70B analysis is available at \url{https://github.com/zangzelin/code_info_fusion_biotree}.

\section{Acknowledgements} 
This work was supported by the National Key R\&D Program of China~(No.2022ZD0115100), the National Natural Science Foundation of China Project~(No. U21A20427), and Project~(No. WU2022A009) from the Center of Synthetic Biology and Integrated Bioengineering of Westlake University. We thank the Westlake University HPC Center for providing computational resources. This work was supported by the InnoHK program. This work was supported by Ant Group through CAAI-Ant Research Fund.

\section{Author Contributions}
Stan Z. Li and Zelin Zang proposed this research. Zelin Zang, Yongjie Xu, and Chenrui Duan collected the information. Zelin Zang, Yongjie Xu, and Chenrui Duan wrote the manuscript. Jinlin Wu, Stan Z. Li, and Zhen Lei provided valuable suggestions on the manuscript. All authors discussed the results, revised the draft manuscript, and read and approved the final manuscript.

\section{Competing Interests}
The authors declare no competing interests.
{
  \footnotesize
\bibliographystyle{abbrvnat}
\bibliography{ 
    bib/ref.bib, 
    bib/intro.bib, 
    bib/niotree_app_bioclassfication.bib,
    bib/app_MolecularPhyloge.bib,
    bib/app_speciesEvolution.bib,
    bib/app_diseases.bib,
    bib/app_biomarker.bib,
    bib/app_agriculture.bib,
    bib/app_cancer.bib,
    bib/dataset.bib,
    bib/prior.bib,
    bib/protein,
    bib/gene_Phylo_tree_methods.bib,
    bib/single_cell_basic.bib,
    bib/classical_gene.bib,
    bib/classical_protein.bib,
    bib/classical_single_cell.bib,
    bib/deep_gene.bib,
    bib/deep_protein.bib,
    bib/future_work.bib,
    bib/deep_general.bib
  }

\begin{thebibliography}{336}
\providecommand{\natexlab}[1]{#1}
\providecommand{\url}[1]{\texttt{#1}}
\expandafter\ifx\csname urlstyle\endcsname\relax
  \providecommand{\doi}[1]{doi: #1}\else
  \providecommand{\doi}{doi: \begingroup \urlstyle{rm}\Url}\fi

\bibitem[Abeer et~al.(2024)Abeer, Jantre, Urban, and Yoon]{abeer_leveraging_2024}
A.~N. M.~N. Abeer, S.~Jantre, N.~M. Urban, and B.-J. Yoon.
\newblock Leveraging {Active} {Subspaces} to {Capture} {Epistemic} {Model} {Uncertainty} in {Deep} {Generative} {Models} for {Molecular} {Design}, Aug. 2024.
\newblock URL \url{http://arxiv.org/abs/2405.00202}.
\newblock arXiv:2405.00202 [cs, q-bio, stat].

\bibitem[Abu-Qasmieh et~al.(2023)Abu-Qasmieh, Al~Fahoum, Alquran, and Zyout]{abu2023innovative}
I.~Abu-Qasmieh, A.~Al~Fahoum, H.~Alquran, and A.~Zyout.
\newblock An innovative bispectral deep learning method for protein family classification.
\newblock \emph{Computers, Materials \& Continua}, 75\penalty0 (2), 2023.

\bibitem[Aebersold and Mann(2003)]{Aebersold2003MassSpectrometry}
R.~Aebersold and M.~Mann.
\newblock Mass spectrometry-based proteomics.
\newblock \emph{Nature}, 422\penalty0 (6928):\penalty0 198--207, 2003.

\bibitem[Agapito and Cannataro(2023)]{agapito2023overview}
G.~Agapito and M.~Cannataro.
\newblock An overview on the challenges and limitations using cloud computing in healthcare corporations.
\newblock \emph{Big Data and Cognitive Computing}, 7\penalty0 (2):\penalty0 68, 2023.

\bibitem[Alamdari et~al.(2023)Alamdari, Thakkar, van~den Berg, Tenenholtz, Strome, Moses, Lu, Fusi, Amini, and Yang]{alamdari2023protein}
S.~Alamdari, N.~Thakkar, R.~van~den Berg, N.~Tenenholtz, B.~Strome, A.~Moses, A.~X. Lu, N.~Fusi, A.~P. Amini, and K.~K. Yang.
\newblock Protein generation with evolutionary diffusion: sequence is all you need.
\newblock \emph{BioRxiv}, pages 2023--09, 2023.

\bibitem[Albahri et~al.(2023)Albahri, Duhaim, Fadhel, Alnoor, Baqer, Alzubaidi, Albahri, Alamoodi, Bai, Salhi, et~al.]{albahri2023systematic}
A.~S. Albahri, A.~M. Duhaim, M.~A. Fadhel, A.~Alnoor, N.~S. Baqer, L.~Alzubaidi, O.~S. Albahri, A.~H. Alamoodi, J.~Bai, A.~Salhi, et~al.
\newblock A systematic review of trustworthy and explainable artificial intelligence in healthcare: Assessment of quality, bias risk, and data fusion.
\newblock \emph{Information Fusion}, 96:\penalty0 156--191, 2023.

\bibitem[Alberts et~al.(2002)Alberts, Johnson, Lewis, Raff, Roberts, and Walter]{Alberts2002}
B.~Alberts, A.~Johnson, J.~Lewis, M.~Raff, K.~Roberts, and P.~Walter.
\newblock \emph{Molecular Biology of the Cell}.
\newblock Garland Science, 4th edition, 2002.
\newblock ISBN 978-0815332183.

\bibitem[Altschul et~al.(1990)Altschul, Gish, Miller, Myers, and Lipman]{Altschul1990}
S.~F. Altschul, W.~Gish, W.~Miller, E.~W. Myers, and D.~J. Lipman.
\newblock Basic local alignment search tool.
\newblock \emph{Journal of Molecular Biology}, 215\penalty0 (3):\penalty0 403--410, 1990.

\bibitem[Angermueller et~al.(2016)Angermueller, P{\"a}rnamaa, Parts, and Stegle]{angermueller2016deep}
C.~Angermueller, T.~P{\"a}rnamaa, L.~Parts, and O.~Stegle.
\newblock Deep learning for computational biology.
\newblock \emph{Molecular systems biology}, 12\penalty0 (7):\penalty0 878, 2016.

\bibitem[Ansorge(2009)]{ansorge2009next}
W.~J. Ansorge.
\newblock Next-generation dna sequencing techniques.
\newblock \emph{New biotechnology}, 25\penalty0 (4):\penalty0 195--203, 2009.

\bibitem[Antonini et~al.(2024)Antonini, Tanzola, Asiain, Ferracutti, Castro, Bjerg, and Ganuza]{antonini2024machine}
A.~S. Antonini, J.~Tanzola, L.~Asiain, G.~R. Ferracutti, S.~M. Castro, E.~A. Bjerg, and M.~L. Ganuza.
\newblock Machine learning model interpretability using shap values: Application to igneous rock classification task.
\newblock \emph{Applied Computing and Geosciences}, page 100178, 2024.

\bibitem[Armingol et~al.(2024)Armingol, Baghdassarian, and Lewis]{armingol2024diversification}
E.~Armingol, H.~M. Baghdassarian, and N.~E. Lewis.
\newblock The diversification of methods for studying cell--cell interactions and communication.
\newblock \emph{Nature Reviews Genetics}, 25\penalty0 (6):\penalty0 381--400, 2024.

\bibitem[Bartlett et~al.(2002)Bartlett, Porter, Borkakoti, and Thornton]{Bartlett2002}
G.~J. Bartlett, C.~T. Porter, N.~Borkakoti, and J.~M. Thornton.
\newblock Analysis of catalytic residues in enzyme active sites.
\newblock \emph{Journal of Molecular Biology}, 324\penalty0 (1):\penalty0 105--121, 2002.

\bibitem[Basra(2024)]{basra2024cotton}
A.~Basra.
\newblock \emph{Cotton fibers: developmental biology, quality improvement, and textile processing}.
\newblock CRC Press, 2024.

\bibitem[Bateman et~al.(2002)Bateman, Coin, Durbin, Finn, et~al.]{Bateman2002}
A.~Bateman, L.~Coin, R.~Durbin, R.~D. Finn, et~al.
\newblock The pfam protein families database.
\newblock \emph{Nucleic Acids Research}, 30\penalty0 (1):\penalty0 276--280, 2002.

\bibitem[Beaini et~al.(2023)Beaini, Huang, Cunha, Li, Moisescu-Pareja, Dymov, Maddrell-Mander, McLean, Wenkel, M{\"u}ller, et~al.]{beaini2023towards}
D.~Beaini, S.~Huang, J.~A. Cunha, Z.~Li, G.~Moisescu-Pareja, O.~Dymov, S.~Maddrell-Mander, C.~McLean, F.~Wenkel, L.~M{\"u}ller, et~al.
\newblock Towards foundational models for molecular learning on large-scale multi-task datasets.
\newblock \emph{arXiv preprint arXiv:2310.04292}, 2023.

\bibitem[Bergen et~al.(2020{\natexlab{a}})Bergen, Lange, Peidli, Wolf, and Theis]{bergen_generalizing_2020}
V.~Bergen, M.~Lange, S.~Peidli, F.~A. Wolf, and F.~J. Theis.
\newblock Generalizing {RNA} velocity to transient cell states through dynamical modeling.
\newblock \emph{Nat Biotechnol}, 38\penalty0 (12):\penalty0 1408--1414, Dec. 2020{\natexlab{a}}.
\newblock ISSN 1546-1696.
\newblock \doi{10.1038/s41587-020-0591-3}.
\newblock URL \url{https://www.nature.com/articles/s41587-020-0591-3}.
\newblock Publisher: Nature Publishing Group.

\bibitem[Bergen et~al.(2020{\natexlab{b}})Bergen, Lange, Peidli, et~al.]{Bergen2020}
V.~Bergen, M.~Lange, S.~Peidli, et~al.
\newblock Generalizing rna velocity to transient cell states through dynamical modeling.
\newblock \emph{Nature Biotechnology}, 38\penalty0 (12):\penalty0 1408--1414, 2020{\natexlab{b}}.

\bibitem[Bergen et~al.(2021)Bergen, Soldatov, Kharchenko, and Theis]{bergen2021rna}
V.~Bergen, R.~A. Soldatov, P.~V. Kharchenko, and F.~J. Theis.
\newblock Rna velocity—current challenges and future perspectives.
\newblock \emph{Molecular systems biology}, 17\penalty0 (8):\penalty0 e10282, 2021.

\bibitem[Berman et~al.(2000)]{PDB}
H.~M. Berman et~al.
\newblock The protein data bank.
\newblock \emph{Nucleic Acids Research}, 28\penalty0 (1):\penalty0 235--242, 2000.

\bibitem[Bernot et~al.(2023)Bernot, Owen, Wolfe, Meland, Olesen, and Crandall]{bernot_major_2023}
J.~P. Bernot, C.~L. Owen, J.~M. Wolfe, K.~Meland, J.~Olesen, and K.~A. Crandall.
\newblock Major {Revisions} in {Pancrustacean} {Phylogeny} and {Evidence} of {Sensitivity} to {Taxon} {Sampling}.
\newblock \emph{Molecular Biology and Evolution}, 40\penalty0 (8):\penalty0 msad175, Aug. 2023.
\newblock ISSN 1537-1719.
\newblock \doi{10.1093/molbev/msad175}.
\newblock URL \url{https://doi.org/10.1093/molbev/msad175}.

\bibitem[Blaimer et~al.(2023)Blaimer, Santos, Cruaud, Gates, Kula, Mikó, Rasplus, Smith, Talamas, Brady, and Buffington]{blaimer_key_2023}
B.~B. Blaimer, B.~F. Santos, A.~Cruaud, M.~W. Gates, R.~R. Kula, I.~Mikó, J.-Y. Rasplus, D.~R. Smith, E.~J. Talamas, S.~G. Brady, and M.~L. Buffington.
\newblock Key innovations and the diversification of {Hymenoptera}.
\newblock \emph{Nature Communications}, 14\penalty0 (1):\penalty0 1212, Mar. 2023.
\newblock ISSN 2041-1723.
\newblock \doi{10.1038/s41467-023-36868-4}.
\newblock URL \url{https://www.nature.com/articles/s41467-023-36868-4}.
\newblock Publisher: Nature Publishing Group.

\bibitem[Blei et~al.(2017)Blei, Kucukelbir, and McAuliffe]{Blei2017}
D.~M. Blei, A.~Kucukelbir, and J.~D. McAuliffe.
\newblock Variational inference: A review for statisticians.
\newblock \emph{Journal of the American Statistical Association}, 112\penalty0 (518):\penalty0 859--877, 2017.

\bibitem[Blum and Francois(2006)]{Blum2006}
M.~G. Blum and O.~Francois.
\newblock Random processes of tree growth and statistical tests of tree imbalance.
\newblock \emph{Evolution}, 60\penalty0 (6):\penalty0 1138--1150, 2006.

\bibitem[Bock(1959)]{bock1959preadaptation}
W.~J. Bock.
\newblock Preadaptation and multiple evolutionary pathways.
\newblock \emph{Evolution}, pages 194--211, 1959.

\bibitem[Bojchevski and G{\"u}nnemann(2018)]{bojchevski2018netgan}
A.~Bojchevski and S.~G{\"u}nnemann.
\newblock Netgan: Generating graphs via random walks.
\newblock In \emph{Proceedings of the 35th International Conference on Machine Learning (ICML)}, 2018.

\bibitem[Bou~Dagher et~al.(2024)Bou~Dagher, Madern, Malbos, and Brochier-Armanet]{bou_dagher_persistent_2024}
L.~Bou~Dagher, D.~Madern, P.~Malbos, and C.~Brochier-Armanet.
\newblock Persistent homology reveals strong phylogenetic signal in {3D} protein structures.
\newblock \emph{PNAS nexus}, 3\penalty0 (4):\penalty0 pgae158, 2024.
\newblock Publisher: Oxford University Press US.

\bibitem[Boukouvalas et~al.(2018)Boukouvalas, Hensman, and Rattray]{boukouvalas_bgp_2018}
A.~Boukouvalas, J.~Hensman, and M.~Rattray.
\newblock {BGP}: identifying gene-specific branching dynamics from single-cell data with a branching {Gaussian} process.
\newblock \emph{Genome Biology}, 19\penalty0 (1):\penalty0 65, May 2018.
\newblock ISSN 1474-760X.
\newblock \doi{10.1186/s13059-018-1440-2}.
\newblock URL \url{https://doi.org/10.1186/s13059-018-1440-2}.

\bibitem[Bourret et~al.(2023)Bourret, Fajardo, Frankel, and Rizzo]{bourret_cataloging_2023}
T.~B. Bourret, S.~N. Fajardo, S.~J. Frankel, and D.~M. Rizzo.
\newblock Cataloging {Phytophthora} {Species} of {Agriculture}, {Forests}, {Horticulture}, and {Restoration} {Outplantings} in {California}, {U}.{S}.{A}.: {A} {Sequence}-{Based} {Meta}-{Analysis}.
\newblock \emph{Plant Disease}, Jan. 2023.
\newblock \doi{10.1094/PDIS-01-22-0187-RE}.
\newblock URL \url{https://apsjournals.apsnet.org/doi/10.1094/PDIS-01-22-0187-RE}.
\newblock Publisher: The American Phytopathological Society TLDR: A meta-analysis of Phytophthora detections within the state was conducted using publicly available sequences as a primary source of data rather than published records to better understand threats to California plant health.

\bibitem[Brylinski(2014)]{brylinski_ematchsite_2014}
M.~Brylinski.
\newblock {eMatchSite}: {Sequence} {Order}-{Independent} {Structure} {Alignments} of {Ligand} {Binding} {Pockets} in {Protein} {Models}.
\newblock \emph{PLoS Computational Biology}, 10\penalty0 (9):\penalty0 e1003829, Sept. 2014.
\newblock ISSN 1553-734X.
\newblock \doi{10.1371/journal.pcbi.1003829}.
\newblock URL \url{https://www.ncbi.nlm.nih.gov/pmc/articles/PMC4168975/}.
\newblock TLDR: eMatchSite is a new method for constructing sequence order-independent alignments of ligand binding sites in protein models that opens up the possibility to investigate drug-protein interaction networks for complete proteomes with prospective systems-level applications in polypharmacology and rational drug repositioning.

\bibitem[Campbell and Yau(2019)]{campbell_descriptive_2019}
K.~R. Campbell and C.~Yau.
\newblock A descriptive marker gene approach to single-cell pseudotime inference.
\newblock \emph{Bioinformatics}, 35\penalty0 (1):\penalty0 28--35, Jan. 2019.
\newblock ISSN 1367-4811.
\newblock \doi{10.1093/bioinformatics/bty498}.

\bibitem[Cao et~al.(2019)Cao, Spielmann, Qiu, Huang, Ibrahim, Hill, Zhang, Mundlos, Christiansen, Steemers, Trapnell, and Shendure]{cao_single-cell_2019}
J.~Cao, M.~Spielmann, X.~Qiu, X.~Huang, D.~M. Ibrahim, A.~J. Hill, F.~Zhang, S.~Mundlos, L.~Christiansen, F.~J. Steemers, C.~Trapnell, and J.~Shendure.
\newblock The single-cell transcriptional landscape of mammalian organogenesis.
\newblock \emph{Nature}, 566\penalty0 (7745):\penalty0 496--502, Feb. 2019.
\newblock ISSN 1476-4687.
\newblock \doi{10.1038/s41586-019-0969-x}.
\newblock URL \url{https://www.nature.com/articles/s41586-019-0969-x}.

\bibitem[Catford et~al.(2022)Catford, Wilson, Py{\v{s}}ek, Hulme, and Duncan]{catford2022addressing}
J.~A. Catford, J.~R. Wilson, P.~Py{\v{s}}ek, P.~E. Hulme, and R.~P. Duncan.
\newblock Addressing context dependence in ecology.
\newblock \emph{Trends in Ecology \& Evolution}, 37\penalty0 (2):\penalty0 158--170, 2022.

\bibitem[Cavalli-Sforza and Edwards(1967)]{Cavalli1967}
L.~Cavalli-Sforza and A.~Edwards.
\newblock Phylogenetic analysis: models and estimation procedures.
\newblock \emph{Evolution}, 21\penalty0 (3):\penalty0 550--570, 1967.

\bibitem[Cech and Steitz(2014)]{Cech2014}
T.~R. Cech and J.~A. Steitz.
\newblock The noncoding rna revolution—trashing old rules to forge new ones.
\newblock \emph{Cell}, 157\penalty0 (1):\penalty0 77--94, 2014.
\newblock \doi{10.1016/j.cell.2014.03.008}.

\bibitem[Chen et~al.(2023{\natexlab{a}})Chen, Wu, Li, Wang, Zeng, Xu, Liu, Feng, Chen, He, et~al.]{chen2023tbtools}
C.~Chen, Y.~Wu, J.~Li, X.~Wang, Z.~Zeng, J.~Xu, Y.~Liu, J.~Feng, H.~Chen, Y.~He, et~al.
\newblock Tbtools-ii: A “one for all, all for one” bioinformatics platform for biological big-data mining.
\newblock \emph{Molecular plant}, 16\penalty0 (11):\penalty0 1733--1742, 2023{\natexlab{a}}.

\bibitem[Chen et~al.(2024{\natexlab{a}})Chen, Ryu, Vinyard, Lerer, and Pinello]{chen2024simba}
H.~Chen, J.~Ryu, M.~E. Vinyard, A.~Lerer, and L.~Pinello.
\newblock Simba: single-cell embedding along with features.
\newblock \emph{Nature Methods}, 21\penalty0 (6):\penalty0 1003--1013, 2024{\natexlab{a}}.

\bibitem[Chen et~al.(2023{\natexlab{b}})Chen, Liu, Liu, Guo, Wang, He, Chen, Liao, Zhang, Gao, Dong, Ren, Yang, Zhang, Qi, Li, Zhao, Wang, Wang, Qiao, Li, Jiang, Liu, Song, Deng, Li, Yan, Dong, Li, Li, Yang, Cui, Wang, Zhou, Zhang, Jia, Lu, Zhi, Tang, and Diao]{chen_pangenome_2023}
J.~Chen, Y.~Liu, M.~Liu, W.~Guo, Y.~Wang, Q.~He, W.~Chen, Y.~Liao, W.~Zhang, Y.~Gao, K.~Dong, R.~Ren, T.~Yang, L.~Zhang, M.~Qi, Z.~Li, M.~Zhao, H.~Wang, J.~Wang, Z.~Qiao, H.~Li, Y.~Jiang, G.~Liu, X.~Song, Y.~Deng, H.~Li, F.~Yan, Y.~Dong, Q.~Li, T.~Li, W.~Yang, J.~Cui, H.~Wang, Y.~Zhou, X.~Zhang, G.~Jia, P.~Lu, H.~Zhi, S.~Tang, and X.~Diao.
\newblock Pangenome analysis reveals genomic variations associated with domestication traits in broomcorn millet.
\newblock \emph{Nature Genetics}, 55\penalty0 (12):\penalty0 2243--2254, Dec. 2023{\natexlab{b}}.
\newblock ISSN 1546-1718.
\newblock \doi{10.1038/s41588-023-01571-z}.
\newblock URL \url{https://www.nature.com/articles/s41588-023-01571-z}.
\newblock Publisher: Nature Publishing Group.

\bibitem[Chen et~al.(2023{\natexlab{c}})Chen, Zhang, Hou, Chen, Gao, Tang, Wangdue, Zhang, Sinding, Liu, Han, Lü, Lei, Marshall, and Liu]{chen_evidence_2023}
N.~Chen, Z.~Zhang, J.~Hou, J.~Chen, X.~Gao, L.~Tang, S.~Wangdue, X.~Zhang, M.-H.~S. Sinding, X.~Liu, J.~Han, H.~Lü, C.~Lei, F.~Marshall, and X.~Liu.
\newblock Evidence for early domestic yak, taurine cattle, and their hybrids on the {Tibetan} {Plateau}.
\newblock \emph{Science Advances}, 9\penalty0 (50):\penalty0 eadi6857, Dec. 2023{\natexlab{c}}.
\newblock \doi{10.1126/sciadv.adi6857}.
\newblock URL \url{https://www.science.org/doi/full/10.1126/sciadv.adi6857}.
\newblock Publisher: American Association for the Advancement of Science.

\bibitem[Chen et~al.(2024{\natexlab{b}})Chen, Yang, Cui, Kim, Talwalkar, and Ma]{chen2024applying}
V.~Chen, M.~Yang, W.~Cui, J.~S. Kim, A.~Talwalkar, and J.~Ma.
\newblock Applying interpretable machine learning in computational biology—pitfalls, recommendations and opportunities for new developments.
\newblock \emph{Nature methods}, 21\penalty0 (8):\penalty0 1454--1461, 2024{\natexlab{b}}.

\bibitem[Chen et~al.(2023{\natexlab{d}})Chen, Xie, Li, Cheng, Leng, and Wang]{chen2023information}
X.~Chen, H.~Xie, Z.~Li, G.~Cheng, M.~Leng, and F.~L. Wang.
\newblock Information fusion and artificial intelligence for smart healthcare: a bibliometric study.
\newblock \emph{Information Processing \& Management}, 60\penalty0 (1):\penalty0 103113, 2023{\natexlab{d}}.

\bibitem[Chen et~al.(2022)Chen, King, Hwang, Gerstein, and Zhang]{chen_deepvelo_2022}
Z.~Chen, W.~C. King, A.~Hwang, M.~Gerstein, and J.~Zhang.
\newblock {DeepVelo}: {Single}-cell transcriptomic deep velocity field learning with neural ordinary differential equations.
\newblock \emph{Science Advances}, 8\penalty0 (48):\penalty0 eabq3745, Nov. 2022.
\newblock \doi{10.1126/sciadv.abq3745}.
\newblock URL \url{https://www.science.org/doi/10.1126/sciadv.abq3745}.
\newblock Publisher: American Association for the Advancement of Science.

\bibitem[Choi and Kim(2020)]{choi_whole-proteome_2020}
J.~Choi and S.-H. Kim.
\newblock Whole-proteome tree of life suggests a deep burst of organism diversity.
\newblock \emph{Proceedings of the National Academy of Sciences}, 117\penalty0 (7):\penalty0 3678--3686, Feb. 2020.
\newblock \doi{10.1073/pnas.1915766117}.
\newblock URL \url{https://www.pnas.org/doi/abs/10.1073/pnas.1915766117}.
\newblock Publisher: Proceedings of the National Academy of Sciences TLDR: The main features of a whole-proteome ToL for 4,023 species with known complete or almost complete genome sequences on grouping and kinship among the groups at deep evolutionary levels are described.

\bibitem[Chothia and Finkelstein(1984)]{Chothia1984}
C.~Chothia and A.~V. Finkelstein.
\newblock Principles that determine the structure of proteins.
\newblock \emph{Annual Review of Biochemistry}, 53\penalty0 (1):\penalty0 537--572, 1984.

\bibitem[Ciampi et~al.(2022)Ciampi, Faraoni, Ballerini, and Meli]{ciampi2022co}
F.~Ciampi, M.~Faraoni, J.~Ballerini, and F.~Meli.
\newblock The co-evolutionary relationship between digitalization and organizational agility: Ongoing debates, theoretical developments and future research perspectives.
\newblock \emph{Technological Forecasting and Social Change}, 176:\penalty0 121383, 2022.

\bibitem[Consortium(2015)]{1000GenomesProject}
.~G.~P. Consortium.
\newblock A global reference for human genetic variation.
\newblock \emph{Nature}, 526:\penalty0 68--74, 2015.

\bibitem[Consortium(2013)]{GTExProject}
G.~Consortium.
\newblock The genotype-tissue expression (gtex) project.
\newblock \emph{Nature Genetics}, 45\penalty0 (6):\penalty0 580--585, 2013.

\bibitem[Consortium(2012)]{HMPConsortium}
H.~M.~P. Consortium.
\newblock Structure, function and diversity of the healthy human microbiome.
\newblock \emph{Nature}, 486:\penalty0 207--214, 2012.

\bibitem[Consortium(2001)]{Consortium2001HumanGenome}
I.~H. G.~S. Consortium.
\newblock Initial sequencing and analysis of the human genome.
\newblock \emph{Nature}, 409:\penalty0 860--921, 2001.

\bibitem[Consortium(2019)]{UniProtConsortium}
U.~Consortium.
\newblock Uniprot: a worldwide hub of protein knowledge.
\newblock \emph{Nucleic Acids Research}, 47\penalty0 (D1):\penalty0 D506--D515, 2019.

\bibitem[Dayhoff et~al.(1978)Dayhoff, Schwartz, and Orcutt]{Dayhoff1978}
M.~O. Dayhoff, R.~M. Schwartz, and B.~C. Orcutt.
\newblock \emph{Atlas of protein sequence and structure}.
\newblock National Biomedical Research Foundation, 1978.

\bibitem[De~Bie et~al.(2006)De~Bie, Cristianini, Demuth, and Hahn]{de_bie_cafe_2006}
T.~De~Bie, N.~Cristianini, J.~P. Demuth, and M.~W. Hahn.
\newblock {CAFE}: a computational tool for the study of gene family evolution.
\newblock \emph{Bioinformatics}, 22\penalty0 (10):\penalty0 1269--1271, 2006.
\newblock Publisher: Oxford University Press.

\bibitem[De~Cao and Kipf(2018)]{de2018molgan}
N.~De~Cao and T.~Kipf.
\newblock Molgan: An implicit generative model for small molecular graphs.
\newblock \emph{arXiv preprint arXiv:1805.11973}, 2018.

\bibitem[Delsuc et~al.(2019)Delsuc, Philippe, and Douzery]{Delsuc2019}
F.~Delsuc, H.~Philippe, and E.~J. Douzery.
\newblock Phylogenomics and the reconstruction of the tree of life.
\newblock \emph{Nature Reviews Genetics}, 20\penalty0 (1):\penalty0 1--12, 2019.
\newblock \doi{10.1038/s41576-018-0029-4}.

\bibitem[Desiere et~al.(2006)]{PeptideAtlas}
F.~Desiere et~al.
\newblock The peptideatlas project.
\newblock \emph{Nucleic Acids Research}, 34\penalty0 (suppl\_1):\penalty0 D655--D658, 2006.

\bibitem[Desper and Gascuel(2004)]{Desper2004}
R.~Desper and O.~Gascuel.
\newblock The balanced minimum evolution method of phylogenetic inference.
\newblock \emph{Molecular Biology and Evolution}, 21\penalty0 (3):\penalty0 587--598, 2004.

\bibitem[Ding and Regev(2021)]{ding2021deep}
J.~Ding and A.~Regev.
\newblock Deep generative model embedding of single-cell rna-seq profiles on hyperspheres and hyperbolic spaces.
\newblock \emph{Nature communications}, 12\penalty0 (1):\penalty0 2554, 2021.

\bibitem[Do et~al.(2005)Do, Mahabhashyam, Brudno, and Batzoglou]{do_probcons_2005}
C.~B. Do, M.~S. Mahabhashyam, M.~Brudno, and S.~Batzoglou.
\newblock {ProbCons}: {Probabilistic} consistency-based multiple sequence alignment.
\newblock \emph{Genome research}, 15\penalty0 (2):\penalty0 330--340, 2005.
\newblock Publisher: Cold Spring Harbor Lab.

\bibitem[Domcke and Shendure(2023)]{domcke2023reference}
S.~Domcke and J.~Shendure.
\newblock A reference cell tree will serve science better than a reference cell atlas.
\newblock \emph{Cell}, 186\penalty0 (6):\penalty0 1103--1114, 2023.

\bibitem[Dong et~al.(2018)Dong, Peng, Zhang, and Yang]{dong_mtm-align_2018}
R.~Dong, Z.~Peng, Y.~Zhang, and J.~Yang.
\newblock {mTM}-align: an algorithm for fast and accurate multiple protein structure alignment.
\newblock \emph{Bioinformatics}, 34\penalty0 (10):\penalty0 1719--1725, May 2018.
\newblock ISSN 1367-4803, 1367-4811.
\newblock \doi{10.1093/bioinformatics/btx828}.
\newblock URL \url{https://academic.oup.com/bioinformatics/article/34/10/1719/4769500}.
\newblock TLDR: The proposed multiple structure alignment algorithm (mTM‐align) was proposed, which is an extension of the highly efficient pairwise structure alignment program TM‐align, and benchmarked on four widely used datasets, showing that mTM‐ align consistently outperforms other algorithms.

\bibitem[Doolittle(1981)]{Doolittle1981}
R.~F. Doolittle.
\newblock Protein evolution.
\newblock \emph{Science}, 214\penalty0 (4517):\penalty0 149--159, 1981.
\newblock \doi{10.1126/science.7280692}.

\bibitem[Drummond et~al.(2012)Drummond, Suchard, Xie, and Rambaut]{Drummond2012}
A.~J. Drummond, M.~A. Suchard, D.~Xie, and A.~Rambaut.
\newblock Bayesian phylogenetics with beauti and the beast 1.7.
\newblock \emph{Molecular Biology and Evolution}, 29\penalty0 (8):\penalty0 1969--1973, 2012.

\bibitem[Du et~al.(2024)Du, Chen, Gao, and Wang]{du_joint_2024}
J.-H. Du, T.~Chen, M.~Gao, and J.~Wang.
\newblock Joint trajectory inference for single-cell genomics using deep learning with a mixture prior.
\newblock \emph{Proceedings of the National Academy of Sciences}, 121\penalty0 (37):\penalty0 e2316256121, Sept. 2024.
\newblock \doi{10.1073/pnas.2316256121}.
\newblock URL \url{https://www.pnas.org/doi/abs/10.1073/pnas.2316256121}.
\newblock Publisher: Proceedings of the National Academy of Sciences.

\bibitem[Duan et~al.()Duan, Zang, Li, Xu, and Li]{duanphylogen}
C.~Duan, Z.~Zang, S.~Li, Y.~Xu, and S.~Z. Li.
\newblock Phylogen: Language model-enhanced phylogenetic inference via graph structure generation.
\newblock In \emph{The Thirty-eighth Annual Conference on Neural Information Processing Systems}.

\bibitem[duVerle et~al.(2016)duVerle, Yotsukura, Nomura, Aburatani, and Tsuda]{duverle_celltree_2016}
D.~A. duVerle, S.~Yotsukura, S.~Nomura, H.~Aburatani, and K.~Tsuda.
\newblock {CellTree}: an {R}/bioconductor package to infer the hierarchical structure of cell populations from single-cell {RNA}-seq data.
\newblock \emph{BMC Bioinformatics}, 17\penalty0 (1):\penalty0 363, Sept. 2016.
\newblock ISSN 1471-2105.
\newblock \doi{10.1186/s12859-016-1175-6}.

\bibitem[Dylus et~al.(2024{\natexlab{a}})Dylus, Altenhoff, Majidian, Sedlazeck, and Dessimoz]{dylus2024inference}
D.~Dylus, A.~Altenhoff, S.~Majidian, F.~J. Sedlazeck, and C.~Dessimoz.
\newblock Inference of phylogenetic trees directly from raw sequencing reads using read2tree.
\newblock \emph{Nature Biotechnology}, 42\penalty0 (1):\penalty0 139--147, 2024{\natexlab{a}}.

\bibitem[Dylus et~al.(2024{\natexlab{b}})Dylus, Altenhoff, Majidian, Sedlazeck, and Dessimoz]{dylus_inference_2024}
D.~Dylus, A.~Altenhoff, S.~Majidian, F.~J. Sedlazeck, and C.~Dessimoz.
\newblock Inference of phylogenetic trees directly from raw sequencing reads using {Read2Tree}.
\newblock \emph{Nature Biotechnology}, 42\penalty0 (1):\penalty0 139--147, Jan. 2024{\natexlab{b}}.
\newblock ISSN 1546-1696.
\newblock \doi{10.1038/s41587-023-01753-4}.
\newblock URL \url{https://www.nature.com/articles/s41587-023-01753-4}.
\newblock Publisher: Nature Publishing Group.

\bibitem[Ebert et~al.(2021)Ebert, Audano, Zhu, Rodriguez-Martin, Porubsky, Bonder, Sulovari, Ebler, Zhou, Serra~Mari, Yilmaz, Zhao, Hsieh, Lee, Kumar, Lin, Rausch, Chen, Ren, Santamarina, Höps, Ashraf, Chuang, Yang, Munson, Lewis, Fairley, Tallon, Clarke, Basile, Byrska-Bishop, Corvelo, Evani, Lu, Chaisson, Chen, Li, Brand, Wenger, Ghareghani, Harvey, Raeder, Hasenfeld, Regier, Abel, Hall, Flicek, Stegle, Gerstein, Tubio, Mu, Li, Shi, Hastie, Ye, Chong, Sanders, Zody, Talkowski, Mills, Devine, Lee, Korbel, Marschall, and Eichler]{ebert_haplotype-resolved_2021}
P.~Ebert, P.~A. Audano, Q.~Zhu, B.~Rodriguez-Martin, D.~Porubsky, M.~J. Bonder, A.~Sulovari, J.~Ebler, W.~Zhou, R.~Serra~Mari, F.~Yilmaz, X.~Zhao, P.~Hsieh, J.~Lee, S.~Kumar, J.~Lin, T.~Rausch, Y.~Chen, J.~Ren, M.~Santamarina, W.~Höps, H.~Ashraf, N.~T. Chuang, X.~Yang, K.~M. Munson, A.~P. Lewis, S.~Fairley, L.~J. Tallon, W.~E. Clarke, A.~O. Basile, M.~Byrska-Bishop, A.~Corvelo, U.~S. Evani, T.-Y. Lu, M.~J.~P. Chaisson, J.~Chen, C.~Li, H.~Brand, A.~M. Wenger, M.~Ghareghani, W.~T. Harvey, B.~Raeder, P.~Hasenfeld, A.~A. Regier, H.~J. Abel, I.~M. Hall, P.~Flicek, O.~Stegle, M.~B. Gerstein, J.~M.~C. Tubio, Z.~Mu, Y.~I. Li, X.~Shi, A.~R. Hastie, K.~Ye, Z.~Chong, A.~D. Sanders, M.~C. Zody, M.~E. Talkowski, R.~E. Mills, S.~E. Devine, C.~Lee, J.~O. Korbel, T.~Marschall, and E.~E. Eichler.
\newblock Haplotype-resolved diverse human genomes and integrated analysis of structural variation.
\newblock \emph{Science}, 372\penalty0 (6537):\penalty0 eabf7117, Apr. 2021.
\newblock \doi{10.1126/science.abf7117}.
\newblock URL \url{https://www.science.org/doi/10.1126/science.abf7117}.
\newblock Publisher: American Association for the Advancement of Science.

\bibitem[Edgar et~al.(2002)Edgar, Domrachev, and Lash]{GEO}
R.~Edgar, M.~Domrachev, and A.~E. Lash.
\newblock Gene expression omnibus: Ncbi gene expression and hybridization array data repository.
\newblock \emph{Nucleic Acids Research}, 30\penalty0 (1):\penalty0 207--210, 2002.

\bibitem[Edogbanya et~al.(2021)Edogbanya, Tejada‐Martinez, Jones, Jaiswal, Bell, Cordeiro, van Dam, Rigden, and de~Magalhães]{edogbanya_evolution_2021}
J.~Edogbanya, D.~Tejada‐Martinez, N.~J. Jones, A.~Jaiswal, S.~Bell, R.~Cordeiro, S.~van Dam, D.~J. Rigden, and J.~P. de~Magalhães.
\newblock Evolution, structure and emerging roles of {C1ORF112} in {DNA} replication, {DNA} damage responses, and cancer.
\newblock \emph{Cellular and Molecular Life Sciences}, 78\penalty0 (9):\penalty0 4365--4376, May 2021.
\newblock ISSN 1420-9071.
\newblock \doi{10.1007/s00018-021-03789-8}.
\newblock URL \url{https://doi.org/10.1007/s00018-021-03789-8}.
\newblock TLDR: Gene expression data show that, among human tissues, C1ORF112 is highly expressed in the testes and overexpressed in various cancers when compared to healthy tissues, and protein models suggest that C1ORN112 is an alpha-helical protein.

\bibitem[Eid et~al.(2009)Eid, Fehr, Gray, Luong, Lyle, Otto, Peluso, Rank, Baybayan, Bettman, et~al.]{Eid2009}
J.~Eid, A.~Fehr, J.~Gray, K.~Luong, J.~Lyle, G.~Otto, P.~Peluso, D.~Rank, P.~Baybayan, B.~Bettman, et~al.
\newblock Real-time dna sequencing from single polymerase molecules.
\newblock \emph{Science}, 323\penalty0 (5910):\penalty0 133--138, 2009.

\bibitem[Elhamod et~al.(2023)Elhamod, Khurana, Manogaran, Uyeda, Balk, Dahdul, Bakis, Bart~Jr, Mabee, Lapp, et~al.]{elhamod2023discovering}
M.~Elhamod, M.~Khurana, H.~B. Manogaran, J.~C. Uyeda, M.~A. Balk, W.~Dahdul, Y.~Bakis, H.~L. Bart~Jr, P.~M. Mabee, H.~Lapp, et~al.
\newblock Discovering novel biological traits from images using phylogeny-guided neural networks.
\newblock In \emph{Proceedings of the 29th ACM SIGKDD Conference on Knowledge Discovery and Data Mining}, pages 3966--3978, 2023.

\bibitem[Eme et~al.(2023)Eme, Tamarit, Caceres, Stairs, De~Anda, Schön, Seitz, Dombrowski, Lewis, Homa, Saw, Lombard, Nunoura, Li, Hua, Chen, Banfield, John, Reysenbach, Stott, Schramm, Kjeldsen, Teske, Baker, and Ettema]{eme_inference_2023}
L.~Eme, D.~Tamarit, E.~F. Caceres, C.~W. Stairs, V.~De~Anda, M.~E. Schön, K.~W. Seitz, N.~Dombrowski, W.~H. Lewis, F.~Homa, J.~H. Saw, J.~Lombard, T.~Nunoura, W.-J. Li, Z.-S. Hua, L.-X. Chen, J.~F. Banfield, E.~S. John, A.-L. Reysenbach, M.~B. Stott, A.~Schramm, K.~U. Kjeldsen, A.~P. Teske, B.~J. Baker, and T.~J.~G. Ettema.
\newblock Inference and reconstruction of the heimdallarchaeial ancestry of eukaryotes.
\newblock \emph{Nature}, 618\penalty0 (7967):\penalty0 992--999, June 2023.
\newblock ISSN 1476-4687.
\newblock \doi{10.1038/s41586-023-06186-2}.
\newblock URL \url{https://www.nature.com/articles/s41586-023-06186-2}.
\newblock Publisher: Nature Publishing Group.

\bibitem[Esser-Skala and Fortelny(2023)]{esser2023reliable}
W.~Esser-Skala and N.~Fortelny.
\newblock Reliable interpretability of biology-inspired deep neural networks.
\newblock \emph{NPJ Systems Biology and Applications}, 9\penalty0 (1):\penalty0 50, 2023.

\bibitem[Faith(1992)]{Faith1992}
D.~P. Faith.
\newblock Conservation evaluation and phylogenetic diversity.
\newblock \emph{Biological Conservation}, 61\penalty0 (1):\penalty0 1--10, 1992.

\bibitem[Felsenstein(1981)]{Felsenstein1981}
J.~Felsenstein.
\newblock Evolutionary trees from dna sequences: A maximum likelihood approach.
\newblock \emph{Journal of Molecular Evolution}, 17\penalty0 (6):\penalty0 368--376, 1981.

\bibitem[Felsenstein(1985)]{Felsenstein1985}
J.~Felsenstein.
\newblock \emph{Phylogenies and the Comparative Method}, volume 125.
\newblock University of Chicago Press, 1985.

\bibitem[Felsenstein(2004)]{Felsenstein2004}
J.~Felsenstein.
\newblock \emph{Inferring phylogenies}.
\newblock Sinauer Associates, 2004.

\bibitem[Feng et~al.(2024)Feng, Zheng, Irisarri, Yu, Zheng, Ali, de~Vries, Keller, Fürst-Jansen, Dadras, Zegers, Rieseberg, Dhabalia~Ashok, Darienko, Bierenbroodspot, Gramzow, Petroll, Haas, Fernandez-Pozo, Nousias, Li, Fitzek, Grayburn, Rittmeier, Permann, Rümpler, Archibald, Theißen, Mower, Lorenz, Buschmann, von Schwartzenberg, Boston, Hayes, Daum, Barry, Grigoriev, Wang, Li, Rensing, Ben~Ari, Keren, Mosquna, Holzinger, Delaux, Zhang, Huang, Mutwil, de~Vries, and Yin]{feng_genomes_2024}
X.~Feng, J.~Zheng, I.~Irisarri, H.~Yu, B.~Zheng, Z.~Ali, S.~de~Vries, J.~Keller, J.~M.~R. Fürst-Jansen, A.~Dadras, J.~M.~S. Zegers, T.~P. Rieseberg, A.~Dhabalia~Ashok, T.~Darienko, M.~J. Bierenbroodspot, L.~Gramzow, R.~Petroll, F.~B. Haas, N.~Fernandez-Pozo, O.~Nousias, T.~Li, E.~Fitzek, W.~S. Grayburn, N.~Rittmeier, C.~Permann, F.~Rümpler, J.~M. Archibald, G.~Theißen, J.~P. Mower, M.~Lorenz, H.~Buschmann, K.~von Schwartzenberg, L.~Boston, R.~D. Hayes, C.~Daum, K.~Barry, I.~V. Grigoriev, X.~Wang, F.-W. Li, S.~A. Rensing, J.~Ben~Ari, N.~Keren, A.~Mosquna, A.~Holzinger, P.-M. Delaux, C.~Zhang, J.~Huang, M.~Mutwil, J.~de~Vries, and Y.~Yin.
\newblock Genomes of multicellular algal sisters to land plants illuminate signaling network evolution.
\newblock \emph{Nature Genetics}, 56\penalty0 (5):\penalty0 1018--1031, May 2024.
\newblock ISSN 1546-1718.
\newblock \doi{10.1038/s41588-024-01737-3}.
\newblock URL \url{https://www.nature.com/articles/s41588-024-01737-3}.
\newblock Publisher: Nature Publishing Group.

\bibitem[Finn et~al.(2016)Finn, Coggill, Eberhardt, et~al.]{Finn2016}
R.~D. Finn, P.~Coggill, R.~Y. Eberhardt, et~al.
\newblock The pfam protein families database: towards a more sustainable future.
\newblock \emph{Nucleic Acids Research}, 44\penalty0 (D1):\penalty0 D279--D285, 2016.

\bibitem[Fisk et~al.(2022)Fisk, Mahal, Dornburg, Gaffney, Aneja, Contessa, Rimm, Yu, and Townsend]{fisk_premetastatic_2022}
J.~N. Fisk, A.~R. Mahal, A.~Dornburg, S.~G. Gaffney, S.~Aneja, J.~N. Contessa, D.~Rimm, J.~B. Yu, and J.~P. Townsend.
\newblock Premetastatic shifts of endogenous and exogenous mutational processes support consolidative therapy in {EGFR}-driven lung adenocarcinoma.
\newblock \emph{Cancer Letters}, 526:\penalty0 346--351, Feb. 2022.
\newblock ISSN 0304-3835.
\newblock \doi{10.1016/j.canlet.2021.11.011}.
\newblock URL \url{https://www.sciencedirect.com/science/article/pii/S0304383521005784}.
\newblock TLDR: Mutational signature analyses within clinically annotated cancer chronograms are applied to detect and describe the shifting mutational processes caused by both endogenous and exogenous factors between tumor sampling timepoints to inform therapeutic decision making and retrospective assessment of disease etiology.

\bibitem[Fitch(1971)]{Fitch1971}
W.~M. Fitch.
\newblock Toward defining the course of evolution: minimum change for a specific tree topology.
\newblock \emph{Systematic Zoology}, 20\penalty0 (4):\penalty0 406--416, 1971.

\bibitem[Flouri et~al.(2018)Flouri, Jiao, Rannala, and Yang]{flouri_species_2018}
T.~Flouri, X.~Jiao, B.~Rannala, and Z.~Yang.
\newblock Species {Tree} {Inference} with {BPP} {Using} {Genomic} {Sequences} and the {Multispecies} {Coalescent}.
\newblock \emph{Molecular Biology and Evolution}, 35\penalty0 (10):\penalty0 2585--2593, Oct. 2018.
\newblock ISSN 0737-4038, 1537-1719.
\newblock \doi{10.1093/molbev/msy147}.
\newblock URL \url{https://academic.oup.com/mbe/article/35/10/2585/5057515}.

\bibitem[Forrow and Schiebinger(2021)]{forrow2021lineageot}
A.~Forrow and G.~Schiebinger.
\newblock Lineageot is a unified framework for lineage tracing and trajectory inference.
\newblock \emph{Nature communications}, 12\penalty0 (1):\penalty0 4940, 2021.

\bibitem[Gao and Skolnick(2013)]{gao_apoc_2013}
M.~Gao and J.~Skolnick.
\newblock {APoc}: large-scale identification of similar protein pockets.
\newblock \emph{Bioinformatics}, 29\penalty0 (5):\penalty0 597--604, Mar. 2013.
\newblock ISSN 1367-4811, 1367-4803.
\newblock \doi{10.1093/bioinformatics/btt024}.
\newblock URL \url{https://academic.oup.com/bioinformatics/article/29/5/597/254660}.
\newblock TLDR: This work introduces a computational method, APoc (Alignment of Pockets), for the large-scale, sequence order-independent, structural comparison of protein pockets, and demonstrates that APoc has better performance than the geometric hashing-based method SiteEngine.

\bibitem[Gayoso et~al.(2024)Gayoso, Weiler, Lotfollahi, Klein, Hong, Streets, Theis, and Yosef]{gayoso_deep_2024}
A.~Gayoso, P.~Weiler, M.~Lotfollahi, D.~Klein, J.~Hong, A.~Streets, F.~J. Theis, and N.~Yosef.
\newblock Deep generative modeling of transcriptional dynamics for {RNA} velocity analysis in single cells.
\newblock \emph{Nat Methods}, 21\penalty0 (1):\penalty0 50--59, Jan. 2024.
\newblock ISSN 1548-7105.
\newblock \doi{10.1038/s41592-023-01994-w}.
\newblock URL \url{https://www.nature.com/articles/s41592-023-01994-w}.
\newblock Publisher: Nature Publishing Group.

\bibitem[Ghaly et~al.(2022)Ghaly, Tetu, Penesyan, Qi, Rajabal, and Gillings]{ghaly_discovery_2022}
T.~M. Ghaly, S.~G. Tetu, A.~Penesyan, Q.~Qi, V.~Rajabal, and M.~R. Gillings.
\newblock Discovery of integrons in {Archaea}: {Platforms} for cross-domain gene transfer.
\newblock \emph{Science Advances}, 8\penalty0 (46):\penalty0 eabq6376, Nov. 2022.
\newblock \doi{10.1126/sciadv.abq6376}.
\newblock URL \url{https://www.science.org/doi/full/10.1126/sciadv.abq6376}.
\newblock Publisher: American Association for the Advancement of Science.

\bibitem[Gharaee et~al.(2024)Gharaee, Gong, Pellegrino, Zarubiieva, Haurum, Lowe, McKeown, Ho, McLeod, and et~al.]{BIOSCAN1M}
Z.~Gharaee, Z.~Gong, N.~Pellegrino, I.~Zarubiieva, J.~B. Haurum, S.~Lowe, J.~McKeown, C.~Ho, J.~McLeod, and Y.-Y.~W. et~al.
\newblock A step towards worldwide biodiversity assessment: The bioscan-1m insect dataset.
\newblock \emph{Advances in Neural Information Processing Systems}, 36, 2024.
\newblock URL \url{https://www.bioscan.org/}.
\newblock Accessed: 2024-09-17.

\bibitem[Gonzalez-Reiche et~al.(2023)Gonzalez-Reiche, Alshammary, Schaefer, Patel, Polanco, Carreño, Amoako, Rooker, Cognigni, Floda, van~de Guchte, Khalil, Farrugia, Assad, Zhang, Alburquerque, Sominsky, Gleason, Srivastava, Sebra, Ramirez, Banu, Shrestha, Krammer, Paniz-Mondolfi, Sordillo, Simon, and van Bakel]{gonzalez-reiche_sequential_2023}
A.~S. Gonzalez-Reiche, H.~Alshammary, S.~Schaefer, G.~Patel, J.~Polanco, J.~M. Carreño, A.~A. Amoako, A.~Rooker, C.~Cognigni, D.~Floda, A.~van~de Guchte, Z.~Khalil, K.~Farrugia, N.~Assad, J.~Zhang, B.~Alburquerque, L.~A. Sominsky, C.~Gleason, K.~Srivastava, R.~Sebra, J.~D. Ramirez, R.~Banu, P.~Shrestha, F.~Krammer, A.~Paniz-Mondolfi, E.~M. Sordillo, V.~Simon, and H.~van Bakel.
\newblock Sequential intrahost evolution and onward transmission of {SARS}-{CoV}-2 variants.
\newblock \emph{Nature Communications}, 14\penalty0 (1):\penalty0 3235, June 2023.
\newblock ISSN 2041-1723.
\newblock \doi{10.1038/s41467-023-38867-x}.
\newblock URL \url{https://www.nature.com/articles/s41467-023-38867-x}.
\newblock Publisher: Nature Publishing Group.

\bibitem[Gonçalves et~al.(2024)Gonçalves, Manduchi, Vandenhirtz, and Vogt]{goncalves_structured_2024}
J.~d.~S. Gonçalves, L.~Manduchi, M.~Vandenhirtz, and J.~E. Vogt.
\newblock Structured {Generations}: {Using} {Hierarchical} {Clusters} to guide {Diffusion} {Models}.
\newblock In \emph{ICML 2024 Workshop on Structured Probabilistic Inference $\{$$\backslash$\&$\}$ Generative Modeling}, July 2024.
\newblock URL \url{https://openreview.net/forum?id=WlibPykp0H}.

\bibitem[Goodfellow et~al.(2014)]{goodfellow2014generative}
I.~Goodfellow et~al.
\newblock Generative adversarial nets.
\newblock In \emph{Advances in Neural Information Processing Systems}, volume~27, pages 2672--2680, 2014.

\bibitem[Gorbalenya et~al.(2020)Gorbalenya, Baker, Baric, de~Groot, Drosten, Gulyaeva, Haagmans, Lauber, Leontovich, Neuman, Penzar, Perlman, Poon, Samborskiy, Sidorov, Sola, Ziebuhr, and {Coronaviridae Study Group of the International Committee on Taxonomy of Viruses}]{gorbalenya_species_2020}
A.~E. Gorbalenya, S.~C. Baker, R.~S. Baric, R.~J. de~Groot, C.~Drosten, A.~A. Gulyaeva, B.~L. Haagmans, C.~Lauber, A.~M. Leontovich, B.~W. Neuman, D.~Penzar, S.~Perlman, L.~L.~M. Poon, D.~V. Samborskiy, I.~A. Sidorov, I.~Sola, J.~Ziebuhr, and {Coronaviridae Study Group of the International Committee on Taxonomy of Viruses}.
\newblock The species {Severe} acute respiratory syndrome-related coronavirus: classifying 2019-{nCoV} and naming it {SARS}-{CoV}-2.
\newblock \emph{Nature Microbiology}, 5\penalty0 (4):\penalty0 536--544, Apr. 2020.
\newblock ISSN 2058-5276.
\newblock \doi{10.1038/s41564-020-0695-z}.
\newblock URL \url{https://www.nature.com/articles/s41564-020-0695-z}.
\newblock Publisher: Nature Publishing Group.

\bibitem[Grigoriadis et~al.(2024)Grigoriadis, Huebner, Bunkum, Colliver, Frankell, Hill, Thol, Birkbak, Swanton, Zaccaria, et~al.]{grigoriadis2024conipher}
K.~Grigoriadis, A.~Huebner, A.~Bunkum, E.~Colliver, A.~M. Frankell, M.~S. Hill, K.~Thol, N.~J. Birkbak, C.~Swanton, S.~Zaccaria, et~al.
\newblock Conipher: a computational framework for scalable phylogenetic reconstruction with error correction.
\newblock \emph{Nature Protocols}, 19\penalty0 (1):\penalty0 159--183, 2024.

\bibitem[Gu et~al.(2005)Gu, Zhang, and Huang]{Gu2005}
X.~Gu, Z.~Zhang, and W.~Huang.
\newblock Rapid evolution of expression and regulatory divergences after yeast gene duplication.
\newblock \emph{Proceedings of the National Academy of Sciences}, 102\penalty0 (3):\penalty0 707--712, 2005.

\bibitem[Guindon and Gascuel(2003)]{Guindon2003}
S.~Guindon and O.~Gascuel.
\newblock A simple, fast, and accurate algorithm to estimate large phylogenies by maximum likelihood.
\newblock \emph{Systematic Biology}, 52\penalty0 (5):\penalty0 696--704, 2003.

\bibitem[Guindon et~al.(2010)Guindon, Dufayard, Lefort, Anisimova, Hordijk, and Gascuel]{guindon2010new}
S.~Guindon, J.-F. Dufayard, V.~Lefort, M.~Anisimova, W.~Hordijk, and O.~Gascuel.
\newblock New algorithms and methods to estimate maximum-likelihood phylogenies: assessing the performance of phyml 3.0.
\newblock \emph{Systematic biology}, 59\penalty0 (3):\penalty0 307--321, 2010.

\bibitem[Guo et~al.(2023)Guo, Luo, Gao, Yi, Li, Yang, and Li]{guo_phylogenomics_2023}
C.~Guo, Y.~Luo, L.-M. Gao, T.-S. Yi, H.-T. Li, J.-B. Yang, and D.-Z. Li.
\newblock Phylogenomics and the flowering plant tree of life.
\newblock \emph{Journal of Integrative Plant Biology}, 65\penalty0 (2):\penalty0 299--323, 2023.
\newblock ISSN 1744-7909.
\newblock \doi{10.1111/jipb.13415}.
\newblock URL \url{https://onlinelibrary.wiley.com/doi/abs/10.1111/jipb.13415}.
\newblock \_eprint: https://onlinelibrary.wiley.com/doi/pdf/10.1111/jipb.13415.

\bibitem[Guo et~al.(2025)Guo, Yang, Zhang, Song, Zhang, Xu, Zhu, Ma, Wang, Bi, et~al.]{guo2025deepseek}
D.~Guo, D.~Yang, H.~Zhang, J.~Song, R.~Zhang, R.~Xu, Q.~Zhu, S.~Ma, P.~Wang, X.~Bi, et~al.
\newblock Deepseek-r1: Incentivizing reasoning capability in llms via reinforcement learning.
\newblock \emph{arXiv preprint arXiv:2501.12948}, 2025.

\bibitem[Göbel et~al.(1994)Göbel, Sander, Schneider, and Valencia]{Gobel1994}
U.~Göbel, C.~Sander, R.~Schneider, and A.~Valencia.
\newblock Correlated mutations and residue contacts in proteins.
\newblock \emph{Proteins: Structure, Function, and Bioinformatics}, 18\penalty0 (4):\penalty0 309--317, 1994.

\bibitem[Haghverdi et~al.(2016)Haghverdi, Buettner, and Theis]{Haghverdi2016}
L.~Haghverdi, F.~Buettner, and F.~J. Theis.
\newblock Diffusion pseudotime robustly reconstructs lineage branching.
\newblock \emph{Nature Methods}, 13\penalty0 (10):\penalty0 845--848, 2016.

\bibitem[Hahn(2009)]{Hahn2009}
M.~W. Hahn.
\newblock Distinguishing among evolutionary models for the maintenance of gene duplicates.
\newblock \emph{Journal of Heredity}, 100\penalty0 (5):\penalty0 605--617, 2009.

\bibitem[Han et~al.(2018)Han, Cao, Lv, Lin, Liu, Sun, and Li]{han2018openke}
X.~Han, S.~Cao, X.~Lv, Y.~Lin, Z.~Liu, M.~Sun, and J.~Li.
\newblock Openke: An open toolkit for knowledge embedding.
\newblock In \emph{Proceedings of the 2018 conference on empirical methods in natural language processing: system demonstrations}, pages 139--144, 2018.

\bibitem[Hasin et~al.(2017)Hasin, Seldin, and Lusis]{hasin2017multi}
Y.~Hasin, M.~Seldin, and A.~Lusis.
\newblock Multi-omics approaches to disease.
\newblock \emph{Genome Biology}, 18\penalty0 (1):\penalty0 83, 2017.

\bibitem[Hayes et~al.(2024)Hayes, Rao, Akin, Sofroniew, Oktay, Lin, Verkuil, Tran, Deaton, Wiggert, Badkundri, Shafkat, Gong, Derry, Molina, Thomas, Khan, Mishra, Kim, Bartie, Nemeth, Hsu, Sercu, Candido, and Rives]{hayes_simulating_2024}
T.~Hayes, R.~Rao, H.~Akin, N.~J. Sofroniew, D.~Oktay, Z.~Lin, R.~Verkuil, V.~Q. Tran, J.~Deaton, M.~Wiggert, R.~Badkundri, I.~Shafkat, J.~Gong, A.~Derry, R.~S. Molina, N.~Thomas, Y.~Khan, C.~Mishra, C.~Kim, L.~J. Bartie, M.~Nemeth, P.~D. Hsu, T.~Sercu, S.~Candido, and A.~Rives.
\newblock Simulating 500 million years of evolution with a language model, July 2024.
\newblock URL \url{https://www.biorxiv.org/content/10.1101/2024.07.01.600583v1}.
\newblock Pages: 2024.07.01.600583 Section: New Results TLDR: This work presents ESM3, a frontier multimodal generative language model that reasons over the sequence, structure, and function of proteins, and prompts ESM3 to generate fluorescent proteins with a chain of thought.

\bibitem[Hedges(2002)]{hedges2002origin}
S.~B. Hedges.
\newblock The origin and evolution of model organisms.
\newblock \emph{Nature Reviews Genetics}, 3\penalty0 (11):\penalty0 838--849, 2002.

\bibitem[Hennig(1965)]{Hennig1965}
W.~Hennig.
\newblock Phylogenetic systematics.
\newblock \emph{Annual Review of Entomology}, 10\penalty0 (1):\penalty0 97--116, 1965.

\bibitem[Hennig(1966)]{Hennig1966}
W.~Hennig.
\newblock \emph{Phylogenetic Systematics}.
\newblock University of Illinois Press, 1966.
\newblock ISBN 978-0252068140.

\bibitem[Hillis(2019)]{Hillis2019}
D.~M. Hillis.
\newblock The tree of life: Resolving the relationships of the majority of living species.
\newblock \emph{Systematic Biology}, 68\penalty0 (5):\penalty0 896--900, 2019.

\bibitem[Hillis and Huelsenbeck(1992)]{Hillis1992}
D.~M. Hillis and J.~P. Huelsenbeck.
\newblock Phylogeny and the evolution of hiv.
\newblock \emph{Science}, 257\penalty0 (5079):\penalty0 1159--1163, 1992.

\bibitem[Hohna et~al.(2016)Hohna, Landis, Heath, Boussau, Lartillot, Moore, Huelsenbeck, and Ronquist]{Hohna2016}
S.~Hohna, M.~J. Landis, T.~A. Heath, B.~Boussau, N.~Lartillot, B.~R. Moore, J.~P. Huelsenbeck, and F.~Ronquist.
\newblock Revbayes: Bayesian phylogenetic inference using graphical models and an interactive model-specification language.
\newblock \emph{Systematic biology}, 65\penalty0 (4):\penalty0 726--736, 2016.

\bibitem[H{\o}ie et~al.(2022)H{\o}ie, Cagiada, Frederiksen, Stein, and Lindorff-Larsen]{hoie2022predicting}
M.~H. H{\o}ie, M.~Cagiada, A.~H.~B. Frederiksen, A.~Stein, and K.~Lindorff-Larsen.
\newblock Predicting and interpreting large-scale mutagenesis data using analyses of protein stability and conservation.
\newblock \emph{Cell reports}, 38\penalty0 (2), 2022.

\bibitem[Holm and Sander(1995)]{holm_dali_1995}
L.~Holm and C.~Sander.
\newblock Dali: a network tool for protein structure comparison.
\newblock \emph{Trends in Biochemical Sciences}, 20\penalty0 (11):\penalty0 478--480, Nov. 1995.
\newblock ISSN 0968-0004.
\newblock \doi{10.1016/S0968-0004(00)89105-7}.
\newblock URL \url{https://www.sciencedirect.com/science/article/pii/S0968000400891057}.

\bibitem[Hong et~al.(2023)Hong, Zhang, Li, Li, Yao, Li, Werner, Chanussot, Zipf, and Zhu]{hong2023cross}
D.~Hong, B.~Zhang, H.~Li, Y.~Li, J.~Yao, C.~Li, M.~Werner, J.~Chanussot, A.~Zipf, and X.~X. Zhu.
\newblock Cross-city matters: A multimodal remote sensing benchmark dataset for cross-city semantic segmentation using high-resolution domain adaptation networks.
\newblock \emph{Remote Sensing of Environment}, 299:\penalty0 113856, 2023.

\bibitem[Hu et~al.(2024)Hu, Sun, Nian, Wang, Dang, Li, Feng, Yu, Tao, et~al.]{hu2024self}
X.~Hu, Z.~Sun, Y.~Nian, Y.~Wang, Y.~Dang, F.~Li, J.~Feng, E.~Yu, C.~Tao, et~al.
\newblock Self-explainable graph neural network for alzheimer disease and related dementias risk prediction: Algorithm development and validation study.
\newblock \emph{JMIR aging}, 7\penalty0 (1):\penalty0 e54748, 2024.

\bibitem[Huang et~al.(2020)Huang, Chaudhary, and Garmire]{huang2020fusion}
S.~Huang, K.~Chaudhary, and L.~X. Garmire.
\newblock Fusion of multi-omics data and deep learning for cancer patient survivability prediction.
\newblock \emph{Methods}, 166:\penalty0 28--37, 2020.

\bibitem[Huelsenbeck and Ronquist(2001)]{Huelsenbeck2001}
J.~P. Huelsenbeck and F.~Ronquist.
\newblock Mrbayes: Bayesian inference of phylogenetic trees.
\newblock \emph{Bioinformatics}, 17\penalty0 (8):\penalty0 754--755, 2001.
\newblock \doi{10.1093/bioinformatics/17.8.754}.

\bibitem[Huelsenbeck et~al.(1996)Huelsenbeck, Bull, and Cunningham]{huelsenbeck1996combining}
J.~P. Huelsenbeck, J.~Bull, and C.~W. Cunningham.
\newblock Combining data in phylogenetic analysis.
\newblock \emph{Trends in Ecology \& Evolution}, 11\penalty0 (4):\penalty0 152--158, 1996.

\bibitem[Huguet et~al.(2022)Huguet, Magruder, Tong, Fasina, Kuchroo, Wolf, and Krishnaswamy]{huguet_manifold_2022}
G.~Huguet, D.~S. Magruder, A.~Tong, O.~Fasina, M.~Kuchroo, G.~Wolf, and S.~Krishnaswamy.
\newblock Manifold {Interpolating} {Optimal}-{Transport} {Flows} for {Trajectory} {Inference}.
\newblock \emph{Advances in Neural Information Processing Systems}, 35:\penalty0 29705--29718, Dec. 2022.
\newblock URL \url{https://proceedings.neurips.cc/paper_files/paper/2022/hash/bfc03f077688d8885c0a9389d77616d0-Abstract-Conference.html}.

\bibitem[Ikotun et~al.(2023)Ikotun, Ezugwu, Abualigah, Abuhaija, and Heming]{ikotun2023k}
A.~M. Ikotun, A.~E. Ezugwu, L.~Abualigah, B.~Abuhaija, and J.~Heming.
\newblock K-means clustering algorithms: A comprehensive review, variants analysis, and advances in the era of big data.
\newblock \emph{Information Sciences}, 622:\penalty0 178--210, 2023.

\bibitem[iNaturalist(2021)]{iNat21}
iNaturalist.
\newblock inaturalist 2021 dataset (inat21), 2021.
\newblock URL \url{https://www.inaturalist.org/}.
\newblock Accessed: 2024-09-17.

\bibitem[Izquierdo-Carrasco et~al.(2011)Izquierdo-Carrasco, Smith, and Stamatakis]{izquierdo-carrasco_algorithms_2011}
F.~Izquierdo-Carrasco, S.~A. Smith, and A.~Stamatakis.
\newblock Algorithms, data structures, and numerics for likelihood-based phylogenetic inference of huge trees.
\newblock \emph{BMC Bioinformatics}, 12\penalty0 (1):\penalty0 470, Dec. 2011.
\newblock ISSN 1471-2105.
\newblock \doi{10.1186/1471-2105-12-470}.
\newblock URL \url{https://doi.org/10.1186/1471-2105-12-470}.

\bibitem[Jacomy et~al.(2014)Jacomy, Venturini, Heymann, and Bastian]{jacomy_forceatlas2_2014}
M.~Jacomy, T.~Venturini, S.~Heymann, and M.~Bastian.
\newblock {ForceAtlas2}, a {Continuous} {Graph} {Layout} {Algorithm} for {Handy} {Network} {Visualization} {Designed} for the {Gephi} {Software}.
\newblock \emph{PLOS ONE}, 9\penalty0 (6):\penalty0 e98679, June 2014.
\newblock ISSN 1932-6203.
\newblock \doi{10.1371/journal.pone.0098679}.
\newblock URL \url{https://journals.plos.org/plosone/article?id=10.1371/journal.pone.0098679}.
\newblock Publisher: Public Library of Science.

\bibitem[Jain et~al.(2016)Jain, Fiddes, Miga, Olsen, Paten, and Akeson]{Jain2016}
M.~Jain, I.~T. Fiddes, K.~H. Miga, H.~E. Olsen, B.~Paten, and M.~Akeson.
\newblock The oxford nanopore minion: delivery of nanopore sequencing to the genomics community.
\newblock \emph{Genome Biology}, 17\penalty0 (1):\penalty0 1--11, 2016.

\bibitem[Ji and Ji(2016)]{ji_tscan_2016}
Z.~Ji and H.~Ji.
\newblock {TSCAN}: {Pseudo}-time reconstruction and evaluation in single-cell {RNA}-seq analysis.
\newblock \emph{Nucleic Acids Research}, 44\penalty0 (13):\penalty0 e117, July 2016.
\newblock ISSN 0305-1048.
\newblock \doi{10.1093/nar/gkw430}.
\newblock URL \url{https://doi.org/10.1093/nar/gkw430}.

\bibitem[Jia et~al.(2023)Jia, Wang, Zhao, Zhang, Chen, Xu, and Yi]{jia_origin_2023}
X.~Jia, L.~Wang, H.~Zhao, Y.~Zhang, Z.~Chen, L.~Xu, and K.~Yi.
\newblock The origin and evolution of salicylic acid signaling and biosynthesis in plants.
\newblock \emph{Molecular Plant}, 16\penalty0 (1):\penalty0 245--259, Jan. 2023.
\newblock ISSN 1674-2052.
\newblock \doi{10.1016/j.molp.2022.12.002}.
\newblock URL \url{https://www.cell.com/molecular-plant/abstract/S1674-2052(22)00437-3}.
\newblock Publisher: Elsevier TLDR: 10 core protein families in SA signaling and biosynthesis across green plant lineages are identified and it is revealed that the ancient abnormal inflorescence meristem 1 (AIM1)-based beta-oxidation pathway is crucial for the biosynthesis of SA in chlorophyte algae, and this biosynthesis pathway may have facilitated the adaptation of early-diverging green algae to the high-light-intensity environment on land.

\bibitem[Jiang et~al.(2022)Jiang, Tabaghi, and Mirarab]{jiang_learning_2022}
Y.~Jiang, P.~Tabaghi, and S.~Mirarab.
\newblock Learning {Hyperbolic} {Embedding} for {Phylogenetic} {Tree} {Placement} and {Updates}.
\newblock \emph{Biology}, 11\penalty0 (9):\penalty0 1256, Sept. 2022.
\newblock ISSN 2079-7737.
\newblock \doi{10.3390/biology11091256}.
\newblock URL \url{https://www.mdpi.com/2079-7737/11/9/1256}.
\newblock Number: 9 Publisher: Multidisciplinary Digital Publishing Institute TLDR: It is shown how the conventional (Euclidean) deep learning methods developed for phylogenetics can benefit from using hyperbolic geometry, and the appropriate geometry for faithfully representing tree distances while embedding gene sequences is examined.

\bibitem[Jiang et~al.(2023)Jiang, Wang, Feng, Jin, Liang, Li, Yu, Ma, Su, Zou, et~al.]{jiang2023explainable}
Y.~Jiang, R.~Wang, J.~Feng, J.~Jin, S.~Liang, Z.~Li, Y.~Yu, A.~Ma, R.~Su, Q.~Zou, et~al.
\newblock Explainable deep hypergraph learning modeling the peptide secondary structure prediction.
\newblock \emph{Advanced Science}, 10\penalty0 (11):\penalty0 2206151, 2023.

\bibitem[Jin et~al.(2018{\natexlab{a}})Jin, Barzilay, and Jaakkola]{jin2018junction}
W.~Jin, R.~Barzilay, and T.~Jaakkola.
\newblock Junction tree variational autoencoder for molecular graph generation.
\newblock In \emph{International conference on machine learning}, pages 2323--2332. PMLR, 2018{\natexlab{a}}.

\bibitem[Jin et~al.(2018{\natexlab{b}})Jin, Barzilay, and Jaakkola]{jin_junction_2018}
W.~Jin, R.~Barzilay, and T.~Jaakkola.
\newblock Junction {Tree} {Variational} {Autoencoder} for {Molecular} {Graph} {Generation}.
\newblock In \emph{Proceedings of the 35th {International} {Conference} on {Machine} {Learning}}, pages 2323--2332. PMLR, July 2018{\natexlab{b}}.
\newblock URL \url{https://proceedings.mlr.press/v80/jin18a.html}.
\newblock ISSN: 2640-3498.

\bibitem[Jones et~al.(1992)Jones, Taylor, and Thornton]{Jones1992}
D.~T. Jones, W.~R. Taylor, and J.~M. Thornton.
\newblock The rapid generation of mutation data matrices from protein sequences.
\newblock \emph{Computer Applications in the Biosciences}, 8\penalty0 (3):\penalty0 275--282, 1992.

\bibitem[Jonsson et~al.(2024)Jonsson, Hofmann, Mereiter, Hartley-Tassell, Sakic, Oliveira, Hoffmann, Novatchkova, Schleiffer, and Penninger]{jonsson_clec18a_2024}
G.~Jonsson, M.~Hofmann, S.~Mereiter, L.~Hartley-Tassell, I.~Sakic, T.~Oliveira, D.~Hoffmann, M.~Novatchkova, A.~Schleiffer, and J.~M. Penninger.
\newblock {CLEC18A} interacts with sulfated {GAGs} and controls clear cell renal cell carcinoma progression, Sept. 2024.
\newblock URL \url{https://www.biorxiv.org/content/10.1101/2024.07.08.602586v3}.
\newblock Pages: 2024.07.08.602586 Section: New Results TLDR: A key role is reported of the CLEC18 family of C-type lectins in the progression of clear cell renal cell carcinoma (ccRCC) and the potential benefit of modulating CLEC18 expression in the renal tumor microenvironment is highlighted.

\bibitem[Julca et~al.(2023)Julca, Mutwil-Anderwald, Manoj, Khan, Lai, Yang, Beh, Dziekan, Lim, Lim, Low, Lam, Tjia, Mu, Tan, Nuc, Choo, Khew, Shining, Kam, Tam, Bozdech, Schmidt, Usadel, Kanagasundaram, Alseekh, Fernie, Li, and Mutwil]{julca_genomic_2023}
I.~Julca, D.~Mutwil-Anderwald, V.~Manoj, Z.~Khan, S.~K. Lai, L.~K. Yang, I.~T. Beh, J.~Dziekan, Y.~P. Lim, S.~K. Lim, Y.~W. Low, Y.~I. Lam, S.~Tjia, Y.~Mu, Q.~W. Tan, P.~Nuc, L.~M. Choo, G.~Khew, L.~Shining, A.~Kam, J.~P. Tam, Z.~Bozdech, M.~Schmidt, B.~Usadel, Y.~Kanagasundaram, S.~Alseekh, A.~Fernie, H.~Y. Li, and M.~Mutwil.
\newblock Genomic, transcriptomic, and metabolomic analysis of {Oldenlandia} corymbosa reveals the biosynthesis and mode of action of anti-cancer metabolites.
\newblock \emph{Journal of Integrative Plant Biology}, 65\penalty0 (6):\penalty0 1442--1466, 2023.
\newblock ISSN 1744-7909.
\newblock \doi{10.1111/jipb.13469}.
\newblock URL \url{https://onlinelibrary.wiley.com/doi/abs/10.1111/jipb.13469}.
\newblock \_eprint: https://onlinelibrary.wiley.com/doi/pdf/10.1111/jipb.13469 TLDR: It is revealed that ursolic acid causes mitotic catastrophe in cancer cells and three high-confidence protein binding targets by Cellular Thermal Shift Assay (CETSA) and reverse docking will allow us to further develop this valuable compound.

\bibitem[Jumper et~al.(2021)Jumper, Evans, Pritzel, Green, Figurnov, Ronneberger, Tunyasuvunakool, Bates, {\v{Z}}{\'\i}dek, Potapenko, et~al.]{jumper2021highly}
J.~Jumper, R.~Evans, A.~Pritzel, T.~Green, M.~Figurnov, O.~Ronneberger, K.~Tunyasuvunakool, R.~Bates, A.~{\v{Z}}{\'\i}dek, A.~Potapenko, et~al.
\newblock Highly accurate protein structure prediction with alphafold.
\newblock \emph{nature}, 596\penalty0 (7873):\penalty0 583--589, 2021.

\bibitem[Kabsch and Sander(1983)]{Kabsch1983}
W.~Kabsch and C.~Sander.
\newblock Dictionary of protein secondary structure: pattern recognition of hydrogen-bonded and geometrical features.
\newblock \emph{Biopolymers}, 22\penalty0 (12):\penalty0 2577--2637, 1983.

\bibitem[Kale et~al.(2022)Kale, Schulthess, Padmarasu, Boeven, Schacht, Himmelbach, Steuernagel, Wulff, Reif, Stein, and Mascher]{kale_catalogue_2022}
S.~M. Kale, A.~W. Schulthess, S.~Padmarasu, P.~H.~G. Boeven, J.~Schacht, A.~Himmelbach, B.~Steuernagel, B.~B.~H. Wulff, J.~C. Reif, N.~Stein, and M.~Mascher.
\newblock A catalogue of resistance gene homologs and a chromosome-scale reference sequence support resistance gene mapping in winter wheat.
\newblock \emph{Plant Biotechnology Journal}, 20\penalty0 (9):\penalty0 1730--1742, 2022.
\newblock ISSN 1467-7652.
\newblock \doi{10.1111/pbi.13843}.
\newblock URL \url{https://onlinelibrary.wiley.com/doi/abs/10.1111/pbi.13843}.
\newblock \_eprint: https://onlinelibrary.wiley.com/doi/pdf/10.1111/pbi.13843.

\bibitem[Kan et~al.(2024)Kan, Liao, Lan, Kong, Wang, Nie, Zou, An, and Wu]{kan_cytonuclear_2024}
S.~Kan, X.~Liao, L.~Lan, J.~Kong, J.~Wang, L.~Nie, J.~Zou, H.~An, and Z.~Wu.
\newblock Cytonuclear {Interactions} and {Subgenome} {Dominance} {Shape} the {Evolution} of {Organelle}-{Targeted} {Genes} in the {Brassica} {Triangle} of {U}.
\newblock \emph{Molecular Biology and Evolution}, 41\penalty0 (3):\penalty0 msae043, Mar. 2024.
\newblock ISSN 1537-1719.
\newblock \doi{10.1093/molbev/msae043}.
\newblock URL \url{https://doi.org/10.1093/molbev/msae043}.
\newblock TLDR: This study investigates the evolutionary pattern of organelle-targeted genes in Brassica carinata and 2 varieties of Brassica juncea at the whole-genome level, with particular focus on cytonuclear enzyme complexes and highlights an important role for subgenome dominance in allopolyploid genome evolution, even in genes whose function depends on separately inherited molecules.

\bibitem[Kang et~al.(2020)Kang, Pandey, Lee, Sim, Jeong, Choi, Jung, Ginzburg, Zhao, Won, Oh, Yu, Kim, Lee, Lee, Bashyal, Kim, Lee, Hawkins, Kim, Kim, Ahn, Rhee, and Sohng]{kang_genome-enabled_2020}
S.-H. Kang, R.~P. Pandey, C.-M. Lee, J.-S. Sim, J.-T. Jeong, B.-S. Choi, M.~Jung, D.~Ginzburg, K.~Zhao, S.~Y. Won, T.-J. Oh, Y.~Yu, N.-H. Kim, O.~R. Lee, T.-H. Lee, P.~Bashyal, T.-S. Kim, W.-H. Lee, C.~Hawkins, C.-K. Kim, J.~S. Kim, B.~O. Ahn, S.~Y. Rhee, and J.~K. Sohng.
\newblock Genome-enabled discovery of anthraquinone biosynthesis in {Senna} tora.
\newblock \emph{Nature Communications}, 11\penalty0 (1):\penalty0 5875, Nov. 2020.
\newblock ISSN 2041-1723.
\newblock \doi{10.1038/s41467-020-19681-1}.
\newblock URL \url{https://www.nature.com/articles/s41467-020-19681-1}.
\newblock Publisher: Nature Publishing Group.

\bibitem[Karczewski et~al.(2020)]{gnomAD}
K.~J. Karczewski et~al.
\newblock The mutational constraint spectrum quantified from variation in 141,456 humans.
\newblock \emph{Nature}, 581:\penalty0 434--443, 2020.

\bibitem[Kathail et~al.(2024)Kathail, Shuai, Chung, Ye, Loeb, and Ioannidis]{kathail2024current}
P.~Kathail, R.~W. Shuai, R.~Chung, C.~J. Ye, G.~B. Loeb, and N.~M. Ioannidis.
\newblock Current genomic deep learning models display decreased performance in cell type-specific accessible regions.
\newblock \emph{Genome Biology}, 25\penalty0 (1):\penalty0 202, 2024.

\bibitem[Katoh and Standley(2016)]{katoh_simple_2016}
K.~Katoh and D.~M. Standley.
\newblock A simple method to control over-alignment in the {MAFFT} multiple sequence alignment program.
\newblock \emph{Bioinformatics}, 32\penalty0 (13):\penalty0 1933--1942, July 2016.
\newblock ISSN 1367-4811, 1367-4803.
\newblock \doi{10.1093/bioinformatics/btw108}.
\newblock URL \url{https://academic.oup.com/bioinformatics/article/32/13/1933/1743504}.
\newblock TLDR: A new feature of the MAFFT multiple alignment program for suppressing over-alignment (aligning unrelated segments) by utilizing a variable scoring matrix for different pairs of sequences (or groups) in a single multiple sequence alignment, based on the global similarity of each pair.

\bibitem[Kersey et~al.(2018)]{EnsemblGenomes}
P.~J. Kersey et~al.
\newblock Ensembl genomes 2018: an integrated omics infrastructure for non-vertebrate species.
\newblock \emph{Nucleic Acids Research}, 46\penalty0 (D1):\penalty0 D802--D808, 2018.

\bibitem[Kimura(1980)]{Kimura1980}
M.~Kimura.
\newblock A simple method for estimating evolutionary rates of base substitutions through comparative studies of nucleotide sequences.
\newblock \emph{Journal of Molecular Evolution}, 16\penalty0 (2):\penalty0 111--120, 1980.

\bibitem[Kingma and Welling(2013)]{kingma2013auto}
D.~P. Kingma and M.~Welling.
\newblock Auto-encoding variational bayes.
\newblock \emph{arXiv preprint arXiv:1312.6114}, 2013.

\bibitem[Kipf and Welling(2016)]{kipf2016variational}
T.~N. Kipf and M.~Welling.
\newblock Variational graph auto-encoders.
\newblock \emph{arXiv preprint arXiv:1611.07308}, 2016.

\bibitem[Klimovskaia et~al.(2020)Klimovskaia, Lopez-Paz, Bottou, and Nickel]{klimovskaia2020poincare}
A.~Klimovskaia, D.~Lopez-Paz, L.~Bottou, and M.~Nickel.
\newblock Poincar{\'e} maps for analyzing complex hierarchies in single-cell data.
\newblock \emph{Nature communications}, 11\penalty0 (1):\penalty0 2966, 2020.

\bibitem[Kolodziejczyk et~al.(2015)Kolodziejczyk, Kim, Svensson, Marioni, and Teichmann]{Kolodziejczyk2015}
A.~A. Kolodziejczyk, J.~K. Kim, V.~Svensson, J.~C. Marioni, and S.~A. Teichmann.
\newblock The technology and biology of single-cell rna sequencing.
\newblock \emph{Molecular Cell}, 58\penalty0 (4):\penalty0 610--620, 2015.

\bibitem[Kolora et~al.(2021)Kolora, Owens, Vazquez, Stubbs, Chatla, Jainese, Seeto, McCrea, Sandel, Vianna, Maslenikov, Bachtrog, Orr, Love, and Sudmant]{kolora_origins_2021}
S.~R.~R. Kolora, G.~L. Owens, J.~M. Vazquez, A.~Stubbs, K.~Chatla, C.~Jainese, K.~Seeto, M.~McCrea, M.~W. Sandel, J.~A. Vianna, K.~Maslenikov, D.~Bachtrog, J.~W. Orr, M.~Love, and P.~H. Sudmant.
\newblock Origins and evolution of extreme life span in {Pacific} {Ocean} rockfishes.
\newblock \emph{Science}, 374\penalty0 (6569):\penalty0 842--847, Nov. 2021.
\newblock \doi{10.1126/science.abg5332}.
\newblock URL \url{https://www.science.org/doi/full/10.1126/science.abg5332}.
\newblock Publisher: American Association for the Advancement of Science TLDR: Genomes generated from rockfish species of different life spans elucidates the genetic determinants of aging and highlights the genetic innovations that underlie life history trait adaptations and, in turn, how they shape genomic diversity.

\bibitem[Koptagel et~al.(2022)Koptagel, Kviman, Melin, Safinianaini, and Lagergren]{koptagel_vaiphy_2022}
H.~Koptagel, O.~Kviman, H.~Melin, N.~Safinianaini, and J.~Lagergren.
\newblock {VaiPhy}: a {Variational} {Inference} {Based} {Algorithm} for {Phylogeny}.
\newblock In \emph{Advances in Neural Information Processing Systems}, Oct. 2022.
\newblock URL \url{https://openreview.net/forum?id=TIXwBZB3Jl6}.

\bibitem[Kouvelis et~al.(2023)Kouvelis, Kortsinoglou, and James]{kouvelis2023evolution}
V.~N. Kouvelis, A.~M. Kortsinoglou, and T.~Y. James.
\newblock The evolution of mitochondrial genomes in fungi.
\newblock \emph{Evolution of Fungi and Fungal-Like Organisms}, pages 65--90, 2023.

\bibitem[Kullback and Leibler(1951)]{Kullback1951}
S.~Kullback and R.~A. Leibler.
\newblock On information and sufficiency.
\newblock \emph{The Annals of Mathematical Statistics}, 22\penalty0 (1):\penalty0 79--86, 1951.

\bibitem[Kundu et~al.(2019)Kundu, Gor, Agrawal, and Babu]{kundu_gan-tree_2019}
J.~N. Kundu, M.~Gor, D.~Agrawal, and R.~V. Babu.
\newblock Gan-tree: An incrementally learned hierarchical generative framework for multi-modal data distributions.
\newblock In \emph{Proceedings of the IEEE/CVF International Conference on Computer Vision}, pages 8191--8200, 2019.

\bibitem[Kwon et~al.(2020)Kwon, Gori, Park, Potts, Swift, Wang, Stammnitz, Cannell, Baez-Ortega, Comte, Fox, Harmsen, Huxtable, Jones, Kreiss, Lawrence, Lazenby, Peck, Pye, Woods, Zimmermann, Wedge, Pemberton, Stratton, Hamede, and Murchison]{kwon_evolution_2020}
Y.~M. Kwon, K.~Gori, N.~Park, N.~Potts, K.~Swift, J.~Wang, M.~R. Stammnitz, N.~Cannell, A.~Baez-Ortega, S.~Comte, S.~Fox, C.~Harmsen, S.~Huxtable, M.~Jones, A.~Kreiss, C.~Lawrence, B.~Lazenby, S.~Peck, R.~Pye, G.~Woods, M.~Zimmermann, D.~C. Wedge, D.~Pemberton, M.~R. Stratton, R.~Hamede, and E.~P. Murchison.
\newblock Evolution and lineage dynamics of a transmissible cancer in {Tasmanian} devils.
\newblock \emph{PLOS Biology}, 18\penalty0 (11):\penalty0 e3000926, Nov. 2020.
\newblock ISSN 1545-7885.
\newblock \doi{10.1371/journal.pbio.3000926}.
\newblock URL \url{https://journals.plos.org/plosbiology/article?id=10.1371/journal.pbio.3000926}.
\newblock Publisher: Public Library of Science TLDR: Overall, DFT1 is a remarkably stable lineage whose genome illustrates how cancer cells adapt to diverse environments and persist in a parasitic niche.

\bibitem[La~Manno et~al.(2018{\natexlab{a}})La~Manno, Soldatov, Zeisel, Braun, Hochgerner, Petukhov, Lidschreiber, Kastriti, L{\"o}nnerberg, Furlan, et~al.]{la2018rna}
G.~La~Manno, R.~Soldatov, A.~Zeisel, E.~Braun, H.~Hochgerner, V.~Petukhov, K.~Lidschreiber, M.~E. Kastriti, P.~L{\"o}nnerberg, A.~Furlan, et~al.
\newblock Rna velocity of single cells.
\newblock \emph{Nature}, 560\penalty0 (7719):\penalty0 494--498, 2018{\natexlab{a}}.

\bibitem[La~Manno et~al.(2018{\natexlab{b}})La~Manno, Soldatov, Zeisel, et~al.]{LaManno2018}
G.~La~Manno, R.~Soldatov, A.~Zeisel, et~al.
\newblock Rna velocity of single cells.
\newblock \emph{Nature}, 560\penalty0 (7719):\penalty0 494--498, 2018{\natexlab{b}}.

\bibitem[Lange et~al.(2022)Lange, Bergen, Klein, Setty, Reuter, Bakhti, Lickert, Ansari, Schniering, Schiller, Pe'er, and Theis]{lange_cellrank_2022}
M.~Lange, V.~Bergen, M.~Klein, M.~Setty, B.~Reuter, M.~Bakhti, H.~Lickert, M.~Ansari, J.~Schniering, H.~B. Schiller, D.~Pe'er, and F.~J. Theis.
\newblock {CellRank} for directed single-cell fate mapping.
\newblock \emph{Nat Methods}, 19\penalty0 (2):\penalty0 159--170, Feb. 2022.
\newblock ISSN 1548-7105.
\newblock \doi{10.1038/s41592-021-01346-6}.
\newblock URL \url{https://www.nature.com/articles/s41592-021-01346-6}.
\newblock Publisher: Nature Publishing Group.

\bibitem[Lax and Keeling(2023)]{lax_molecular_2023}
G.~Lax and P.~J. Keeling.
\newblock Molecular phylogenetics of sessile {Dolium} sedentarium, a petalomonad euglenid.
\newblock \emph{Journal of Eukaryotic Microbiology}, 70\penalty0 (5):\penalty0 e12991, 2023.
\newblock ISSN 1550-7408.
\newblock \doi{10.1111/jeu.12991}.
\newblock URL \url{https://onlinelibrary.wiley.com/doi/abs/10.1111/jeu.12991}.
\newblock \_eprint: https://onlinelibrary.wiley.com/doi/pdf/10.1111/jeu.12991.

\bibitem[Leguia et~al.(2023)Leguia, Garcia-Glaessner, Muñoz-Saavedra, Juarez, Barrera, Calvo-Mac, Jara, Silva, Ploog, Amaro, Colchao-Claux, Johnson, Uhart, Nelson, and Lescano]{leguia_highly_2023}
M.~Leguia, A.~Garcia-Glaessner, B.~Muñoz-Saavedra, D.~Juarez, P.~Barrera, C.~Calvo-Mac, J.~Jara, W.~Silva, K.~Ploog, L.~Amaro, P.~Colchao-Claux, C.~K. Johnson, M.~M. Uhart, M.~I. Nelson, and J.~Lescano.
\newblock Highly pathogenic avian influenza {A} ({H5N1}) in marine mammals and seabirds in {Peru}.
\newblock \emph{Nature Communications}, 14\penalty0 (1):\penalty0 5489, Sept. 2023.
\newblock ISSN 2041-1723.
\newblock \doi{10.1038/s41467-023-41182-0}.
\newblock URL \url{https://www.nature.com/articles/s41467-023-41182-0}.
\newblock Publisher: Nature Publishing Group.

\bibitem[Leone and Powell(2020)]{leone2020metabolism}
R.~D. Leone and J.~D. Powell.
\newblock Metabolism of immune cells in cancer.
\newblock \emph{Nature reviews cancer}, 20\penalty0 (9):\penalty0 516--531, 2020.

\bibitem[Lewanski et~al.(2024)Lewanski, Grundler, and Bradburd]{lewanski2024era}
A.~L. Lewanski, M.~C. Grundler, and G.~S. Bradburd.
\newblock The era of the arg: An introduction to ancestral recombination graphs and their significance in empirical evolutionary genomics.
\newblock \emph{PLoS Genetics}, 20\penalty0 (1):\penalty0 e1011110, 2024.

\bibitem[Leypold and Speicher(2021)]{leypold2021evolutionary}
N.~A. Leypold and M.~R. Speicher.
\newblock Evolutionary conservation in noncoding genomic regions.
\newblock \emph{Trends in Genetics}, 37\penalty0 (10):\penalty0 903--918, 2021.

\bibitem[Li et~al.(2024{\natexlab{a}})Li, Zhang, Hong, Zhou, Vivone, Li, and Chanussot]{li2024casformer}
C.~Li, B.~Zhang, D.~Hong, J.~Zhou, G.~Vivone, S.~Li, and J.~Chanussot.
\newblock Casformer: Cascaded transformers for fusion-aware computational hyperspectral imaging.
\newblock \emph{Information Fusion}, 108:\penalty0 102408, 2024{\natexlab{a}}.

\bibitem[Li et~al.(2022{\natexlab{a}})Li, Velazquez, Ding, Hislop, Ebrahimkhani, and Bar-Joseph]{li2022trasig}
D.~Li, J.~J. Velazquez, J.~Ding, J.~Hislop, M.~R. Ebrahimkhani, and Z.~Bar-Joseph.
\newblock Trasig: inferring cell-cell interactions from pseudotime ordering of scrna-seq data.
\newblock \emph{Genome biology}, 23\penalty0 (1):\penalty0 73, 2022{\natexlab{a}}.

\bibitem[Li(2017)]{Li2017}
H.~Li.
\newblock Minimap2: pairwise alignment for nucleotide sequences.
\newblock \emph{Bioinformatics}, 34\penalty0 (18):\penalty0 3094--3100, 2017.
\newblock \doi{10.1093/bioinformatics/bty191}.

\bibitem[Li et~al.(2022{\natexlab{b}})Li, Li, Xiong, and Hoi]{li2022blip}
J.~Li, D.~Li, C.~Xiong, and S.~Hoi.
\newblock Blip: Bootstrapping language-image pre-training for unified vision-language understanding and generation.
\newblock \emph{arXiv preprint arXiv:2201.12086}, 2022{\natexlab{b}}.

\bibitem[Li et~al.(2023)Li, Li, Savarese, and Hoi]{li2023blip}
J.~Li, D.~Li, S.~Savarese, and S.~Hoi.
\newblock Blip-2: Bootstrapping language-image pre-training with frozen image encoders and large language models.
\newblock In \emph{International conference on machine learning}, pages 19730--19742. PMLR, 2023.

\bibitem[Li et~al.(2024{\natexlab{b}})Li, Pan, Yuan, and Shen]{li_tfvelo_2024}
J.~Li, X.~Pan, Y.~Yuan, and H.-B. Shen.
\newblock {TFvelo}: gene regulation inspired {RNA} velocity estimation.
\newblock \emph{Nat Commun}, 15\penalty0 (1):\penalty0 1387, Feb. 2024{\natexlab{b}}.
\newblock ISSN 2041-1723.
\newblock \doi{10.1038/s41467-024-45661-w}.
\newblock URL \url{https://www.nature.com/articles/s41467-024-45661-w}.
\newblock Publisher: Nature Publishing Group.

\bibitem[Li(2023)]{li_sctour_2023}
Q.~Li.
\newblock {scTour}: a deep learning architecture for robust inference and accurate prediction of cellular dynamics.
\newblock \emph{Genome Biology}, 24\penalty0 (1):\penalty0 149, June 2023.
\newblock ISSN 1474-760X.
\newblock \doi{10.1186/s13059-023-02988-9}.
\newblock URL \url{https://doi.org/10.1186/s13059-023-02988-9}.

\bibitem[Li et~al.(2018)Li, Vinyals, Dyer, Pascanu, and Battaglia]{li2018learning}
Y.~Li, O.~Vinyals, C.~Dyer, R.~Pascanu, and P.~Battaglia.
\newblock Learning deep generative models of graphs.
\newblock In \emph{International Conference on Machine Learning (ICML)}, 2018.

\bibitem[Liang et~al.(2020)Liang, Wang, Han, and Chen]{liang_latent_2020}
S.~Liang, F.~Wang, J.~Han, and K.~Chen.
\newblock Latent periodic process inference from single-cell {RNA}-seq data.
\newblock \emph{Nat Commun}, 11\penalty0 (1):\penalty0 1441, Mar. 2020.
\newblock ISSN 2041-1723.
\newblock \doi{10.1038/s41467-020-15295-9}.
\newblock URL \url{https://www.nature.com/articles/s41467-020-15295-9}.
\newblock Publisher: Nature Publishing Group.

\bibitem[Lin and Bar-Joseph(2019)]{lin_continuous-state_2019}
C.~Lin and Z.~Bar-Joseph.
\newblock Continuous-state {HMMs} for modeling time-series single-cell {RNA}-{Seq} data.
\newblock \emph{Bioinformatics}, 35\penalty0 (22):\penalty0 4707--4715, Nov. 2019.
\newblock ISSN 1367-4803.
\newblock \doi{10.1093/bioinformatics/btz296}.
\newblock URL \url{https://doi.org/10.1093/bioinformatics/btz296}.

\bibitem[Lopez et~al.(2018)Lopez, Regier, Cole, Jordan, and Yosef]{lopez2018deep}
R.~Lopez, J.~Regier, M.~B. Cole, M.~I. Jordan, and N.~Yosef.
\newblock Deep generative modeling for single-cell transcriptomics.
\newblock \emph{Nature methods}, 15\penalty0 (12):\penalty0 1053--1058, 2018.

\bibitem[Lucaci et~al.(2023)Lucaci, Zehr, Enard, Thornton, and Kosakovsky~Pond]{lucaci2023evolutionary}
A.~G. Lucaci, J.~D. Zehr, D.~Enard, J.~W. Thornton, and S.~L. Kosakovsky~Pond.
\newblock Evolutionary shortcuts via multinucleotide substitutions and their impact on natural selection analyses.
\newblock \emph{Molecular Biology and Evolution}, 40\penalty0 (7):\penalty0 msad150, 2023.

\bibitem[Ly-Trong et~al.(2024)Ly-Trong, Albert Matsen~IV, and Minh]{ly2024treeformer}
N.~Ly-Trong, F.~Albert Matsen~IV, and B.~Q. Minh.
\newblock Treeformer: A transformer-based tree rearrangement operation for phylogenetic reconstruction.
\newblock \emph{bioRxiv}, pages 2024--10, 2024.

\bibitem[Ma et~al.(2018)Ma, Yu, Fong, Ono, Sage, Demchak, Sharan, and Ideker]{ma2018using}
J.~Ma, M.~K. Yu, S.~Fong, K.~Ono, E.~Sage, B.~Demchak, R.~Sharan, and T.~Ideker.
\newblock Using deep learning to model the hierarchical structure and function of a cell.
\newblock \emph{Nature methods}, 15\penalty0 (4):\penalty0 290--298, 2018.

\bibitem[Ma et~al.(2023)Ma, Ma, Jiao, Liu, Li, Feng, Yang, et~al.]{ma2023multimodal}
M.~Ma, W.~Ma, L.~Jiao, X.~Liu, L.~Li, Z.~Feng, S.~Yang, et~al.
\newblock A multimodal hyper-fusion transformer for remote sensing image classification.
\newblock \emph{Information Fusion}, 96:\penalty0 66--79, 2023.

\bibitem[Macaulay and Voet(2017)]{Macaulay2017}
I.~C. Macaulay and T.~Voet.
\newblock Single-cell multiomics: multiple measurements from single cells.
\newblock \emph{Trends in Genetics}, 33\penalty0 (2):\penalty0 155--168, 2017.
\newblock \doi{10.1016/j.tig.2016.12.003}.

\bibitem[Maddison and Schulz(2018)]{Maddison2018}
D.~R. Maddison and K.-S. Schulz.
\newblock The tree of life.
\newblock \emph{Systematic Biology}, 67\penalty0 (5):\penalty0 719--729, 2018.

\bibitem[Maddison and Maddison(2007)]{Maddison2007}
W.~P. Maddison and D.~R. Maddison.
\newblock Mesquite: a modular system for evolutionary analysis.
\newblock \emph{Evolutionary Bioinformatics}, 3:\penalty0 47--50, 2007.

\bibitem[Maizels et~al.(2023)Maizels, Snell, and Briscoe]{maizels_deep_2023}
R.~J. Maizels, D.~M. Snell, and J.~Briscoe.
\newblock Deep dynamical modelling of developmental trajectories with temporal transcriptomics, July 2023.
\newblock URL \url{https://www.biorxiv.org/content/10.1101/2023.07.06.547989v1}.
\newblock Pages: 2023.07.06.547989 Section: New Results.

\bibitem[Man et~al.(2020)Man, Gallagher, and Bartlett]{man_structural_2020}
J.~Man, J.~P. Gallagher, and M.~Bartlett.
\newblock Structural evolution drives diversification of the large {LRR}-{RLK} gene family.
\newblock \emph{New Phytologist}, 226\penalty0 (5):\penalty0 1492--1505, 2020.
\newblock ISSN 1469-8137.
\newblock \doi{10.1111/nph.16455}.
\newblock URL \url{https://onlinelibrary.wiley.com/doi/abs/10.1111/nph.16455}.
\newblock \_eprint: https://onlinelibrary.wiley.com/doi/pdf/10.1111/nph.16455.

\bibitem[Manduchi et~al.(2023)Manduchi, Vandenhirtz, Ryser, and Vogt]{manduchi_tree_2023}
L.~Manduchi, M.~Vandenhirtz, A.~Ryser, and J.~E. Vogt.
\newblock Tree {Variational} {Autoencoders}.
\newblock In \emph{Advances in Neural Information Processing Systems}, July 2023.
\newblock URL \url{https://openreview.net/forum?id=8rt7bIDlY2#all}.

\bibitem[Marchler-Bauer et~al.(2011)Marchler-Bauer, Lu, Anderson, et~al.]{Marchler2011}
A.~Marchler-Bauer, S.~Lu, J.~B. Anderson, et~al.
\newblock Cdd: a conserved domain database for the functional annotation of proteins.
\newblock \emph{Nucleic Acids Research}, 39\penalty0 (suppl\_1):\penalty0 D225--D229, 2011.

\bibitem[Marco et~al.(2014)Marco, Karp, Guo, Robson, Hart, Trippa, and Yuan]{marco_bifurcation_2014}
E.~Marco, R.~L. Karp, G.~Guo, P.~Robson, A.~H. Hart, L.~Trippa, and G.-C. Yuan.
\newblock Bifurcation analysis of single-cell gene expression data reveals epigenetic landscape.
\newblock \emph{Proc Natl Acad Sci U S A}, 111\penalty0 (52):\penalty0 E5643--5650, Dec. 2014.
\newblock ISSN 1091-6490.
\newblock \doi{10.1073/pnas.1408993111}.

\bibitem[Mardis(2008)]{Mardis2008}
E.~R. Mardis.
\newblock Next-generation dna sequencing methods.
\newblock \emph{Annual Review of Genomics and Human Genetics}, 9:\penalty0 387--402, 2008.

\bibitem[Margelevičius(2024)]{margelevicius_gtalign_2024}
M.~Margelevičius.
\newblock {GTalign}: spatial index-driven protein structure alignment, superposition, and search.
\newblock \emph{Nature Communications}, 15\penalty0 (1):\penalty0 7305, Aug. 2024.
\newblock ISSN 2041-1723.
\newblock \doi{10.1038/s41467-024-51669-z}.
\newblock URL \url{https://www.nature.com/articles/s41467-024-51669-z}.
\newblock Publisher: Nature Publishing Group.

\bibitem[Marks et~al.(2011)Marks, Hopf, and Sander]{Marks2011}
D.~S. Marks, T.~A. Hopf, and C.~Sander.
\newblock Protein 3d structure computed from evolutionary sequence variation.
\newblock \emph{PloS One}, 6\penalty0 (12):\penalty0 e28766, 2011.

\bibitem[Marletaz et~al.(2023)Marletaz, Calle, Acemel, Paliou, Naranjo, Martinez, Cases, Sleight, Hirschberger, Marcet, Navon, Andrescavage, Skvortsova, Duckett, Gonzalez, Bogdanovic, Gibcus, Yang, Gallardo, and Gomez]{marletaz_little_2023}
F.~Marletaz, E.~Calle, R.~Acemel, C.~Paliou, S.~Naranjo, P.~Martinez, I.~Cases, V.~Sleight, C.~Hirschberger, M.~Marcet, D.~Navon, A.~Andrescavage, K.~Skvortsova, P.~Duckett, A.~Gonzalez, O.~Bogdanovic, J.~Gibcus, L.~Yang, L.~Gallardo, and J.~Gomez.
\newblock The little skate genome and the evolutionary emergence of wing-like fins.
\newblock \emph{Nature}, 616:\penalty0 1--9, 04 2023.
\newblock \doi{10.1038/s41586-023-05868-1}.

\bibitem[Mart{\'\i}nez-Jim{\'e}nez et~al.(2020)Mart{\'\i}nez-Jim{\'e}nez, Mui{\~n}os, Sent{\'\i}s, Deu-Pons, Reyes-Salazar, Arnedo-Pac, Mularoni, Pich, Bonet, Kranas, et~al.]{martinez2020compendium}
F.~Mart{\'\i}nez-Jim{\'e}nez, F.~Mui{\~n}os, I.~Sent{\'\i}s, J.~Deu-Pons, I.~Reyes-Salazar, C.~Arnedo-Pac, L.~Mularoni, O.~Pich, J.~Bonet, H.~Kranas, et~al.
\newblock A compendium of mutational cancer driver genes.
\newblock \emph{Nature Reviews Cancer}, 20\penalty0 (10):\penalty0 555--572, 2020.

\bibitem[Mathur et~al.(2024)Mathur, Mattoo, and Bar-Joseph]{mathur2024constrained}
S.~Mathur, H.~Mattoo, and Z.~Bar-Joseph.
\newblock Constrained pseudo-time ordering for clinical transcriptomics data.
\newblock \emph{IEEE/ACM Transactions on Computational Biology and Bioinformatics}, 2024.

\bibitem[Maurya et~al.(2023)Maurya, Cornejo, Lee, Kim, Hai, and Choudhary]{maurya_molecular_2023}
S.~Maurya, X.~Cornejo, C.~Lee, S.-Y. Kim, D.~V. Hai, and R.~K. Choudhary.
\newblock Molecular phylogenetic tools reveal the phytogeographic history of the genus \textit{{Capparis}} {L}. and suggest its reclassification.
\newblock \emph{Perspectives in Plant Ecology, Evolution and Systematics}, 58:\penalty0 125720, Mar. 2023.
\newblock ISSN 1433-8319.
\newblock \doi{10.1016/j.ppees.2023.125720}.
\newblock URL \url{https://www.sciencedirect.com/science/article/pii/S1433831923000045}.

\bibitem[McCormack et~al.(2013)McCormack, Hird, Zellmer, Carstens, and Brumfield]{Yang2012}
J.~E. McCormack, S.~M. Hird, A.~J. Zellmer, B.~C. Carstens, and R.~T. Brumfield.
\newblock Applications of next-generation sequencing to phylogeography and phylogenetics.
\newblock \emph{Molecular phylogenetics and evolution}, 66\penalty0 (2):\penalty0 526--538, 2013.

\bibitem[Mello et~al.(2022)Mello, Assunção, and Murai]{mello_top-down_2022}
D.~P. M.~d. Mello, R.~M. Assunção, and F.~Murai.
\newblock Top-{Down} {Deep} {Clustering} with {Multi}-{Generator} {GANs}.
\newblock \emph{Proceedings of the AAAI Conference on Artificial Intelligence}, 36\penalty0 (7):\penalty0 7770--7778, June 2022.
\newblock ISSN 2374-3468.
\newblock \doi{10.1609/aaai.v36i7.20745}.
\newblock URL \url{https://ojs.aaai.org/index.php/AAAI/article/view/20745}.
\newblock Number: 7.

\bibitem[Meng et~al.(2019)Meng, Jin, Wang, and Guo]{meng2019gene}
C.~Meng, S.~Jin, L.~Wang, and F.~Guo.
\newblock Gene ontology-based transfer learning for gene function prediction.
\newblock \emph{IEEE Access}, 7:\penalty0 54995--55007, 2019.

\bibitem[Michener and Sokal(1957)]{Michener1957}
C.~D. Michener and R.~R. Sokal.
\newblock A quantitative approach to a problem in classification.
\newblock \emph{Evolution}, 11\penalty0 (2):\penalty0 130--162, 1957.

\bibitem[Mimitou et~al.(2021)Mimitou, Cheng, Montalbano, Hao, Stoeckius, Legut, Roush, Herrera, Papalexi, Ouyang, et~al.]{Mimitou2021}
E.~P. Mimitou, A.~Cheng, A.~Montalbano, Y.~Hao, M.~Stoeckius, M.~Legut, T.~Roush, A.~Herrera, E.~Papalexi, Z.~Ouyang, et~al.
\newblock Multiplexed detection of proteins, transcriptomes, clonotypes and crispr perturbations in single cells.
\newblock \emph{Nature Methods}, 18\penalty0 (5):\penalty0 527--537, 2021.

\bibitem[Mimori and Hamada(2023)]{mimori_geophy_2023}
T.~Mimori and M.~Hamada.
\newblock {GeoPhy}: {Differentiable} {Phylogenetic} {Inference} via {Geometric} {Gradients} of {Tree} {Topologies}.
\newblock In \emph{Advances in Neural Information Processing Systems}, Nov. 2023.
\newblock URL \url{https://openreview.net/forum?id=54z8M7NTbJ&noteId=htP6Pvbgt7}.

\bibitem[Mirarab et~al.(2014)Mirarab, Reaz, Bayzid, Zimmermann, Swenson, and Warnow]{mirarab_astral_2014}
S.~Mirarab, R.~Reaz, M.~S. Bayzid, T.~Zimmermann, M.~S. Swenson, and T.~Warnow.
\newblock {ASTRAL}: genome-scale coalescent-based species tree estimation.
\newblock \emph{Bioinformatics}, 30\penalty0 (17):\penalty0 i541--i548, Sept. 2014.
\newblock ISSN 1367-4811, 1367-4803.
\newblock \doi{10.1093/bioinformatics/btu462}.
\newblock URL \url{https://academic.oup.com/bioinformatics/article/30/17/i541/200803}.

\bibitem[Moi et~al.(2022)Moi, Nishio, Li, Valansi, Langleib, Brukman, Flyak, Dessimoz, de~Sanctis, Tunyasuvunakool, Jumper, Graña, Romero, Aguilar, Jovine, and Podbilewicz]{moi_discovery_2022}
D.~Moi, S.~Nishio, X.~Li, C.~Valansi, M.~Langleib, N.~G. Brukman, K.~Flyak, C.~Dessimoz, D.~de~Sanctis, K.~Tunyasuvunakool, J.~Jumper, M.~Graña, H.~Romero, P.~S. Aguilar, L.~Jovine, and B.~Podbilewicz.
\newblock Discovery of archaeal fusexins homologous to eukaryotic {HAP2}/{GCS1} gamete fusion proteins.
\newblock \emph{Nature Communications}, 13\penalty0 (1):\penalty0 3880, July 2022.
\newblock ISSN 2041-1723.
\newblock \doi{10.1038/s41467-022-31564-1}.
\newblock URL \url{https://www.nature.com/articles/s41467-022-31564-1}.
\newblock Publisher: Nature Publishing Group.

\bibitem[Moi et~al.(2023)Moi, Bernard, Steinegger, Nevers, Langleib, and Dessimoz]{moi_structural_2023}
D.~Moi, C.~Bernard, M.~Steinegger, Y.~Nevers, M.~Langleib, and C.~Dessimoz.
\newblock Structural phylogenetics unravels the evolutionary diversification of communication systems in gram-positive bacteria and their viruses, Sept. 2023.
\newblock URL \url{https://www.biorxiv.org/content/10.1101/2023.09.19.558401v1}.
\newblock Pages: 2023.09.19.558401 Section: New Results TLDR: It is demonstrated that structure-informed phylogenies can outperform sequence-only ones not only for distantly related proteins but also, remarkably, for more closely related ones.

\bibitem[Mount(2004)]{Mount2004}
D.~Mount.
\newblock \emph{Bioinformatics: Sequence and Genome Analysis}.
\newblock Cold Spring Harbor Laboratory Press, 2004.

\bibitem[M{\"u}ller-Dott et~al.(2023)M{\"u}ller-Dott, Tsirvouli, Vazquez, Ramirez~Flores, Badia-i Mompel, Fallegger, T{\"u}rei, L{\ae}greid, and Saez-Rodriguez]{muller2023expanding}
S.~M{\"u}ller-Dott, E.~Tsirvouli, M.~Vazquez, R.~O. Ramirez~Flores, P.~Badia-i Mompel, R.~Fallegger, D.~T{\"u}rei, A.~L{\ae}greid, and J.~Saez-Rodriguez.
\newblock Expanding the coverage of regulons from high-confidence prior knowledge for accurate estimation of transcription factor activities.
\newblock \emph{Nucleic acids research}, 51\penalty0 (20):\penalty0 10934--10949, 2023.

\bibitem[Murzin et~al.(1995)Murzin, Brenner, Hubbard, and Chothia]{Murzin1995}
A.~G. Murzin, S.~E. Brenner, T.~J. Hubbard, and C.~Chothia.
\newblock Scop: a structural classification of proteins database for the investigation of sequences and structures.
\newblock \emph{Journal of Molecular Biology}, 247\penalty0 (4):\penalty0 536--540, 1995.

\bibitem[Nagalakshmi et~al.(2008)Nagalakshmi, Wang, Waern, Shou, Raha, Gerstein, and Snyder]{nagalakshmi2008transcriptional}
U.~Nagalakshmi, Z.~Wang, K.~Waern, C.~Shou, D.~Raha, M.~Gerstein, and M.~Snyder.
\newblock The transcriptional landscape of the yeast genome defined by rna sequencing.
\newblock \emph{Science}, 320\penalty0 (5881):\penalty0 1344--1349, 2008.
\newblock \doi{10.1126/science.1158441}.

\bibitem[Needleman and Wunsch(1970)]{needleman_general_1970}
S.~B. Needleman and C.~D. Wunsch.
\newblock A general method applicable to the search for similarities in the amino acid sequence of two proteins.
\newblock \emph{Journal of Molecular Biology}, 48\penalty0 (3):\penalty0 443--453, 1970.
\newblock \doi{10.1016/0022-2836(70)90057-4}.

\bibitem[Nei(1987)]{Nei1987}
M.~Nei.
\newblock \emph{Molecular Evolutionary Genetics}.
\newblock Columbia University Press, 1987.

\bibitem[Nesterenko et~al.(2022)Nesterenko, Boussau, and Jacob]{nesterenko_phyloformer_2022}
L.~Nesterenko, B.~Boussau, and L.~Jacob.
\newblock Phyloformer: towards fast and accurate phylogeny estimation with self-attention networks, June 2022.
\newblock URL \url{https://www.biorxiv.org/content/10.1101/2022.06.24.496975v1}.
\newblock Pages: 2022.06.24.496975 Section: New Results TLDR: This work presents a radically different approach with a transformer-based network architecture that, given a multiple sequence alignment, predicts all the pairwise evolutionary distances between the sequences, which in turn allow us to accurately reconstruct the tree topology with standard distance-based algorithms.

\bibitem[Network(2013)]{TCGA}
C.~G. A.~R. Network.
\newblock The cancer genome atlas pan-cancer analysis project.
\newblock \emph{Nature Genetics}, 45\penalty0 (10):\penalty0 1113--1120, 2013.

\bibitem[Nguyen et~al.(2015)Nguyen, Schmidt, Von~Haeseler, and Minh]{Nguyen2015}
L.-T. Nguyen, H.~A. Schmidt, A.~Von~Haeseler, and B.~Q. Minh.
\newblock Iq-tree: A fast and effective stochastic algorithm for estimating maximum-likelihood phylogenies.
\newblock \emph{Molecular Biology and Evolution}, 32\penalty0 (1):\penalty0 268--274, 2015.

\bibitem[Notredame(2007)]{Notredame2007}
C.~Notredame.
\newblock Recent evolutions of multiple sequence alignment algorithms.
\newblock \emph{PLoS Computational Biology}, 3\penalty0 (8):\penalty0 e123, 2007.

\bibitem[of~Life~(EOL)(2024)]{EOL}
E.~of~Life~(EOL).
\newblock Encyclopedia of life (eol) dataset, 2024.
\newblock URL \url{https://eol.org/}.
\newblock Accessed: 2024-09-17.

\bibitem[Ogilvie et~al.(2017)Ogilvie, Bouckaert, and Drummond]{ogilvie_starbeast2_2017}
H.~A. Ogilvie, R.~R. Bouckaert, and A.~J. Drummond.
\newblock {StarBEAST2} {Brings} {Faster} {Species} {Tree} {Inference} and {Accurate} {Estimates} of {Substitution} {Rates}.
\newblock \emph{Molecular Biology and Evolution}, 34\penalty0 (8):\penalty0 2101--2114, Aug. 2017.
\newblock ISSN 0737-4038, 1537-1719.
\newblock \doi{10.1093/molbev/msx126}.
\newblock URL \url{https://academic.oup.com/mbe/article/34/8/2101/3738283}.

\bibitem[Ozsolak and Milos(2011)]{ozsolak2011rna}
F.~Ozsolak and P.~M. Milos.
\newblock Rna sequencing: advances, challenges and opportunities.
\newblock \emph{Nature reviews genetics}, 12\penalty0 (2):\penalty0 87--98, 2011.

\bibitem[Papadopoulos et~al.(2019)Papadopoulos, Gonzalo, and Söding]{papadopoulos_prosstt_2019}
N.~Papadopoulos, P.~R. Gonzalo, and J.~Söding.
\newblock {PROSSTT}: probabilistic simulation of single-cell {RNA}-seq data for complex differentiation processes.
\newblock \emph{Bioinformatics}, 35\penalty0 (18):\penalty0 3517--3519, Sept. 2019.
\newblock ISSN 1367-4803, 1367-4811.
\newblock \doi{10.1093/bioinformatics/btz078}.
\newblock URL \url{https://academic.oup.com/bioinformatics/article/35/18/3517/5305637}.

\bibitem[Papili~Gao et~al.(2020)Papili~Gao, Hartmann, Fang, and Gunawan]{papili_gao_calista_2020}
N.~Papili~Gao, T.~Hartmann, T.~Fang, and R.~Gunawan.
\newblock {CALISTA}: {Clustering} and {LINEAGE} {Inference} in {Single}-{Cell} {Transcriptional} {Analysis}.
\newblock \emph{Front. Bioeng. Biotechnol.}, 8, Feb. 2020.
\newblock ISSN 2296-4185.
\newblock \doi{10.3389/fbioe.2020.00018}.
\newblock URL \url{https://www.frontiersin.org/journals/bioengineering-and-biotechnology/articles/10.3389/fbioe.2020.00018/full}.
\newblock Publisher: Frontiers.

\bibitem[Park et~al.(2023)Park, Ivanovic, Chu, Shen, and Warnow]{park_upp2_2023}
M.~Park, S.~Ivanovic, G.~Chu, C.~Shen, and T.~Warnow.
\newblock {UPP2}: fast and accurate alignment of datasets with fragmentary sequences.
\newblock \emph{Bioinformatics}, 39\penalty0 (1):\penalty0 btad007, Jan. 2023.
\newblock ISSN 1367-4811.
\newblock \doi{10.1093/bioinformatics/btad007}.
\newblock URL \url{https://academic.oup.com/bioinformatics/article/doi/10.1093/bioinformatics/btad007/6982552}.
\newblock TLDR: UPP2 is presented, a direct improvement on UPP that produces more accurate alignments compared to leading MSA methods on datasets exhibiting substantial sequence length heterogeneity and is among the most accurate otherwise.

\bibitem[Peng et~al.(2023)Peng, He, Peng, Li, and Zhang]{peng2023stgnnks}
L.~Peng, X.~He, X.~Peng, Z.~Li, and L.~Zhang.
\newblock Stgnnks: identifying cell types in spatial transcriptomics data based on graph neural network, denoising auto-encoder, and k-sums clustering.
\newblock \emph{Computers in Biology and Medicine}, 166:\penalty0 107440, 2023.

\bibitem[Perez et~al.(2022)Perez, Gordon, Subramaniam, Kim, Hartoularos, Targ, Sun, Ogorodnikov, Bueno, Lu, et~al.]{perez2022single}
R.~K. Perez, M.~G. Gordon, M.~Subramaniam, M.~C. Kim, G.~C. Hartoularos, S.~Targ, Y.~Sun, A.~Ogorodnikov, R.~Bueno, A.~Lu, et~al.
\newblock Single-cell rna-seq reveals cell type--specific molecular and genetic associations to lupus.
\newblock \emph{Science}, 376\penalty0 (6589):\penalty0 eabf1970, 2022.

\bibitem[Plass et~al.(2018)Plass, Solana, Wolf, et~al.]{Plass2018}
M.~Plass, J.~Solana, F.~A. Wolf, et~al.
\newblock Cell type atlas and lineage tree of a whole complex animal by single-cell transcriptomics.
\newblock \emph{Science}, 360\penalty0 (6391):\penalty0 eaaq1723, 2018.

\bibitem[Poulin et~al.(2020)Poulin, Gaertner, Moreno-Ramos, and Awatramani]{poulin2020classification}
J.-F. Poulin, Z.~Gaertner, O.~A. Moreno-Ramos, and R.~Awatramani.
\newblock Classification of midbrain dopamine neurons using single-cell gene expression profiling approaches.
\newblock \emph{Trends in neurosciences}, 43\penalty0 (3):\penalty0 155--169, 2020.

\bibitem[Qian et~al.(2023)Qian, Huang, Xu, Shu, and Ding]{qian2023survey}
W.~Qian, J.~Huang, F.~Xu, W.~Shu, and W.~Ding.
\newblock A survey on multi-label feature selection from perspectives of label fusion.
\newblock \emph{Information Fusion}, 100:\penalty0 101948, 2023.

\bibitem[Qiu et~al.(2017{\natexlab{a}})Qiu, Hill, Packer, et~al.]{Qiu2017}
X.~Qiu, A.~Hill, J.~Packer, et~al.
\newblock Single-cell mrna quantification and differential analysis with census.
\newblock \emph{Nature Methods}, 14\penalty0 (3):\penalty0 309--315, 2017{\natexlab{a}}.

\bibitem[Qiu et~al.(2017{\natexlab{b}})Qiu, Mao, Tang, Wang, Chawla, Pliner, and Trapnell]{qiu_reversed_2017}
X.~Qiu, Q.~Mao, Y.~Tang, L.~Wang, R.~Chawla, H.~A. Pliner, and C.~Trapnell.
\newblock Reversed graph embedding resolves complex single-cell trajectories.
\newblock \emph{Nat Methods}, 14\penalty0 (10):\penalty0 979--982, Oct. 2017{\natexlab{b}}.
\newblock ISSN 1548-7105.
\newblock \doi{10.1038/nmeth.4402}.
\newblock URL \url{https://www.nature.com/articles/nmeth.4402}.

\bibitem[Qu et~al.(2024)Qu, Cheng, Sefik, Stanley~III, Landa, Strino, Platt, Garritano, Odell, Coifman, Flavell, Myung, and Kluger]{qu_gene_2024}
R.~Qu, X.~Cheng, E.~Sefik, J.~S. Stanley~III, B.~Landa, F.~Strino, S.~Platt, J.~Garritano, I.~D. Odell, R.~Coifman, R.~A. Flavell, P.~Myung, and Y.~Kluger.
\newblock Gene trajectory inference for single-cell data by optimal transport metrics.
\newblock \emph{Nat Biotechnol}, pages 1--11, Apr. 2024.
\newblock ISSN 1546-1696.
\newblock \doi{10.1038/s41587-024-02186-3}.
\newblock URL \url{https://www.nature.com/articles/s41587-024-02186-3}.
\newblock Publisher: Nature Publishing Group.

\bibitem[Radford et~al.(2021)Radford, Kim, Hallacy, Ramesh, Goh, Agarwal, Sastry, Askell, Mishkin, et~al.]{radford2021learning}
A.~Radford, J.~W. Kim, C.~Hallacy, A.~Ramesh, G.~Goh, S.~Agarwal, G.~Sastry, A.~Askell, P.~Mishkin, et~al.
\newblock Learning transferable visual models from natural language supervision.
\newblock In \emph{International Conference on Machine Learning}, pages 8748--8763. PMLR, 2021.

\bibitem[Rao et~al.(2021)Rao, Barkley, Franca, and Yanai]{rao2021deep}
A.~Rao, D.~Barkley, G.~S. Franca, and I.~Yanai.
\newblock Deep learning for spatially resolved data in single-cell omics.
\newblock \emph{Annual Review of Biomedical Data Science}, 4:\penalty0 123--142, 2021.

\bibitem[Regev et~al.(2017)Regev, Teichmann, et~al.]{HCAConsortium}
A.~Regev, S.~A. Teichmann, et~al.
\newblock The human cell atlas.
\newblock \emph{eLife}, 6:\penalty0 e27041, 2017.

\bibitem[Rhodes(2006)]{Rhodes2006Crystallography}
G.~Rhodes.
\newblock \emph{Crystallography Made Crystal Clear: A Guide for Users of Macromolecular Models}.
\newblock Academic Press, 2006.

\bibitem[Riba et~al.(2022)Riba, Oravecz, Durik, Jiménez, Alunni, Cerciat, Jung, Keime, Keyes, and Molina]{riba_cell_2022}
A.~Riba, A.~Oravecz, M.~Durik, S.~Jiménez, V.~Alunni, M.~Cerciat, M.~Jung, C.~Keime, W.~M. Keyes, and N.~Molina.
\newblock Cell cycle gene regulation dynamics revealed by {RNA} velocity and deep-learning.
\newblock \emph{Nat Commun}, 13\penalty0 (1):\penalty0 2865, May 2022.
\newblock ISSN 2041-1723.
\newblock \doi{10.1038/s41467-022-30545-8}.
\newblock URL \url{https://www.nature.com/articles/s41467-022-30545-8}.
\newblock Publisher: Nature Publishing Group.

\bibitem[Rieppel(1988)]{Rieppel1988}
O.~Rieppel.
\newblock Fundamentals of comparative biology.
\newblock \emph{The Quarterly Review of Biology}, 63\penalty0 (3):\penalty0 319--320, 1988.
\newblock \doi{10.1086/416708}.

\bibitem[Ronquist and Huelsenbeck(2003)]{Ronquist2003}
F.~Ronquist and J.~P. Huelsenbeck.
\newblock Mrbayes 3: Bayesian phylogenetic inference under mixed models.
\newblock \emph{Bioinformatics}, 19\penalty0 (12):\penalty0 1572--1574, 2003.

\bibitem[Ronquist et~al.(2012)Ronquist, Teslenko, van~der Mark, Ayres, Darling, Höhna, Larget, Liu, Suchard, and Huelsenbeck]{Ronquist2012}
F.~Ronquist, M.~Teslenko, P.~van~der Mark, D.~L. Ayres, A.~Darling, S.~Höhna, B.~Larget, L.~Liu, M.~A. Suchard, and J.~P. Huelsenbeck.
\newblock Mrbayes 3.2: efficient bayesian phylogenetic inference and model choice across a large model space.
\newblock \emph{Systematic Biology}, 61\penalty0 (3):\penalty0 539--542, 2012.

\bibitem[Rzhetsky and Nei(1992)]{Rzhetsky1992}
A.~Rzhetsky and M.~Nei.
\newblock The minimum evolution approach to distance-based phylogenetic analysis: Theory and practice.
\newblock \emph{Molecular Biology and Evolution}, 9\penalty0 (5):\penalty0 945--967, 1992.

\bibitem[Saitou and Nei(1987)]{Saitou1987}
N.~Saitou and M.~Nei.
\newblock The neighbor-joining method: a new method for reconstructing phylogenetic trees.
\newblock \emph{Molecular Biology and Evolution}, 4\penalty0 (4):\penalty0 406--425, 1987.

\bibitem[Sali and Blundell(1994)]{Sali1994}
A.~Sali and T.~L. Blundell.
\newblock Comparative protein modeling by satisfaction of spatial restraints.
\newblock \emph{Journal of Molecular Biology}, 234\penalty0 (3):\penalty0 779--815, 1994.

\bibitem[Sanger et~al.(1977)Sanger, Nicklen, and Coulson]{Sanger1977}
F.~Sanger, S.~Nicklen, and A.~R. Coulson.
\newblock Dna sequencing with chain-terminating inhibitors.
\newblock \emph{Proceedings of the National Academy of Sciences}, 74\penalty0 (12):\penalty0 5463--5467, 1977.

\bibitem[Schiebinger et~al.(2019)Schiebinger, Shu, Tabaka, Cleary, Subramanian, Solomon, Gould, Liu, Lin, Berube, et~al.]{schiebinger2019optimal}
G.~Schiebinger, J.~Shu, M.~Tabaka, B.~Cleary, V.~Subramanian, A.~Solomon, J.~Gould, S.~Liu, S.~Lin, P.~Berube, et~al.
\newblock Optimal-transport analysis of single-cell gene expression identifies developmental trajectories in reprogramming.
\newblock \emph{Cell}, 176\penalty0 (4):\penalty0 928--943, 2019.

\bibitem[Schi{\o}tz et~al.(2024)Schi{\o}tz, Kaiser, Klumpe, Morado, Poege, Schneider, Beck, Klebl, Thompson, and Plitzko]{schiotz2024serial}
O.~H. Schi{\o}tz, C.~J. Kaiser, S.~Klumpe, D.~R. Morado, M.~Poege, J.~Schneider, F.~Beck, D.~P. Klebl, C.~Thompson, and J.~M. Plitzko.
\newblock Serial lift-out: sampling the molecular anatomy of whole organisms.
\newblock \emph{Nature Methods}, 21\penalty0 (9):\penalty0 1684--1692, 2024.

\bibitem[Schmidt et~al.(2023)Schmidt, Sashittal, and Raphael]{schmidt_zero-agnostic_2023}
H.~Schmidt, P.~Sashittal, and B.~J. Raphael.
\newblock A zero-agnostic model for copy number evolution in cancer.
\newblock \emph{PLOS Computational Biology}, 19\penalty0 (11):\penalty0 e1011590, Nov. 2023.
\newblock ISSN 1553-7358.
\newblock \doi{10.1371/journal.pcbi.1011590}.
\newblock URL \url{https://journals.plos.org/ploscompbiol/article?id=10.1371/journal.pcbi.1011590}.
\newblock Publisher: Public Library of Science TLDR: The zero-agnostic copy number transformation (ZCNT) model is introduced, a simplification of the CNT model that allows the amplification or deletion of regions with zero copies and an algorithm, Lazac, is developed for solving the large parsimony problem on copy number profiles.

\bibitem[Semple and Steel(2003)]{Semple2003}
C.~Semple and M.~Steel.
\newblock \emph{Phylogenetics}.
\newblock Oxford University Press, 2003.

\bibitem[Servellita et~al.(2023)Servellita, Sotomayor~Gonzalez, Lamson, Foresythe, Huh, Bazinet, Bergman, Bull, Garcia, Goodrich, Lovett, Parker, Radune, Hatada, Pan, Rizzo, Bertumen, Morales, Oluniyi, Nguyen, Tan, Stryke, Jaber, Leslie, Lyons, Hedman, Parashar, Sullivan, Wroblewski, Oberste, Tate, Baker, Sugerman, Potts, Lu, Chhabra, Ingram, Shiau, Britt, Gutierrez~Sanchez, Ciric, Rostad, Vinjé, Kirking, Wadford, Raborn, St.~George, and Chiu]{servellita_adeno-associated_2023}
V.~Servellita, A.~Sotomayor~Gonzalez, D.~M. Lamson, A.~Foresythe, H.~J. Huh, A.~L. Bazinet, N.~H. Bergman, R.~L. Bull, K.~Y. Garcia, J.~S. Goodrich, S.~P. Lovett, K.~Parker, D.~Radune, A.~Hatada, C.-Y. Pan, K.~Rizzo, J.~B. Bertumen, C.~Morales, P.~E. Oluniyi, J.~Nguyen, J.~Tan, D.~Stryke, R.~Jaber, M.~T. Leslie, Z.~Lyons, H.~D. Hedman, U.~Parashar, M.~Sullivan, K.~Wroblewski, M.~S. Oberste, J.~E. Tate, J.~M. Baker, D.~Sugerman, C.~Potts, X.~Lu, P.~Chhabra, L.~A. Ingram, H.~Shiau, W.~Britt, L.~H. Gutierrez~Sanchez, C.~Ciric, C.~A. Rostad, J.~Vinjé, H.~L. Kirking, D.~A. Wadford, R.~T. Raborn, K.~St.~George, and C.~Y. Chiu.
\newblock Adeno-associated virus type 2 in {US} children with acute severe hepatitis.
\newblock \emph{Nature}, 617\penalty0 (7961):\penalty0 574--580, May 2023.
\newblock ISSN 1476-4687.
\newblock \doi{10.1038/s41586-023-05949-1}.
\newblock URL \url{https://www.nature.com/articles/s41586-023-05949-1}.
\newblock Publisher: Nature Publishing Group.

\bibitem[Sharma et~al.(2017)Sharma, Li, Sengupta, Prabhakar, and {Jayadeva}]{sharma_forks_2017}
M.~Sharma, H.~Li, D.~Sengupta, S.~Prabhakar, and {Jayadeva}.
\newblock {FORKS}: {Finding} {Orderings} {Robustly} using k-means and {Steiner} trees, June 2017.
\newblock URL \url{https://www.biorxiv.org/content/10.1101/132811v3}.

\bibitem[Sharp(1985)]{Sharp1985}
P.~A. Sharp.
\newblock On the origin of rna splicing and introns.
\newblock \emph{Cell}, 42\penalty0 (2):\penalty0 397--400, 1985.
\newblock \doi{10.1016/S0092-8674(85)80151-5}.

\bibitem[Shatsky et~al.(2002)Shatsky, Nussinov, and Wolfson]{shatsky_multiprot_2002}
M.~Shatsky, R.~Nussinov, and H.~J. Wolfson.
\newblock {MultiProt} — {A} {Multiple} {Protein} {Structural} {Alignment} {Algorithm}.
\newblock In R.~Guigó and D.~Gusfield, editors, \emph{Algorithms in {Bioinformatics}}, pages 235--250, Berlin, Heidelberg, 2002. Springer.
\newblock ISBN 978-3-540-45784-8.
\newblock \doi{10.1007/3-540-45784-4_18}.
\newblock TLDR: A fully automated highly efficient technique which detects the multiple structural alignments of protein structures and presents new multiple structural alignment results of protein families from the All beta proteins class in the SCOP classification.

\bibitem[Shen et~al.(2020)Shen, Li, Hittinger, Chen, and Rokas]{shen2020investigation}
X.-X. Shen, Y.~Li, C.~T. Hittinger, X.-x. Chen, and A.~Rokas.
\newblock An investigation of irreproducibility in maximum likelihood phylogenetic inference.
\newblock \emph{Nature communications}, 11\penalty0 (1):\penalty0 6096, 2020.

\bibitem[Shen et~al.(2024)Shen, Song, Tan, Li, Lu, and Zhuang]{shen2024hugginggpt}
Y.~Shen, K.~Song, X.~Tan, D.~Li, W.~Lu, and Y.~Zhuang.
\newblock Hugginggpt: Solving ai tasks with chatgpt and its friends in hugging face.
\newblock \emph{Advances in Neural Information Processing Systems}, 36, 2024.

\bibitem[Sherry et~al.(2001)]{dbSNP}
S.~T. Sherry et~al.
\newblock dbsnp: the ncbi database of genetic variation.
\newblock \emph{Nucleic Acids Research}, 29\penalty0 (1):\penalty0 308--311, 2001.

\bibitem[Shulman-Peleg et~al.(2004)Shulman-Peleg, Nussinov, and Wolfson]{shulman-peleg_recognition_2004}
A.~Shulman-Peleg, R.~Nussinov, and H.~J. Wolfson.
\newblock Recognition of {Functional} {Sites} in {Protein} {Structures}.
\newblock \emph{Journal of Molecular Biology}, 339\penalty0 (3):\penalty0 607--633, June 2004.
\newblock ISSN 0022-2836.
\newblock \doi{10.1016/j.jmb.2004.04.012}.
\newblock URL \url{https://www.sciencedirect.com/science/article/pii/S0022283604004139}.
\newblock TLDR: A novel method is described, SiteEngine, that assumes no sequence or fold similarities and is able to recognize proteins that have similar binding sites and may perform similar functions, and which may aid in assigning a function and in classification of binding patterns.

\bibitem[Sievers and Higgins(2014)]{sievers_clustal_2014}
F.~Sievers and D.~G. Higgins.
\newblock Clustal {Omega}, accurate alignment of very large numbers of sequences.
\newblock \emph{Methods in molecular biology}, 1079:\penalty0 105--116, 2014.
\newblock Publisher: Springer.

\bibitem[Smith and Hahn(2023)]{smith_phylogenetic_2023}
M.~L. Smith and M.~W. Hahn.
\newblock Phylogenetic inference using generative adversarial networks.
\newblock \emph{Bioinformatics}, 39\penalty0 (9):\penalty0 btad543, Sept. 2023.
\newblock ISSN 1367-4811.
\newblock \doi{10.1093/bioinformatics/btad543}.
\newblock URL \url{https://doi.org/10.1093/bioinformatics/btad543}.
\newblock TLDR: PhyloGAN is developed, a GAN that infers phylogenetic relationships among species and uses an evolutionary model as the generator, and infers a phylogenetic tree either considering or ignoring gene tree heterogeneity.

\bibitem[Smith and Waterman(1981{\natexlab{a}})]{Smith1981}
T.~F. Smith and M.~S. Waterman.
\newblock Identification of common molecular subsequences.
\newblock \emph{Journal of Molecular Biology}, 147\penalty0 (1):\penalty0 195--197, 1981{\natexlab{a}}.

\bibitem[Smith and Waterman(1981{\natexlab{b}})]{smith_identification_1981}
T.~F. Smith and M.~S. Waterman.
\newblock Identification of common molecular subsequences.
\newblock \emph{Journal of Molecular Biology}, 147\penalty0 (1):\penalty0 195--197, 1981{\natexlab{b}}.
\newblock \doi{10.1016/0022-2836(81)90087-5}.

\bibitem[Spain et~al.(2023)Spain, Coulton, Lobon, Rowan, Schnidrig, Shepherd, Shum, Byrne, Goicoechea, Piperni, et~al.]{spain2023late}
L.~Spain, A.~Coulton, I.~Lobon, A.~Rowan, D.~Schnidrig, S.~T. Shepherd, B.~Shum, F.~Byrne, M.~Goicoechea, E.~Piperni, et~al.
\newblock Late-stage metastatic melanoma emerges through a diversity of evolutionary pathways.
\newblock \emph{Cancer discovery}, 13\penalty0 (6):\penalty0 1364--1385, 2023.

\bibitem[Stahl et~al.(2023)Stahl, Graziadei, Dau, Brock, and Rappsilber]{stahl2023protein}
K.~Stahl, A.~Graziadei, T.~Dau, O.~Brock, and J.~Rappsilber.
\newblock Protein structure prediction with in-cell photo-crosslinking mass spectrometry and deep learning.
\newblock \emph{Nature Biotechnology}, 41\penalty0 (12):\penalty0 1810--1819, 2023.

\bibitem[Stamatakis(2014)]{Stamatakis2014}
A.~Stamatakis.
\newblock Raxml version 8: a tool for phylogenetic analysis and post-analysis of large phylogenies.
\newblock \emph{Bioinformatics}, 30\penalty0 (9):\penalty0 1312--1313, 2014.

\bibitem[Stassen et~al.(2021)Stassen, Yip, Wong, Ho, and Tsia]{stassen_generalized_2021}
S.~V. Stassen, G.~G.~K. Yip, K.~K.~Y. Wong, J.~W.~K. Ho, and K.~K. Tsia.
\newblock Generalized and scalable trajectory inference in single-cell omics data with {VIA}.
\newblock \emph{Nat Commun}, 12\penalty0 (1):\penalty0 5528, Sept. 2021.
\newblock ISSN 2041-1723.
\newblock \doi{10.1038/s41467-021-25773-3}.
\newblock URL \url{https://www.nature.com/articles/s41467-021-25773-3}.
\newblock Publisher: Nature Publishing Group.

\bibitem[Stefan Van~Dongen and Winnepenninckx(1996)]{stefan1996multiple}
T.~Stefan Van~Dongen and B.~Winnepenninckx.
\newblock Multiple upgma and neighbor-joining trees and the performance of some computer packages.
\newblock \emph{Mol. Biol. Evol}, 13\penalty0 (2):\penalty0 309--313, 1996.

\bibitem[Stevens et~al.(2024)Stevens, Wu, Thompson, Campolongo, Song, Carlyn, Dong, Dahdul, Stewart, Berger-Wolf, Chao, and Su]{Stevens2024BioCLIP}
S.~Stevens, J.~Wu, M.~J. Thompson, E.~G. Campolongo, C.~H. Song, D.~E. Carlyn, L.~Dong, W.~M. Dahdul, C.~Stewart, T.~Berger-Wolf, W.-L. Chao, and Y.~Su.
\newblock Bioclip: A vision foundation model for the tree of life.
\newblock \emph{arXiv preprint arXiv:2311.18803}, 2024.
\newblock URL \url{https://imageomics.github.io/bioclip}.
\newblock Accessed: 2024-09-17.

\bibitem[Stoeckius et~al.(2017)Stoeckius, Hafemeister, Stephenson, Houck-Loomis, Chattopadhyay, Swerdlow, Satija, and Smibert]{Stoeckius2017}
M.~Stoeckius, C.~Hafemeister, W.~Stephenson, B.~Houck-Loomis, P.~K. Chattopadhyay, H.~Swerdlow, R.~Satija, and P.~Smibert.
\newblock Simultaneous epitope and transcriptome measurement in single cells.
\newblock \emph{Nature Methods}, 14\penalty0 (9):\penalty0 865--868, 2017.

\bibitem[Street et~al.(2018)Street, Risso, Fletcher, Das, Ngai, Yosef, Purdom, and Dudoit]{street_slingshot_2018}
K.~Street, D.~Risso, R.~B. Fletcher, D.~Das, J.~Ngai, N.~Yosef, E.~Purdom, and S.~Dudoit.
\newblock Slingshot: cell lineage and pseudotime inference for single-cell transcriptomics.
\newblock \emph{BMC Genomics}, 19\penalty0 (1):\penalty0 477, June 2018.
\newblock ISSN 1471-2164.
\newblock \doi{10.1186/s12864-018-4772-0}.
\newblock URL \url{https://doi.org/10.1186/s12864-018-4772-0}.

\bibitem[Stuart and Satija(2019)]{stuart2019integrative}
T.~Stuart and R.~Satija.
\newblock Integrative single-cell analysis.
\newblock \emph{Nature reviews genetics}, 20\penalty0 (5):\penalty0 257--272, 2019.

\bibitem[Stuart et~al.(2019{\natexlab{a}})Stuart, Butler, Hoffman, Hafemeister, Papalexi, Mauck, Hao, Stoeckius, Smibert, and Satija]{stuart2019comprehensive}
T.~Stuart, A.~Butler, P.~Hoffman, C.~Hafemeister, E.~Papalexi, W.~M. Mauck, Y.~Hao, M.~Stoeckius, P.~Smibert, and R.~Satija.
\newblock Comprehensive integration of single-cell data.
\newblock \emph{cell}, 177\penalty0 (7):\penalty0 1888--1902, 2019{\natexlab{a}}.

\bibitem[Stuart et~al.(2019{\natexlab{b}})Stuart, Butler, Hoffman, Hafemeister, Papalexi, Mauck, Hao, Stoeckius, Smibert, and Satija]{Stuart2019}
T.~Stuart, A.~Butler, P.~Hoffman, C.~Hafemeister, E.~Papalexi, W.~M.~I. Mauck, Y.~Hao, M.~Stoeckius, P.~Smibert, and R.~Satija.
\newblock Comprehensive integration of single-cell data.
\newblock \emph{Cell}, 177\penalty0 (7):\penalty0 1888--1902.e21, 2019{\natexlab{b}}.

\bibitem[Subramanian et~al.(2020)Subramanian, Verma, Kumar, and et~al.]{subramanian2020multiomics}
I.~Subramanian, S.~Verma, S.~Kumar, and et~al.
\newblock Multi-omics data integration, interpretation, and its application.
\newblock \emph{Bioinformatics Reviews}, 36\penalty0 (9):\penalty0 2605--2614, 2020.
\newblock \doi{10.1093/bioinformatics/btaa005}.

\bibitem[Suchard and Redelings(2006)]{suchard_bali-phy_2006}
M.~A. Suchard and B.~D. Redelings.
\newblock {BAli}-{Phy}: simultaneous {Bayesian} inference of alignment and phylogeny.
\newblock \emph{Bioinformatics}, 22\penalty0 (16):\penalty0 2047--2048, 2006.
\newblock Publisher: Oxford University Press.

\bibitem[Suvorov et~al.(2020)Suvorov, Hochuli, and Schrider]{suvorov_accurate_2020}
A.~Suvorov, J.~Hochuli, and D.~R. Schrider.
\newblock Accurate {Inference} of {Tree} {Topologies} from {Multiple} {Sequence} {Alignments} {Using} {Deep} {Learning}.
\newblock \emph{Systematic Biology}, 69\penalty0 (2):\penalty0 221--233, Mar. 2020.
\newblock ISSN 1063-5157, 1076-836X.
\newblock \doi{10.1093/sysbio/syz060}.
\newblock URL \url{https://academic.oup.com/sysbio/article/69/2/221/5559282}.

\bibitem[Svensson et~al.(2018)Svensson, Vento-Tormo, and Teichmann]{svensson2018single}
V.~Svensson, R.~Vento-Tormo, and S.~A. Teichmann.
\newblock Exponential scaling of single-cell rna-seq in the past decade.
\newblock \emph{Nature Protocols}, 13\penalty0 (4):\penalty0 599--604, 2018.
\newblock \doi{10.1038/nprot.2017.149}.

\bibitem[Swofford et~al.(1996)Swofford, Olsen, Waddell, and Hillis]{Swofford1996}
D.~L. Swofford, G.~J. Olsen, P.~J. Waddell, and D.~M. Hillis.
\newblock \emph{Phylogenetic inference}, pages 407--514.
\newblock Sinauer Associates, 1996.

\bibitem[Sza{\l}ata et~al.(2024)Sza{\l}ata, Hrovatin, Becker, Tejada-Lapuerta, Cui, Wang, and Theis]{szalata2024transformers}
A.~Sza{\l}ata, K.~Hrovatin, S.~Becker, A.~Tejada-Lapuerta, H.~Cui, B.~Wang, and F.~J. Theis.
\newblock Transformers in single-cell omics: a review and new perspectives.
\newblock \emph{Nature Methods}, 21\penalty0 (8):\penalty0 1430--1443, 2024.

\bibitem[Szklarczyk et~al.(2019)]{STRING}
D.~Szklarczyk et~al.
\newblock String v11: protein--protein association networks with increased coverage, supporting functional discovery in genome-wide experimental datasets.
\newblock \emph{Nucleic Acids Research}, 47\penalty0 (D1):\penalty0 D607--D613, 2019.

\bibitem[Szöllősi et~al.(2020)Szöllősi, Davín, Tannier, Daubin, and Boussau]{Szollosi2020}
G.~J. Szöllősi, A.~A. Davín, E.~Tannier, V.~Daubin, and B.~Boussau.
\newblock Genome-scale phylogenetic analysis finds extensive gene transfer among fungi.
\newblock \emph{Nature Ecology \& Evolution}, 4\penalty0 (8):\penalty0 1160--1165, 2020.
\newblock \doi{10.1038/s41559-020-1241-9}.

\bibitem[Tamura et~al.(2023)Tamura, Ito, Uriu, Zahradnik, Kida, Anraku, Nasser, Shofa, Oda, Lytras, Nao, Itakura, Deguchi, Suzuki, Wang, Begum, Kita, Yajima, Sasaki, Sasaki-Tabata, Shimizu, Tsuda, Kosugi, Fujita, Pan, Sauter, Yoshimatsu, Suzuki, Asakura, Nagashima, Sadamasu, Yoshimura, Yamamoto, Nagamoto, Schreiber, Maenaka, Hashiguchi, Ikeda, Fukuhara, Saito, Tanaka, Matsuno, Takayama, and Sato]{tamura_virological_2023}
T.~Tamura, J.~Ito, K.~Uriu, J.~Zahradnik, I.~Kida, Y.~Anraku, H.~Nasser, M.~Shofa, Y.~Oda, S.~Lytras, N.~Nao, Y.~Itakura, S.~Deguchi, R.~Suzuki, L.~Wang, M.~M. Begum, S.~Kita, H.~Yajima, J.~Sasaki, K.~Sasaki-Tabata, R.~Shimizu, M.~Tsuda, Y.~Kosugi, S.~Fujita, L.~Pan, D.~Sauter, K.~Yoshimatsu, S.~Suzuki, H.~Asakura, M.~Nagashima, K.~Sadamasu, K.~Yoshimura, Y.~Yamamoto, T.~Nagamoto, G.~Schreiber, K.~Maenaka, T.~Hashiguchi, T.~Ikeda, T.~Fukuhara, A.~Saito, S.~Tanaka, K.~Matsuno, K.~Takayama, and K.~Sato.
\newblock Virological characteristics of the {SARS}-{CoV}-2 {XBB} variant derived from recombination of two {Omicron} subvariants.
\newblock \emph{Nature Communications}, 14\penalty0 (1):\penalty0 2800, May 2023.
\newblock ISSN 2041-1723.
\newblock \doi{10.1038/s41467-023-38435-3}.
\newblock URL \url{https://www.nature.com/articles/s41467-023-38435-3}.
\newblock Publisher: Nature Publishing Group.

\bibitem[Tan et~al.(2024)Tan, Tian, and Li]{tan2024glime}
Z.~Tan, Y.~Tian, and J.~Li.
\newblock Glime: general, stable and local lime explanation.
\newblock \emph{Advances in Neural Information Processing Systems}, 36, 2024.

\bibitem[Tavar{\'e}(1986)]{tavare1986some}
S.~Tavar{\'e}.
\newblock Some probabilistic and statistical problems on the analysis of dna sequence.
\newblock \emph{Lecture of Mathematics for Life Science}, 17:\penalty0 57, 1986.

\bibitem[Tegally et~al.(2022)Tegally, Moir, Everatt, Giovanetti, Scheepers, Wilkinson, Subramoney, Makatini, Moyo, Amoako, Baxter, Althaus, Anyaneji, Kekana, Viana, Giandhari, Lessells, Maponga, Maruapula, Choga, Matshaba, Mbulawa, Msomi, Naidoo, Pillay, Sanko, San, Scott, Singh, Magini, Smith-Lawrence, Stevens, Dor, Tshiabuila, Wolter, Preiser, Treurnicht, Venter, Chiloane, McIntyre, O'Toole, Ruis, Peacock, Roemer, Kosakovsky~Pond, Williamson, Pybus, Bhiman, Glass, Martin, Jackson, Rambaut, Laguda-Akingba, Gaseitsiwe, von Gottberg, and de~Oliveira]{tegally_emergence_2022}
H.~Tegally, M.~Moir, J.~Everatt, M.~Giovanetti, C.~Scheepers, E.~Wilkinson, K.~Subramoney, Z.~Makatini, S.~Moyo, D.~G. Amoako, C.~Baxter, C.~L. Althaus, U.~J. Anyaneji, D.~Kekana, R.~Viana, J.~Giandhari, R.~J. Lessells, T.~Maponga, D.~Maruapula, W.~Choga, M.~Matshaba, M.~B. Mbulawa, N.~Msomi, Y.~Naidoo, S.~Pillay, T.~J. Sanko, J.~E. San, L.~Scott, L.~Singh, N.~A. Magini, P.~Smith-Lawrence, W.~Stevens, G.~Dor, D.~Tshiabuila, N.~Wolter, W.~Preiser, F.~K. Treurnicht, M.~Venter, G.~Chiloane, C.~McIntyre, A.~O'Toole, C.~Ruis, T.~P. Peacock, C.~Roemer, S.~L. Kosakovsky~Pond, C.~Williamson, O.~G. Pybus, J.~N. Bhiman, A.~Glass, D.~P. Martin, B.~Jackson, A.~Rambaut, O.~Laguda-Akingba, S.~Gaseitsiwe, A.~von Gottberg, and T.~de~Oliveira.
\newblock Emergence of {SARS}-{CoV}-2 {Omicron} lineages {BA}.4 and {BA}.5 in {South} {Africa}.
\newblock \emph{Nature Medicine}, 28\penalty0 (9):\penalty0 1785--1790, Sept. 2022.
\newblock ISSN 1546-170X.
\newblock \doi{10.1038/s41591-022-01911-2}.
\newblock URL \url{https://www.nature.com/articles/s41591-022-01911-2}.
\newblock Publisher: Nature Publishing Group.

\bibitem[Thompson et~al.(1994{\natexlab{a}})Thompson, Higgins, and Gibson]{Thompson1994}
J.~D. Thompson, D.~G. Higgins, and T.~J. Gibson.
\newblock Clustal w: improving the sensitivity of progressive multiple sequence alignment through sequence weighting, position-specific gap penalties and weight matrix choice.
\newblock \emph{Nucleic acids research}, 22\penalty0 (22):\penalty0 4673--4680, 1994{\natexlab{a}}.

\bibitem[Thompson et~al.(1994{\natexlab{b}})Thompson, Higgins, and Gibson]{thompson_clustal_1994}
J.~D. Thompson, D.~G. Higgins, and T.~J. Gibson.
\newblock {CLUSTAL} {W}: improving the sensitivity of progressive multiple sequence alignment through sequence weighting, position-specific gap penalties and weight matrix choice.
\newblock \emph{Nucleic acids research}, 22\penalty0 (22):\penalty0 4673--4680, 1994{\natexlab{b}}.
\newblock Publisher: Oxford University Press.

\bibitem[Thornton et~al.(2000)Thornton, Orengo, Todd, and Pearl]{Thornton2000}
J.~M. Thornton, C.~A. Orengo, A.~E. Todd, and F.~M. Pearl.
\newblock From sequence to function: methods and applications.
\newblock \emph{Current Opinion in Structural Biology}, 10\penalty0 (3):\penalty0 374--380, 2000.

\bibitem[Thrall et~al.(2011)Thrall, Oakeshott, Fitt, Southerton, Burdon, Sheppard, Russell, Zalucki, Heino, and Ford~Denison]{thrall_evolution_2011}
P.~H. Thrall, J.~G. Oakeshott, G.~Fitt, S.~Southerton, J.~J. Burdon, A.~Sheppard, R.~J. Russell, M.~Zalucki, M.~Heino, and R.~Ford~Denison.
\newblock Evolution in agriculture: the application of evolutionary approaches to the management of biotic interactions in agro-ecosystems.
\newblock \emph{Evolutionary Applications}, 4\penalty0 (2):\penalty0 200--215, 2011.
\newblock ISSN 1752-4571.
\newblock \doi{10.1111/j.1752-4571.2010.00179.x}.
\newblock URL \url{https://onlinelibrary.wiley.com/doi/abs/10.1111/j.1752-4571.2010.00179.x}.
\newblock \_eprint: https://onlinelibrary.wiley.com/doi/pdf/10.1111/j.1752-4571.2010.00179.x TLDR: Biotic interactions involving pests and pathogens are focused on as exemplars of situations where integration of agronomic, ecological and evolutionary perspectives has practical value and the use of predictive frameworks based on evolutionary models as pre emptive management tools are advocated.

\bibitem[Tian et~al.(2023)Tian, Zhong, Lin, Wei, and Hakonarson]{tian2023complex}
T.~Tian, C.~Zhong, X.~Lin, Z.~Wei, and H.~Hakonarson.
\newblock Complex hierarchical structures in single-cell genomics data unveiled by deep hyperbolic manifold learning.
\newblock \emph{Genome Research}, 33\penalty0 (2):\penalty0 232--246, 2023.

\bibitem[Tirosh et~al.(2016)Tirosh, Venteicher, Hebert, et~al.]{Tirosh2016}
I.~Tirosh, A.~S. Venteicher, C.~Hebert, et~al.
\newblock Single-cell rna-seq supports a developmental hierarchy in human oligodendroglioma.
\newblock \emph{Nature}, 539\penalty0 (7628):\penalty0 309--313, 2016.

\bibitem[Tisza et~al.(2020)Tisza, Pastrana, Welch, Stewart, Peretti, Starrett, Pang, Krishnamurthy, Pesavento, McDermott, Murphy, Whited, Miller, Brenchley, Rosshart, Rehermann, Doorbar, Ta'ala, Pletnikova, Troncoso, Resnick, Bolduc, Sullivan, Varsani, Segall, and Buck]{tisza_discovery_2020}
M.~J. Tisza, D.~V. Pastrana, N.~L. Welch, B.~Stewart, A.~Peretti, G.~J. Starrett, Y.-Y.~S. Pang, S.~R. Krishnamurthy, P.~A. Pesavento, D.~H. McDermott, P.~M. Murphy, J.~L. Whited, B.~Miller, J.~Brenchley, S.~P. Rosshart, B.~Rehermann, J.~Doorbar, B.~A. Ta'ala, O.~Pletnikova, J.~C. Troncoso, S.~M. Resnick, B.~Bolduc, M.~B. Sullivan, A.~Varsani, A.~M. Segall, and C.~B. Buck.
\newblock Discovery of several thousand highly diverse circular {DNA} viruses.
\newblock \emph{eLife}, 9:\penalty0 e51971, Feb. 2020.
\newblock ISSN 2050-084X.
\newblock \doi{10.7554/eLife.51971}.
\newblock URL \url{https://doi.org/10.7554/eLife.51971}.
\newblock Publisher: eLife Sciences Publications, Ltd.

\bibitem[Tran et~al.(2021)Tran, Nguyen, Tran, La~Vecchia, Luu, and Nguyen]{tran2021fast}
D.~Tran, H.~Nguyen, B.~Tran, C.~La~Vecchia, H.~N. Luu, and T.~Nguyen.
\newblock Fast and precise single-cell data analysis using a hierarchical autoencoder.
\newblock \emph{Nature communications}, 12\penalty0 (1):\penalty0 1029, 2021.

\bibitem[Trapnell et~al.(2014{\natexlab{a}})Trapnell, Cacchiarelli, Grimsby, Pokharel, Li, Morse, Lennon, Livak, Mikkelsen, and Rinn]{trapnell2014dynamics}
C.~Trapnell, D.~Cacchiarelli, J.~Grimsby, P.~Pokharel, S.~Li, M.~Morse, N.~J. Lennon, K.~J. Livak, T.~S. Mikkelsen, and J.~L. Rinn.
\newblock The dynamics and regulators of cell fate decisions are revealed by pseudotemporal ordering of single cells.
\newblock \emph{Nature biotechnology}, 32\penalty0 (4):\penalty0 381--386, 2014{\natexlab{a}}.

\bibitem[Trapnell et~al.(2014{\natexlab{b}})Trapnell, Cacchiarelli, Grimsby, Pokharel, Li, Morse, Lennon, Livak, Mikkelsen, and Rinn]{trapnell_dynamics_2014}
C.~Trapnell, D.~Cacchiarelli, J.~Grimsby, P.~Pokharel, S.~Li, M.~Morse, N.~J. Lennon, K.~J. Livak, T.~S. Mikkelsen, and J.~L. Rinn.
\newblock The dynamics and regulators of cell fate decisions are revealed by pseudotemporal ordering of single cells.
\newblock \emph{Nat Biotechnol}, 32\penalty0 (4):\penalty0 381--386, Apr. 2014{\natexlab{b}}.
\newblock ISSN 1546-1696.
\newblock \doi{10.1038/nbt.2859}.
\newblock URL \url{https://www.nature.com/articles/nbt.2859}.

\bibitem[Trapnell et~al.(2014{\natexlab{c}})Trapnell, Cacchiarelli, Grimsby, Pokharel, Li, Morse, Lennon, Livak, Mikkelsen, and Rinn]{Trapnell2014}
C.~Trapnell, D.~Cacchiarelli, J.~Grimsby, P.~Pokharel, S.-R. Li, M.~Morse, N.~J. Lennon, K.~J. Livak, T.~S. Mikkelsen, and J.~L. Rinn.
\newblock The dynamics and regulators of cell fate decisions are revealed by pseudotemporal ordering of single cells.
\newblock \emph{Nature Biotechnology}, 32\penalty0 (4):\penalty0 381--386, 2014{\natexlab{c}}.

\bibitem[Tsai et~al.(2019)]{tsai2019multimodal}
Y.-H.~H. Tsai et~al.
\newblock Multimodal transformer for unaligned multimodal language sequences.
\newblock In \emph{Proceedings of the 57th Annual Meeting of the Association for Computational Linguistics}, pages 6558--6569, 2019.

\bibitem[van Kempen et~al.(2023)van Kempen, Kim, Tumescheit, Mirdita, Lee, Gilchrist, Söding, and Steinegger]{van_kempen_fast_2023}
M.~van Kempen, S.~S. Kim, C.~Tumescheit, M.~Mirdita, J.~Lee, C.~L. Gilchrist, J.~Söding, and M.~Steinegger.
\newblock Fast and accurate protein structure search with {Foldseek}.
\newblock \emph{Nature Biotechnology}, 42\penalty0 (February), 2023.
\newblock ISSN 15461696.
\newblock \doi{10.1038/s41587-023-01773-0}.
\newblock Publisher: Springer US.

\bibitem[Vandenhirtz et~al.(2023)Vandenhirtz, Barkmann, Manduchi, Vogt, and Boeva]{vandenhirtz_sctree_2024}
M.~Vandenhirtz, F.~Barkmann, L.~Manduchi, J.~E. Vogt, and V.~Boeva.
\newblock sctree: Discovering cellular hierarchies in the presence of batch effects in scrna-seq data.
\newblock \emph{arXiv preprint arXiv:2304.12345}, 2023.

\bibitem[Vazquez et~al.(2022)Vazquez, Pena, Muhammad, Kraft, Adams, and Lynch]{vazquez_parallel_2022}
J.~M. Vazquez, M.~T. Pena, B.~Muhammad, M.~Kraft, L.~B. Adams, and V.~J. Lynch.
\newblock Parallel evolution of reduced cancer risk and tumor suppressor duplications in {Xenarthra}.
\newblock \emph{eLife}, 11:\penalty0 e82558, Dec. 2022.
\newblock ISSN 2050-084X.
\newblock \doi{10.7554/eLife.82558}.
\newblock URL \url{https://doi.org/10.7554/eLife.82558}.
\newblock Publisher: eLife Sciences Publications, Ltd.

\bibitem[Wagner et~al.(2020)Wagner, Regev, and Yosef]{Wagner2020}
A.~Wagner, A.~Regev, and N.~Yosef.
\newblock Revealing the vectors of cellular identity with single-cell genomics.
\newblock \emph{Nature Biotechnology}, 38\penalty0 (12):\penalty0 1401--1414, 2020.
\newblock \doi{10.1038/s41587-020-00710-4}.

\bibitem[Waits and Paetkau(2005)]{waits2005noninvasive}
L.~P. Waits and D.~Paetkau.
\newblock Noninvasive genetic sampling tools for wildlife biologists: a review of applications and recommendations for accurate data collection.
\newblock \emph{The Journal of Wildlife Management}, 69\penalty0 (4):\penalty0 1419--1433, 2005.

\bibitem[Wang et~al.(2022)Wang, Hu, Zhang, Li, and Yu]{wang_hierarchical_2022}
B.~Wang, X.~Hu, C.~Zhang, P.~Li, and P.~S. Yu.
\newblock Hierarchical {GAN}-{Tree} and {Bi}-{Directional} {Capsules} for multi-label image classification.
\newblock \emph{Knowledge-Based Systems}, 238:\penalty0 107882, Feb. 2022.
\newblock ISSN 0950-7051.
\newblock \doi{10.1016/j.knosys.2021.107882}.
\newblock URL \url{https://www.sciencedirect.com/science/article/pii/S0950705121010510}.

\bibitem[Wang and Gu(2018)]{wang2018vasc}
D.~Wang and J.~Gu.
\newblock Vasc: dimension reduction and visualization of single-cell rna-seq data by deep variational autoencoder.
\newblock \emph{Genomics, Proteomics and Bioinformatics}, 16\penalty0 (5):\penalty0 320--331, 2018.

\bibitem[Wang et~al.(2023{\natexlab{a}})Wang, Chitsaz, Derbyshire, Gonzales, Gwadz, Lu, Marchler, Song, Thanki, Yamashita, et~al.]{wang2023conserved}
J.~Wang, F.~Chitsaz, M.~K. Derbyshire, N.~R. Gonzales, M.~Gwadz, S.~Lu, G.~H. Marchler, J.~S. Song, N.~Thanki, R.~A. Yamashita, et~al.
\newblock The conserved domain database in 2023.
\newblock \emph{Nucleic Acids Research}, 51\penalty0 (D1):\penalty0 D384--D388, 2023{\natexlab{a}}.

\bibitem[Wang et~al.(2024)Wang, Hou, Wang, Zhai, Lu, Zi, Zhai, He, Curtis, Zhou, and Hu]{wang_phylovelo_2024}
K.~Wang, L.~Hou, X.~Wang, X.~Zhai, Z.~Lu, Z.~Zi, W.~Zhai, X.~He, C.~Curtis, D.~Zhou, and Z.~Hu.
\newblock {PhyloVelo} enhances transcriptomic velocity field mapping using monotonically expressed genes.
\newblock \emph{Nature Biotechnology}, 42\penalty0 (5):\penalty0 778--789, May 2024.
\newblock ISSN 1546-1696.
\newblock \doi{10.1038/s41587-023-01887-5}.
\newblock URL \url{https://www.nature.com/articles/s41587-023-01887-5}.
\newblock Publisher: Nature Publishing Group TLDR: Applying PhyloVelo to seven lineage-traced scRNA-seq datasets, generated using CRISPR-Cas9 editing, lentiviral barcoding or immune repertoire profiling, demonstrates its high accuracy and robustness in inferring complex lineage trajectories while outperforming RNA velocity.

\bibitem[Wang et~al.(2023{\natexlab{b}})Wang, Zhang, Khodaverdian, and Yosef]{wang2023theoretical}
R.~Wang, R.~Zhang, A.~Khodaverdian, and N.~Yosef.
\newblock Theoretical guarantees for phylogeny inference from single-cell lineage tracing.
\newblock \emph{Proceedings of the National Academy of Sciences}, 120\penalty0 (12):\penalty0 e2203352120, 2023{\natexlab{b}}.

\bibitem[Wang et~al.(2013)Wang, Ma, Peng, and Xu]{wang_protein_2013}
S.~Wang, J.~Ma, J.~Peng, and J.~Xu.
\newblock Protein structure alignment beyond spatial proximity.
\newblock \emph{Scientific Reports}, 3\penalty0 (1):\penalty0 1448, Mar. 2013.
\newblock ISSN 2045-2322.
\newblock \doi{10.1038/srep01448}.
\newblock URL \url{https://www.nature.com/articles/srep01448}.
\newblock Publisher: Nature Publishing Group TLDR: Experimental results show that DeepAlign can generate structure alignments much more consistent with manually-curated alignments than other automatic tools especially when proteins under consideration are remote homologs, implying that in addition to geometric similarity, evolutionary information and hydrogen-bonding similarity are essential to aligning two protein structures.

\bibitem[Wang et~al.(2019)Wang, Karikomi, MacLean, and Nie]{wang_cell_2019}
S.~Wang, M.~Karikomi, A.~L. MacLean, and Q.~Nie.
\newblock Cell lineage and communication network inference via optimization for single-cell transcriptomics.
\newblock \emph{Nucleic Acids Research}, 47\penalty0 (11):\penalty0 e66, June 2019.
\newblock ISSN 0305-1048.
\newblock \doi{10.1093/nar/gkz204}.
\newblock URL \url{https://doi.org/10.1093/nar/gkz204}.

\bibitem[Wang et~al.(2021)Wang, Guan, Zhang, Liu, Wang, Fan, Du, Yan, Zhang, Chen, et~al.]{wang2021structural}
Y.~Wang, X.~Guan, S.~Zhang, Y.~Liu, S.~Wang, P.~Fan, X.~Du, S.~Yan, P.~Zhang, H.-Y. Chen, et~al.
\newblock Structural-profiling of low molecular weight rnas by nanopore trapping/translocation using mycobacterium smegmatis porin a.
\newblock \emph{Nature communications}, 12\penalty0 (1):\penalty0 3368, 2021.

\bibitem[Webb and Sali(2016)]{webb_comparative_nodate_2016}
B.~Webb and A.~Sali.
\newblock Comparative protein structure modeling using modeller.
\newblock \emph{Current protocols in bioinformatics}, 54\penalty0 (1):\penalty0 5--6, 2016.

\bibitem[Wei and Mei(2024)]{wei2024towards}
F.~Wei and K.~Mei.
\newblock Towards self-explainable graph convolutional neural network with frequency adaptive inception.
\newblock \emph{Pattern Recognition}, 146:\penalty0 109991, 2024.

\bibitem[Wiens(2006)]{wiens2006missing}
J.~J. Wiens.
\newblock Missing data and the design of phylogenetic analyses.
\newblock \emph{Journal of biomedical informatics}, 39\penalty0 (1):\penalty0 34--42, 2006.

\bibitem[Wolf et~al.(2018)Wolf, Angerer, and Theis]{wolf2018scanpy}
F.~A. Wolf, P.~Angerer, and F.~J. Theis.
\newblock Scanpy: large-scale single-cell gene expression data analysis.
\newblock \emph{Genome biology}, 19:\penalty0 1--5, 2018.

\bibitem[Wolf et~al.(2019)Wolf, Hamey, Plass, Solana, Dahlin, Göttgens, Rajewsky, Simon, and Theis]{wolf_paga_2019}
F.~A. Wolf, F.~K. Hamey, M.~Plass, J.~Solana, J.~S. Dahlin, B.~Göttgens, N.~Rajewsky, L.~Simon, and F.~J. Theis.
\newblock {PAGA}: graph abstraction reconciles clustering with trajectory inference through a topology preserving map of single cells.
\newblock \emph{Genome Biology}, 20\penalty0 (1):\penalty0 59, Mar. 2019.
\newblock ISSN 1474-760X.
\newblock \doi{10.1186/s13059-019-1663-x}.
\newblock URL \url{https://doi.org/10.1186/s13059-019-1663-x}.

\bibitem[Wolf et~al.(2022)Wolf, Cowley, Breister, Matatov, Lucio, Polak, Ridlon, Gaskins, and Anantharaman]{wolf_diversity_2022}
P.~G. Wolf, E.~S. Cowley, A.~Breister, S.~Matatov, L.~Lucio, P.~Polak, J.~M. Ridlon, H.~R. Gaskins, and K.~Anantharaman.
\newblock Diversity and distribution of sulfur metabolic genes in the human gut microbiome and their association with colorectal cancer.
\newblock \emph{Microbiome}, 10\penalty0 (1):\penalty0 64, Apr. 2022.
\newblock ISSN 2049-2618.
\newblock \doi{10.1186/s40168-022-01242-x}.
\newblock URL \url{https://doi.org/10.1186/s40168-022-01242-x}.

\bibitem[Xia et~al.(2019)Xia, Fan, Emanuel, Hao, and Zhuang]{xia2019spatial}
C.~Xia, J.~Fan, G.~Emanuel, J.~Hao, and X.~Zhuang.
\newblock Spatial transcriptomics creates a multi-omic atlas of human disease.
\newblock \emph{Nature Biotechnology}, 37\penalty0 (10):\penalty0 1088--1094, 2019.
\newblock \doi{10.1038/s41587-019-0236-z}.

\bibitem[Xie and Zhang(2023)]{xie_artree_2023}
T.~Xie and C.~Zhang.
\newblock {ARTree}: {A} {Deep} {Autoregressive} {Model} for {Phylogenetic} {Inference}.
\newblock In \emph{Advances in Neural Information Processing Systems}, Nov. 2023.
\newblock URL \url{https://openreview.net/forum?id=SoLebIqHgZ}.

\bibitem[Xie et~al.(2024)Xie, Matsen~IV, Suchard, and Zhang]{xie_variational_2024}
T.~Xie, F.~A. Matsen~IV, M.~A. Suchard, and C.~Zhang.
\newblock Variational {Bayesian} {Phylogenetic} {Inference} with {Semi}-implicit {Branch} {Length} {Distributions}, Aug. 2024.
\newblock URL \url{http://arxiv.org/abs/2408.05058}.
\newblock arXiv:2408.05058 [cs, stat].

\bibitem[Xu et~al.(2023)Xu, He, Wang, Zhu, and Ding]{xu2023dm}
G.~Xu, C.~He, H.~Wang, H.~Zhu, and W.~Ding.
\newblock Dm-fusion: Deep model-driven network for heterogeneous image fusion.
\newblock \emph{IEEE transactions on neural networks and learning systems}, 2023.

\bibitem[Yamamoto et~al.(2023)Yamamoto, Zhang, and Mizushima]{yamamoto2023autophagy}
H.~Yamamoto, S.~Zhang, and N.~Mizushima.
\newblock Autophagy genes in biology and disease.
\newblock \emph{Nature Reviews Genetics}, 24\penalty0 (6):\penalty0 382--400, 2023.

\bibitem[Yeung et~al.(2023)Yeung, Zhou, Mathew, Gravel, Taujale, O'Boyle, Salcedo, Venkat, Lanzilotta, Li, and Kannan]{yeung_tree_2023}
W.~Yeung, Z.~Zhou, L.~Mathew, N.~Gravel, R.~Taujale, B.~O'Boyle, M.~Salcedo, A.~Venkat, W.~Lanzilotta, S.~Li, and N.~Kannan.
\newblock Tree visualizations of protein sequence embedding space enable improved functional clustering of diverse protein superfamilies.
\newblock \emph{Briefings in Bioinformatics}, 24\penalty0 (1):\penalty0 bbac619, Jan. 2023.
\newblock ISSN 1467-5463, 1477-4054.
\newblock \doi{10.1093/bib/bbac619}.
\newblock URL \url{https://academic.oup.com/bib/article/doi/10.1093/bib/bbac619/6987820}.
\newblock TLDR: This work develops workflows and visualization methods for the classification of protein families using sequence embedding derived from protein language models and proposes a new hierarchical classification for the S-Adenosyl-L-Methionine enzyme superfamily which has been difficult to classify using traditional alignment-based approaches.

\bibitem[You et~al.(2018{\natexlab{a}})You, Liu, Ying, Pande, and Leskovec]{you2018graph}
J.~You, B.~Liu, R.~Ying, V.~Pande, and J.~Leskovec.
\newblock Graph convolutional policy network for goal-directed molecular graph generation.
\newblock In \emph{Advances in Neural Information Processing Systems (NeurIPS)}, 2018{\natexlab{a}}.

\bibitem[You et~al.(2018{\natexlab{b}})You, Ying, Ren, Hamilton, and Leskovec]{you2018graphrnn}
J.~You, R.~Ying, X.~Ren, W.~Hamilton, and J.~Leskovec.
\newblock Graphrnn: Generating realistic graphs with deep auto-regressive models.
\newblock In \emph{Proceedings of the 35th International Conference on Machine Learning (ICML)}, 2018{\natexlab{b}}.

\bibitem[Yuan et~al.(2023)Yuan, Zheng, Li, Ma, Shu, Qu, Ye, Li, Tang, and Chen]{yuan_chromosome-level_2023}
R.~Yuan, B.~Zheng, Z.~Li, X.~Ma, X.~Shu, Q.~Qu, X.~Ye, S.~Li, P.~Tang, and X.~Chen.
\newblock The chromosome-level genome of {Chinese} praying mantis {Tenodera} sinensis ({Mantodea}: {Mantidae}) reveals its biology as a predator.
\newblock \emph{GigaScience}, 12:\penalty0 giad090, Jan. 2023.
\newblock ISSN 2047-217X.
\newblock \doi{10.1093/gigascience/giad090}.
\newblock URL \url{https://doi.org/10.1093/gigascience/giad090}.
\newblock TLDR: The high-quality genome assembly of the praying mantis provides a valuable repository for studying the evolutionary patterns of the mantis genomes and the gene expression profiles of insect predators.

\bibitem[Zang et~al.(2019)Zang, Wang, Song, Lu, Li, Wang, and Zhao]{zang_hybrid_2019}
Z.~Zang, W.~Wang, Y.~Song, L.~Lu, W.~Li, Y.~Wang, and Y.~Zhao.
\newblock Hybrid {Deep} {Neural} {Network} {Scheduler} for {Job}-{Shop} {Problem} {Based} on {Convolution} {Two}-{Dimensional} {Transformation}.
\newblock \emph{Computational Intelligence and Neuroscience}, 2019\penalty0 (Research Article):\penalty0 1--19, 2019.
\newblock ISSN 1687-5265.
\newblock \doi{10.1155/2019/7172842}.
\newblock TLDR: A hybrid deep neural network scheduler (HDNNS) is proposed to solve job-shop scheduling problems (JSSPs) and the results show that the MAKESPAN index of HDNNS is 9\% better than that of HNN and the index is also 4\% betterthan that of ANN in ZLP dataset.

\bibitem[Zang et~al.(2021)Zang, Li, Wu, Guo, Xu, and Li]{zang_unsupervised_2021}
Z.~Zang, S.~Li, D.~Wu, J.~Guo, Y.~Xu, and S.~Z. Li.
\newblock Unsupervised {Deep} {Manifold} {Attributed} {Graph} {Embedding}.
\newblock \emph{arXiv:2104.13048 [cs]}, Apr. 2021.
\newblock URL \url{http://arxiv.org/abs/2104.13048}.
\newblock arXiv: 2104.13048 version: 1.

\bibitem[Zang et~al.(2022{\natexlab{a}})Zang, Cheng, Xia, Li, Sun, Xu, Shang, Sun, and Li]{zang2022dmt}
Z.~Zang, S.~Cheng, H.~Xia, L.~Li, Y.~Sun, Y.~Xu, L.~Shang, B.~Sun, and S.~Z. Li.
\newblock Dmt-ev: An explainable deep network for dimension reduction.
\newblock \emph{IEEE Transactions on Visualization and Computer Graphics}, 30\penalty0 (3):\penalty0 1710--1727, 2022{\natexlab{a}}.

\bibitem[Zang et~al.(2022{\natexlab{b}})Zang, Li, Wu, Wang, Wang, Shang, Sun, Li, and Li]{zang_dlme_2022}
Z.~Zang, S.~Li, D.~Wu, G.~Wang, K.~Wang, L.~Shang, B.~Sun, H.~Li, and S.~Z. Li.
\newblock Dlme: {Deep} local-flatness manifold embedding.
\newblock pages 576--592. Springer, Cham, 2022{\natexlab{b}}.

\bibitem[Zang et~al.(2023{\natexlab{a}})Zang, Shang, Yang, Wang, Sun, Xie, and Li]{zang_boosting_2023}
Z.~Zang, L.~Shang, S.~Yang, F.~Wang, B.~Sun, X.~Xie, and S.~Z. Li.
\newblock Boosting {Novel} {Category} {Discovery} {Over} {Domains} with {Soft} {Contrastive} {Learning} and {All} in {One} {Classifier}.
\newblock pages 11824--11833. IEEE Computer Society, Oct. 2023{\natexlab{a}}.
\newblock ISBN 9798350307184.
\newblock \doi{10.1109/ICCV51070.2023.01089}.
\newblock URL \url{https://www.computer.org/csdl/proceedings-article/iccv/2023/071800l1824/1TJgmnlAVGw}.
\newblock TLDR: A framework named Soft-contrastive All-in-one Network (SAN) is proposed for ODA and UNDA tasks, which includes a novel data-augmentation-based soft contrastive learning (SCL) loss to fine-tune the backbone for feature transfer and a more human-intuitive classifier to improve new class discovery capability.

\bibitem[Zang et~al.(2023{\natexlab{b}})Zang, Xu, Lu, Geng, Yang, and Li]{zang2023udrn}
Z.~Zang, Y.~Xu, L.~Lu, Y.~Geng, S.~Yang, and S.~Z. Li.
\newblock Udrn: unified dimensional reduction neural network for feature selection and feature projection.
\newblock \emph{Neural Networks}, 161:\penalty0 626--637, 2023{\natexlab{b}}.

\bibitem[Zang et~al.(2023{\natexlab{c}})Zang, Xu, Lu, Geng, Yang, and Li]{zang_udrn_2023}
Z.~Zang, Y.~Xu, L.~Lu, Y.~Geng, S.~Yang, and S.~Z. Li.
\newblock Udrn: unified dimensional reduction neural network for feature selection and feature projection.
\newblock \emph{Neural Networks}, 161:\penalty0 626--637, 2023{\natexlab{c}}.
\newblock ISSN 0893-6080.
\newblock Publisher: Pergamon.

\bibitem[Zang et~al.(2024{\natexlab{a}})Zang, Cheng, Xia, Li, Sun, Xu, Shang, Sun, and Li]{zang_dmt-ev_2024}
Z.~Zang, S.~Cheng, H.~Xia, L.~Li, Y.~Sun, Y.~Xu, L.~Shang, B.~Sun, and S.~Z. Li.
\newblock {DMT}-{EV}: {An} {Explainable} {Deep} {Network} for {Dimension} {Reduction}.
\newblock \emph{IEEE transactions on visualization and computer graphics}, 30\penalty0 (3):\penalty0 1710--1727, Mar. 2024{\natexlab{a}}.
\newblock ISSN 1941-0506.
\newblock \doi{10.1109/TVCG.2022.3223399}.
\newblock TLDR: A deep neural network method called DMT-EV is developed, which provides not only excellent performance in structural maintainability but also explainability to the DR therein, and consistently outperforms the state-of-the-art methods in both performance measures and explainability.

\bibitem[Zang et~al.(2024{\natexlab{b}})Zang, Luo, Wang, Zhang, Wang, Li, and You]{zang_diffaug_2024}
Z.~Zang, H.~Luo, K.~Wang, P.~Zhang, F.~Wang, S.~Z. Li, and Y.~You.
\newblock {DiffAug}: {Enhance} {Unsupervised} {Contrastive} {Learning} with {Domain}-{Knowledge}-{Free} {Diffusion}-based {Data} {Augmentation}.
\newblock In \emph{International Conference on Machine Learning}, June 2024{\natexlab{b}}.
\newblock URL \url{https://openreview.net/forum?id=s0UDX7Kswl}.

\bibitem[Zhang(2020)]{zhang_improved_2020}
C.~Zhang.
\newblock Improved {Variational} {Bayesian} {Phylogenetic} {Inference} with {Normalizing} {Flows}.
\newblock In \emph{Advances in {Neural} {Information} {Processing} {Systems}}, volume~33, pages 18760--18771. Curran Associates, Inc., 2020.
\newblock URL \url{https://proceedings.neurips.cc/paper/2020/hash/d96409bf894217686ba124d7356686c9-Abstract.html}.

\bibitem[Zhang and Iv(2018)]{zhang_variational_2018}
C.~Zhang and F.~A.~M. Iv.
\newblock Variational {Bayesian} {Phylogenetic} {Inference}.
\newblock In \emph{Advances in Neural Information Processing Systems}, Sept. 2018.
\newblock URL \url{https://openreview.net/forum?id=SJVmjjR9FX}.

\bibitem[Zhang et~al.(2018)Zhang, Fu, Hu, Cao, Liu, and Tian]{zhang2018multi}
C.~Zhang, H.~Fu, Q.~Hu, X.~Cao, Q.~Liu, and Q.~Tian.
\newblock Multi-view multiple clusterings via deep matrix factorization.
\newblock \emph{IEEE Transactions on Pattern Analysis and Machine Intelligence}, 41\penalty0 (1):\penalty0 90--103, 2018.

\bibitem[Zhang et~al.(2024)Zhang, Tan, Tang, Li, Zhang, Sun, Guo, Gao, Cai, Sun, Wang, Fu, Ma, Wu, Hu, Zhang, Gee, Yan, Zhao, Chen, Guo, Wang, and Zhang]{zhang_heterologous_2024}
T.~Zhang, S.~Tan, N.~Tang, Y.~Li, C.~Zhang, J.~Sun, Y.~Guo, H.~Gao, Y.~Cai, W.~Sun, C.~Wang, L.~Fu, H.~Ma, Y.~Wu, X.~Hu, X.~Zhang, P.~Gee, W.~Yan, Y.~Zhao, Q.~Chen, B.~Guo, H.~Wang, and Y.~E. Zhang.
\newblock Heterologous survey of 130 {DNA} transposons in human cells highlights their functional divergence and expands the genome engineering toolbox.
\newblock \emph{Cell}, 187\penalty0 (14):\penalty0 3741--3760.e30, July 2024.
\newblock ISSN 0092-8674, 1097-4172.
\newblock \doi{10.1016/j.cell.2024.05.007}.
\newblock URL \url{https://www.cell.com/cell/abstract/S0092-8674(24)00516-6}.
\newblock Publisher: Elsevier TLDR: It is found that the Tc1/mariner superfamily exhibits elevated activity, potentially explaining their pervasive horizontal transfers and highlights the varied transposition features and evolutionary dynamics of DNA TEs and increases the TE toolbox diversity.

\bibitem[Zhang and Skolnick(2005)]{zhang_tm-align_2005}
Y.~Zhang and J.~Skolnick.
\newblock {TM}-align: a protein structure alignment algorithm based on the {TM}-score.
\newblock \emph{Nucleic Acids Research}, 33\penalty0 (7):\penalty0 2302--2309, 2005.
\newblock Publisher: Oxford University Press.

\bibitem[Zhang et~al.(2023)Zhang, Tran, Nguyen, Dascalu, and Harris]{zhang_robust_2023}
Y.~Zhang, D.~Tran, T.~Nguyen, S.~M. Dascalu, and F.~C. Harris.
\newblock A robust and accurate single-cell data trajectory inference method using ensemble pseudotime.
\newblock \emph{BMC Bioinformatics}, 24\penalty0 (1):\penalty0 55, Feb. 2023.
\newblock ISSN 1471-2105.
\newblock \doi{10.1186/s12859-023-05179-2}.
\newblock URL \url{https://doi.org/10.1186/s12859-023-05179-2}.

\bibitem[Zhao et~al.(2023)Zhao, Sun, and Liu]{zhao2023understanding}
A.~Zhao, J.~Sun, and Y.~Liu.
\newblock Understanding bacterial biofilms: From definition to treatment strategies.
\newblock \emph{Frontiers in cellular and infection microbiology}, 13:\penalty0 1137947, 2023.

\bibitem[Zheng et~al.(2017{\natexlab{a}})]{TenXGenomics}
G.-C. Zheng et~al.
\newblock Massively parallel digital transcriptional profiling of single cells.
\newblock \emph{Nature Communications}, 8\penalty0 (1):\penalty0 14049, 2017{\natexlab{a}}.

\bibitem[Zheng et~al.(2017{\natexlab{b}})Zheng, Terry, Belgrader, Ryvkin, Bent, Wilson, Ziraldo, Wheeler, McDermott, Zhu, et~al.]{Zheng2017}
G.~X. Zheng, J.~M. Terry, P.~Belgrader, P.~Ryvkin, Z.~W. Bent, R.~Wilson, S.~B. Ziraldo, T.~D. Wheeler, G.~P. McDermott, J.~Zhu, et~al.
\newblock Massively parallel digital transcriptional profiling of single cells.
\newblock \emph{Nature Communications}, 8\penalty0 (1):\penalty0 14049, 2017{\natexlab{b}}.

\bibitem[Zheng et~al.(2024)Zheng, Liu, Yang, Dong, Zhang, Tian, Yu, Ren, Hou, Zhu, et~al.]{zheng2024multi}
Y.~Zheng, Y.~Liu, J.~Yang, L.~Dong, R.~Zhang, S.~Tian, Y.~Yu, L.~Ren, W.~Hou, F.~Zhu, et~al.
\newblock Multi-omics data integration using ratio-based quantitative profiling with quartet reference materials.
\newblock \emph{Nature biotechnology}, 42\penalty0 (7):\penalty0 1133--1149, 2024.

\bibitem[Zhou et~al.(2023{\natexlab{a}})Zhou, Yan, Layne, Malkin, Zhang, Jain, Blanchette, and Bengio]{zhou2023phylogfn}
M.~Zhou, Z.~Yan, E.~Layne, N.~Malkin, D.~Zhang, M.~Jain, M.~Blanchette, and Y.~Bengio.
\newblock Phylogfn: Phylogenetic inference with generative flow networks.
\newblock \emph{arXiv preprint arXiv:2310.08774}, 2023{\natexlab{a}}.

\bibitem[Zhou et~al.(2023{\natexlab{b}})Zhou, Yan, Layne, Malkin, Zhang, Jain, Blanchette, and Bengio]{zhou_phylogfn_2023}
M.~Y. Zhou, Z.~Yan, E.~Layne, N.~Malkin, D.~Zhang, M.~Jain, M.~Blanchette, and Y.~Bengio.
\newblock {PhyloGFN}: {Phylogenetic} inference with generative flow networks.
\newblock In \emph{The Twelfth International Conference on Learning Representations}, Oct. 2023{\natexlab{b}}.
\newblock URL \url{https://openreview.net/forum?id=hB7SlfEmze}.

\bibitem[Zhu et~al.(2023)Zhu, Wu, Li, Fang, Zhang, Chen, Chen, Cheng, Zhu, Wu, Li, Fang, Zhang, Chen, Chen, and Cheng]{zhu_genome-wide_2023}
L.~Zhu, J.~Wu, M.~Li, H.~Fang, J.~Zhang, Y.~Chen, J.~Chen, T.~Cheng, L.~Zhu, J.~Wu, M.~Li, H.~Fang, J.~Zhang, Y.~Chen, J.~Chen, and T.~Cheng.
\newblock Genome-wide discovery of {CBL} genes in \textit{{Nitraria} tangutorum} {Bobr}. and functional analysis of \textit{{NtCBL1}-1} under drought and salt stress.
\newblock \emph{Forestry Research}, 3\penalty0 (1), Dec. 2023.
\newblock ISSN 2767-3812.
\newblock \doi{10.48130/FR-2023-0028}.
\newblock URL \url{https://www.maxapress.com/article/doi/10.48130/FR-2023-0028}.
\newblock Bandiera\_abtest: a Cc\_license\_type: cc\_by Cg\_type: Maximum Academic Press Number: FR-2023-0028 Primary\_atype: Forestry Research Publisher: Maximum Academic Press Subject\_term: ARTICLE Subject\_term\_id: ARTICLE.

\end{thebibliography}


\begin{thebibliography}{6}
\providecommand{\natexlab}[1]{#1}
\providecommand{\url}[1]{\texttt{#1}}
\expandafter\ifx\csname urlstyle\endcsname\relax
  \providecommand{\doi}[1]{doi: #1}\else
  \providecommand{\doi}{doi: \begingroup \urlstyle{rm}\Url}\fi

\bibitem[Huang et~al.(2020)Huang, Chaudhary, and Garmire]{huang2020fusion}
S.~Huang, K.~Chaudhary, and L.~X. Garmire.
\newblock Fusion of multi-omics data and deep learning for cancer patient survivability prediction.
\newblock \emph{Methods}, 166:\penalty0 28--37, 2020.

\bibitem[Kingma and Welling(2013)]{kingma2013auto}
D.~P. Kingma and M.~Welling.
\newblock Auto-encoding variational bayes.
\newblock \emph{arXiv preprint arXiv:1312.6114}, 2013.

\bibitem[Meng et~al.(2019)Meng, Jin, Wang, and Guo]{meng2019gene}
C.~Meng, S.~Jin, L.~Wang, and F.~Guo.
\newblock Gene ontology-based transfer learning for gene function prediction.
\newblock \emph{IEEE Access}, 7:\penalty0 54995--55007, 2019.

\bibitem[Rao et~al.(2021)Rao, Barkley, Franca, and Yanai]{rao2021deep}
A.~Rao, D.~Barkley, G.~S. Franca, and I.~Yanai.
\newblock Deep learning for spatially resolved data in single-cell omics.
\newblock \emph{Annual Review of Biomedical Data Science}, 4:\penalty0 123--142, 2021.

\bibitem[Tsai et~al.(2019)]{tsai2019multimodal}
Y.-H.~H. Tsai et~al.
\newblock Multimodal transformer for unaligned multimodal language sequences.
\newblock In \emph{Proceedings of the 57th Annual Meeting of the Association for Computational Linguistics}, pages 6558--6569, 2019.

\bibitem[Zhang et~al.(2018)Zhang, Fu, Hu, Cao, Liu, and Tian]{zhang2018multi}
C.~Zhang, H.~Fu, Q.~Hu, X.~Cao, Q.~Liu, and Q.~Tian.
\newblock Multi-view multiple clusterings via deep matrix factorization.
\newblock \emph{IEEE Transactions on Pattern Analysis and Machine Intelligence}, 41\penalty0 (1):\penalty0 90--103, 2018.

\end{thebibliography}
  }

\end{document}